\documentclass[twocolumn,showpacs,preprintnumbers,nofootinbib,prd,superscriptaddress,10pt]{revtex4-1}
\usepackage{amsmath,amssymb}
\usepackage[normalem]{ulem}
\usepackage{textcomp}
\usepackage{hyperref}
\usepackage{bm}
\usepackage{graphicx}
\usepackage{psfrag}
\usepackage[usenames,dvipsnames]{xcolor}
\usepackage[utf8]{inputenc}
\usepackage{multirow}
\usepackage{tabularx}

\graphicspath{{./figures/}}
\allowdisplaybreaks[4]

\newcolumntype{C}[1]{>{\centering\arraybackslash}p{#1}}

\begin{document}

\title{Nonlinear interacting cosmological models after Planck 2018 legacy release and the $H_0$ tension}

\author{Supriya Pan}
\email{supriya.maths@presiuniv.ac.in}
\affiliation{Department of Mathematics, Presidency University, 86/1 College Street, Kolkata 700073, India}

\author{Weiqiang Yang}
\email{d11102004@163.com}
\affiliation{Department of Physics, Liaoning Normal University, Dalian, 116029, P. R.  China}

\author{Andronikos Paliathanasis}
\email{anpaliat@phys.uoa.gr}
\affiliation{Institute of Systems Science, Durban University of Technology, PO Box 1334,
Durban 4000, Republic of South Africa}

\begin{abstract}
Interacting dark energy models are widely renowned for giving an explanation to the cosmic coincidence problem as well as several observational issues. 
According to the recent observational data, and so far we are concerned with the literature, the choice of the interaction function
between dark matter and dark energy is always questionable since there is no such
underlying theory that could derive it. Thus, in this work we have raised
this issue by proposing two new nonlinear interaction functions and
constrain them using cosmic microwave background (CMB) from Planck 2018, baryon acoustic oscillations (BAO), dark energy survey and a measurement of the Hubble constant $H_0$ from Hubble Space Telescope (HST) 2019. The dark
energy equation of state is considered to be constant throughout the work
and the geometry of the universe is assumed to be homogeneous and isotropic
with zero spatial curvature. Our analyses report that a non-zero
interaction is always allowed by the observational data and 
the dark energy equation of state is bent towards the phantom
regime. In particular, when $H_0$ from HST is added to Planck 2018+BAO, we find an evidence for a non-zero coupling at more than $2\sigma$ confidence level.
Our analyses also report that for both the models, $H_0$ is close to its local measurements and thus alleviating the $H_0$ tension. In particular, one of the interacting models perfectly  solves the $H_0$ tension. 	
\end{abstract}

\maketitle

\section{Introduction}

Cosmological models where dark matter (DM) and dark energy (DE), two heavy components
of the universe, interact with each other in a non-gravitational way, are
the most general ones with respect to the minimally coupled fields, since
the latter can be seen as a special case of the interaction models. The
dynamics of such interacting models offers some interesting results that have
been explored greatly in the last couple of years \cite%
{Wetterich-ide1,Amendola-ide1,Billyard:2000bh,Zimdahl:2001ar,Amendola-ide2,Pavon:2005yx,Barrow:2006hia, He:2008tn, Valiviita:2008iv,delCampo:2008sr,delCampo:2008jx,Gavela:2009cy, Majerotto:2009np,Clemson:2011an}
and also recently \cite{Avelino:2012tc,Harko:2012za,Sun:2013pda,
Pan:2013rha, Yang:2014gza, yang:2014vza,
Nunes:2014qoa, Pan:2014afa, Yang:2014hea, Pan:2012ki,
Nunes:2016dlj, Kumar:2016zpg, Yang:2016evp, Caprini:2016qxs, Pan:2016ngu,
Sharov:2017iue, Kumar:2017dnp, Yang:2017yme, Yang:2017ccc, Shahalam:2017fqt,
Kumar:2017bpv,DiValentino:2017iww,Yang:2017zjs,Pan:2017ent,Yang:2018euj,Yang:2018xlt, Yang:2018uae, Yang:2018ubt,Yang:2018pej,vonMarttens:2018iav,vonMarttens:2018bvz,Paliathanasis:2019hbi,Yang:2019bpr,Yang:2019vni,Yang:2019uzo,Pan:2019gop,Yang:2019uog,DiValentino:2019ffd,DiValentino:2019jae,vonMarttens:2019ixw,Papagiannopoulos:2019kar}
(also see \cite{Bolotin:2013jpa,Wang:2016lxa}, for extensive reviews on interacting
dark fluids). Since the nature of the interaction is not known, a very
common approach is adopted similar to what we do with DE and/or
modified gravity models. That means, we specify a phenomenological
interaction rate and test the dynamics of the universe using the
cosmological data. One can realize that such a procedure actually enables us
to reconstruct the expansion history of the universe. The usual choices of
the interaction rates are generally assumed to obey some simple and some
complicated power law relations between the energy density of
DM and DE. However, looking at the conservation law of this
joint dark sector, $\nabla _{\nu }(T_{\mathrm{DM}}^{\mu \nu }+T_{\mathrm{DE}%
}^{\mu \nu })=0$, implying $\nabla _{\nu }T_{\mathrm{DM}}^{\mu \nu
}=-Q(t)=\nabla _{\nu }T_{\mathrm{DE}}^{\mu \nu }$, where $Q(t)$ determines
the interaction rate (or, the energy flow) between these dark sectors, one thing
is clear that the exact functional form constraining the flow of energy
between the dark sectors is actually unknown since the nature of the dark
sectors is not really known. We mention that the terms $T_{\mathrm{DM}%
}^{\mu \nu}$, $T_{\mathrm{DE}}^{\mu \nu }$ respectively denote the
energy-momentum tensors of DM and DE. The  general choices
that have been studied widely in the literature include $Q(t)\propto \rho _{c},$
(where $\rho _{c}$ is the energy density of the pressureless DM), $%
Q(t)\propto \rho _{x},$ ($\rho _{x}$ denotes the energy density of DE), $%
Q(t)\propto (\rho _{c}+\rho _{x})$, $Q (t) \propto (\rho_c \rho_x)/(\rho_c+\rho_x)$, $Q\propto \dot{\rho}_{x}$ (here dot is
the derivative with respect to the cosmic time) and some others where
mainly we assume some phenomenologically simple or 
complicated functions related to the powers
of $\rho _{c}$, $\rho _{x}$ \cite{Yang:2017zjs}. Although we always stress that the interaction models are mostly phenomenological, but a recent investigation by \cite{Pan:2020zza} argues that various interaction models have a strong field theoretical origin.   In particular, some well known linear and non-linear interaction functions of $\rho_c$ and/or $\rho_x$, can  be derived following the field theoretical arguments \cite{Pan:2020zza}. However,  it is very usual
and actually natural to ask why the interaction functions involving only the powers of $\rho_c$ and/or $\rho_x$ should be preferred as there is no such objection to consider other possible choices following the arguments in \cite{Pan:2020zza}. So, one can 
look for some other kind of interaction rates that may allow some 
more complicated functions of $\rho_c$ and $\rho_x$. 
In fact, similar to DE and modified gravity models, the basic approach remains same, that means one may allow 
various type of interaction models and examine their viabilities in light of the recent observational data.

Thus, in the current work, we introduce some new type of interaction models
that have never been investigated without any proper justifications. In order 
to investigate the observational acceptance of the models, we perform 
global fittings of the models using  the latest observational data with diverse origins.
Moreover, we also address several theoretical issues of the models both at background and perturbative levels. For the numerical simulations, we consider the Markov chain
Monte Carlo package \texttt{\small COSMOMC} \cite{Lewis:1999bs, Lewis:2002ah}.

The work has been organized in the following way. In section \ref{sec-2} we
describe the gravitational equations of an interacting universe at the level
of background and perturbations as well the interaction models that we have
studied. After that in section \ref{sec-data} we provide the details of the
observational data that we have used to analyze the models. In the next
section \ref{sec-results} we give the details of the analyses for two
interaction scenarios. Finally, in section \ref{sec-conclu} we close the
present work with a brief summary of the entire findings.

\section{Interacting cosmology at Background and perturbative levels}
\label{sec-2}

In the large scale our Universe is almost homogeneous and isotropic and such geometric configuration is well described by a spatially flat Friedmann-Lema\^{i}tre-Robertson-Walker (FLRW) universe characterized by the line element 
\begin{equation}
ds^{2}=-dt^{2}+a^{2}(t)(dx_{1}^{2}+dx_{2}^{2}+dx_{3}^{2})
\end{equation}%
where $a(t)$ being the scale factor of the universe and $(x_1, x_2, x_3)$ are the co-moving coordinates. In 
such a universe we consider 
that the total energy content of the universe is shared by 
radiation (photon + neutrinos), baryons, DM and DE \footnote{Let us note that we fix the total neutrino mass to $M_{\nu} =0.06\; {\rm eV}$ according to Planck's baseline analyses.}. Although all species interact with other at the gravitational level, however,  in the present work we focus only on the non-gravitational interaction between DM and DE, that means, a matter flow between DM and DE exists. 
We further assume that DM is pressureless (also known as cold DM, abbreviated as CDM) while DE has a constant barotropic
state parameter. Finally, the entire matter sector is also assumed to be minimally coupled to gravity. 

Thus, in such a model of General Relativity, the conservation equations for
the interacting DM and DE take the forms

\begin{eqnarray}
\dot{\rho}_{c}+ 3H \rho _{c} &=&-Q(t),  \label{cons-cdm} \\
\dot{\rho}_{x}+ 3H (1+w_{x})\rho _{x} &=&Q(t),  \label{cons-de}
\end{eqnarray}%
where $\rho_c$, $\rho_x$, as already quoted in the introduction,
are the energy densities of pressureless DM
and DE, respectively; the quantity 
$w_{x}=p_{x}/\rho _{x}\;(< -1/3)$ refers to the equation of state for DE in which 
$p_x$ is the pressure of the DE fluid and $Q$ is the interaction rate between these dark sectors. The positivity in the interaction rate, that means $Q >0$ denotes the energy transfer from pressureless DM to DE while its negative sign (i.e. $Q <0$) denotes the opposite case, that means the energy flow occurs from DE to pressureless DM. 
The quantity $H\equiv \dot{a}/a$, is the Hubble rate of the FLRW universe which provides the constraint equation on the dynamics as  
\begin{equation}
H^{2}=\frac{8\pi G}{3}(\rho _{r}+\rho _{b}+\rho _{c}+\rho _{x}),
\label{Hubble}
\end{equation}
where $\rho_r$ and $\rho_b$ are respectively the energy density 
of radiation and baryons. 
Hence, if one describes the interaction rate $Q(t)$, then the set of
equations (\ref{cons-cdm}), (\ref{cons-de}) and (\ref{Hubble}) can determine
the entire dynamics of the universe. However, as discussed earlier, in this
work we aim to discuss some particular choices for the interaction rates
beyond the general. In what follows, we introduce two interaction rates

\begin{equation}
Q(t)=3H\xi \rho _{x}\sin \left( \frac{\rho _{x}}{\rho _{c}}-1\right),
\label{Q1}
\end{equation}%
and 
\begin{equation}
Q(t)=3H\xi \rho _{x}\left[ 1+\sin \left( \frac{\rho _{x}}{\rho _{c}}%
-1\right) \right],  \label{Q2}
\end{equation}%
where $\xi $ in both (\ref{Q1}) and (\ref{Q2}) refers to the coupling
parameter of the interaction rates.  We note that the interaction models featured with trigonometric functions, so far we are aware of the literature, have not been investigated.  One could easily realize that since $ -1 \leq \sin \theta \leq 1$ ($\theta$ being a real number), thus, specifically model (\ref{Q1}) could allow both positive and negative values. That means it seems to possess an oscillating character. 
In case of model (\ref{Q2}), although an oscillating trigonometric function is present, but since $ 0 \leq (1+ \sin \theta) \leq 2$, thus, for this model $Q$ could take take either positive or negative values depending on the sign of the coupling parameter.  
Since the models are new in the literature, we expect that it might be interesting to see what new ingredients the models can add to the existing literature of IDE models.  
For convenience, we label the
interacting scenario with the interaction rate (\ref{Q1}) as IDE1 while the
interacting scenario driven by the interaction model (\ref{Q2}) is labeled
as IDE2. 

Considering the Taylor expansion of the interaction models around $\frac{%
\rho _{x}}{\rho _{c}}=1$, it is easy to find that the interaction function (\ref%
{Q1}) in its first approximation takes the form of a nonlinear model

\begin{eqnarray}
Q (t) \approx 3 H \xi \rho_x \left(\frac{\rho_x}{\rho_c}-1\right) = 3 H \xi
\left(\frac{\rho_x^2}{\rho_c}- \rho_x\right),
\end{eqnarray}
and similarly the first approximation of the model (\ref{Q2})
around $\frac{\rho_x}{\rho_c} = 1$, leads to another nonlinear interaction
rate

\begin{eqnarray}
Q (t) \approx 3 H \xi \, \frac{\rho_x^2}{\rho_c}.
\end{eqnarray}
Thus, one can see that the models have some known nonlinear structure \cite%
{Yang:2017zjs}.

We digress for a moment from the above discussion and go back to the conservation equations (\ref%
{cons-cdm}) and (\ref{cons-de}) by recasting them in terms of the effective
equations of state for the dark fluids as follows

\begin{eqnarray}
\dot{\rho}_{c}+3H\left( 1+w_{c}^{\mathtt{eff}}\right) \rho _{c} &=&0,  \notag
\\
\dot{\rho}_{x}+3H\left( 1+w_{x}^{\mathtt{eff}}\right) \rho _{x} &=&0,  \notag
\end{eqnarray}%
where $w_{c}^{\mathtt{eff}}$, $w_{x}^{\mathtt{eff}}$ are respectively identified as the
effective equation of state parameters for CDM and DE having the following expressions
\begin{eqnarray}
w_{c}^{\mathtt{eff}} &=&\frac{Q(t)}{3H\rho _{c}},  \label{eff-eos-cdm} \\
w_{x}^{\mathtt{eff}} &=&w_{x}-\frac{Q(t)}{3H\rho _{x}},  \label{eff-eos-de}
\end{eqnarray}%
and both of which, as one may note, clearly describe that the effective fluid could deviate from the original nature. That means, the effective equation of
state for CDM does not behave like a dust if we have a nonzero interaction in the dark sector, i.e. for $Q(t) \neq 0$. One may further note that in terms of the effective fluids description, we now have a non-interacting cosmological description. Thus, in spite of having an interaction in the dark sectors, it is possible to view it as a non-interacting cosmological description in which both the dark fluids might be exotic depending on the signal and strength of $Q$.

In fact, $w_{c}^{\mathtt{eff}}$ could be negative and it may even behave like a DE
fluid, equivalently, $w_{c}^{\mathtt{eff}}<-1/3$, depending on the strength as well as sign (of course for $Q <0$) of the interaction function. On the other hand, the effective DE fluid may exhibit different
characters. For instance, if the DE fluid is the vacuum, i.e., $%
w_{x}=-1$, then depending on the sign of the interaction function, $Q$,
the nature of the effective fluid could be different. For instance, 
if $Q(t)>0$, the effective DE fluid behaves like a
phantom fluid (i.e. $w_{x}<-1$) while for $Q(t)<0$, the equation of state parameter of the effective DE 
fluid will be constrained as $w_x > -1$. 
In a similar fashion, if the DE is of quintessence or phantom, 
then depending on the sign of $Q$, alternatively, depending on the 
energy transfer between DE 
and pressureless DM, one could classify the nature of the effective dark fluid. 
Thus, one could clearly realize  that the
presence of interaction (i.e. $Q \neq 0$) in the dark sectors may change the qualitative
behaviour of the fluids under interaction. 

Let us now come to the exact effective equations of state for the above
interaction rates. For model IDE1, one can derive the effective equations of
state as,

\begin{eqnarray}
w_{c}^{\mathtt{eff}} &=&\frac{\xi }{r}\sin \left( \frac{1}{r}-1\right), \label{eff-eos-dm-ide1}\\
w_{x}^{\mathtt{eff}} &=&w_{x}-\xi \sin \left( \frac{1}{r}-1\right),\label{eff-eos-de-ide1}
\end{eqnarray}%
where we have used $r=\rho _{c}/\rho _{x}$, known as the coincidence
parameter. Similarly, for model IDE2, one can find that 
\begin{eqnarray}
w_{c}^{\mathtt{eff}} &=&\frac{\xi }{r}\left[ 1+\sin \left( \frac{1}{r}%
-1\right) \right]  \\
w_{x}^{\mathtt{eff}} &=&w_{x}-\xi \left[ 1+\sin \left( \frac{1}{r}-1\right) %
\right] .
\end{eqnarray}%
From the above couple of equations, one could determine the nature of the effective equation of state parameters. One could easily realize that for IDE1, the evolution of 
$w_{c}^{\mathtt{eff}}$ and $w_{x}^{\mathtt{eff}}$ are oscillating with respect to the coincidence parameter, $r$. 
Additionally, the expression of $w_{c}^{\mathtt{eff}}$ for IDE1 directs that the effective equation of state for CDM could be
exotic in nature.

\begin{figure*}
\includegraphics[width=0.4\textwidth]{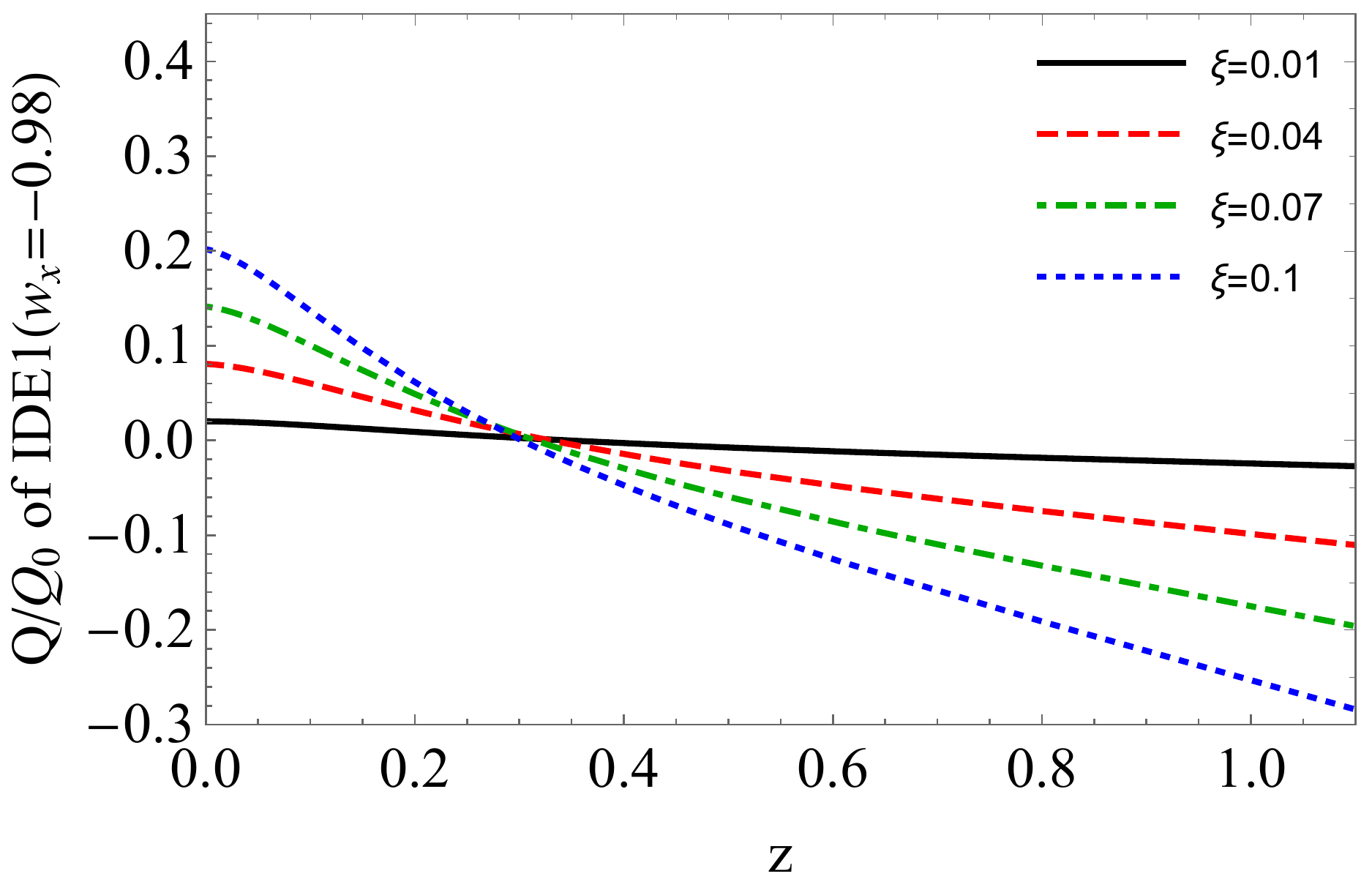} %
\includegraphics[width=0.4\textwidth]{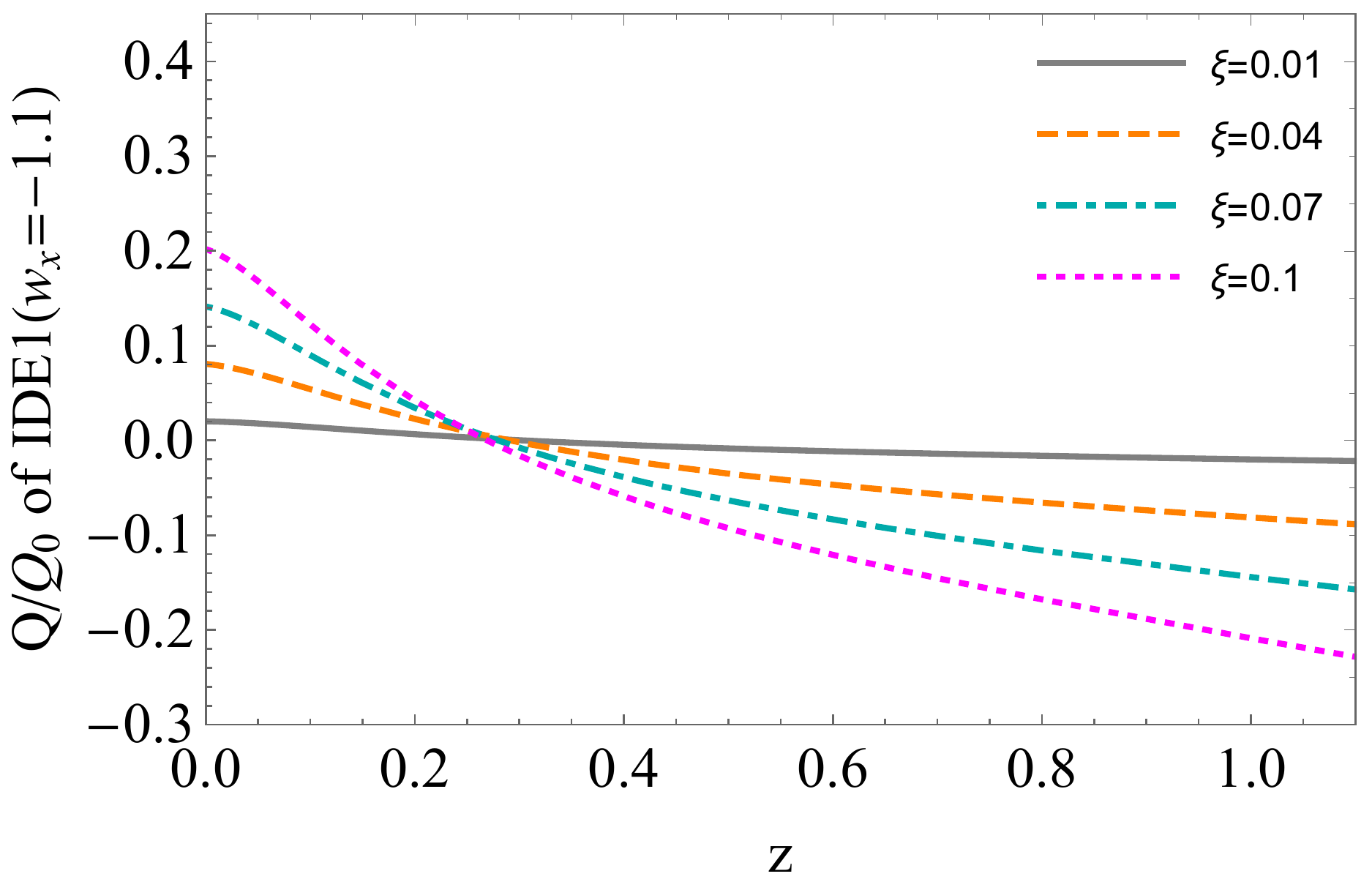}
\includegraphics[width=0.4\textwidth]{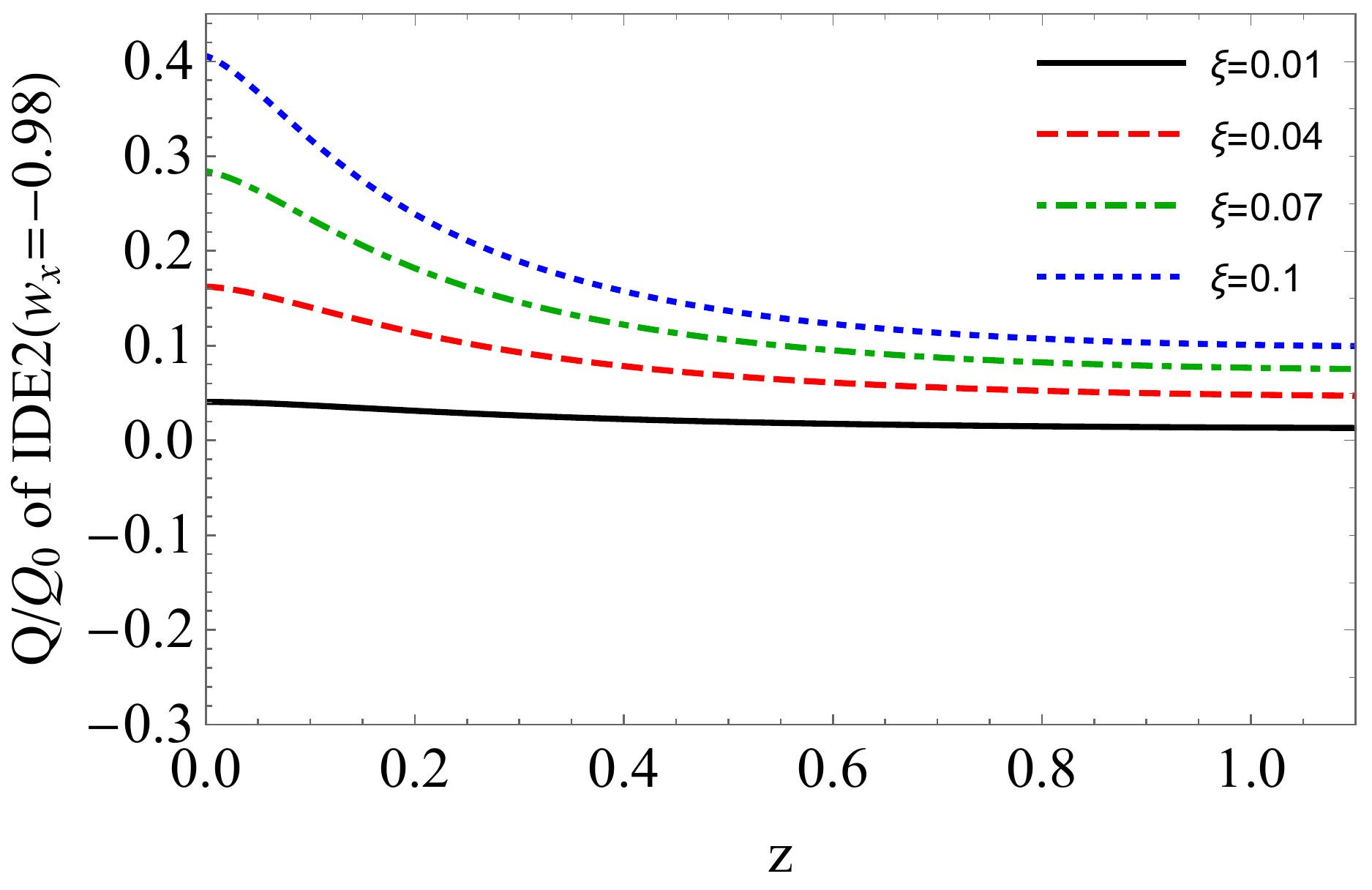} %
\includegraphics[width=0.4\textwidth]{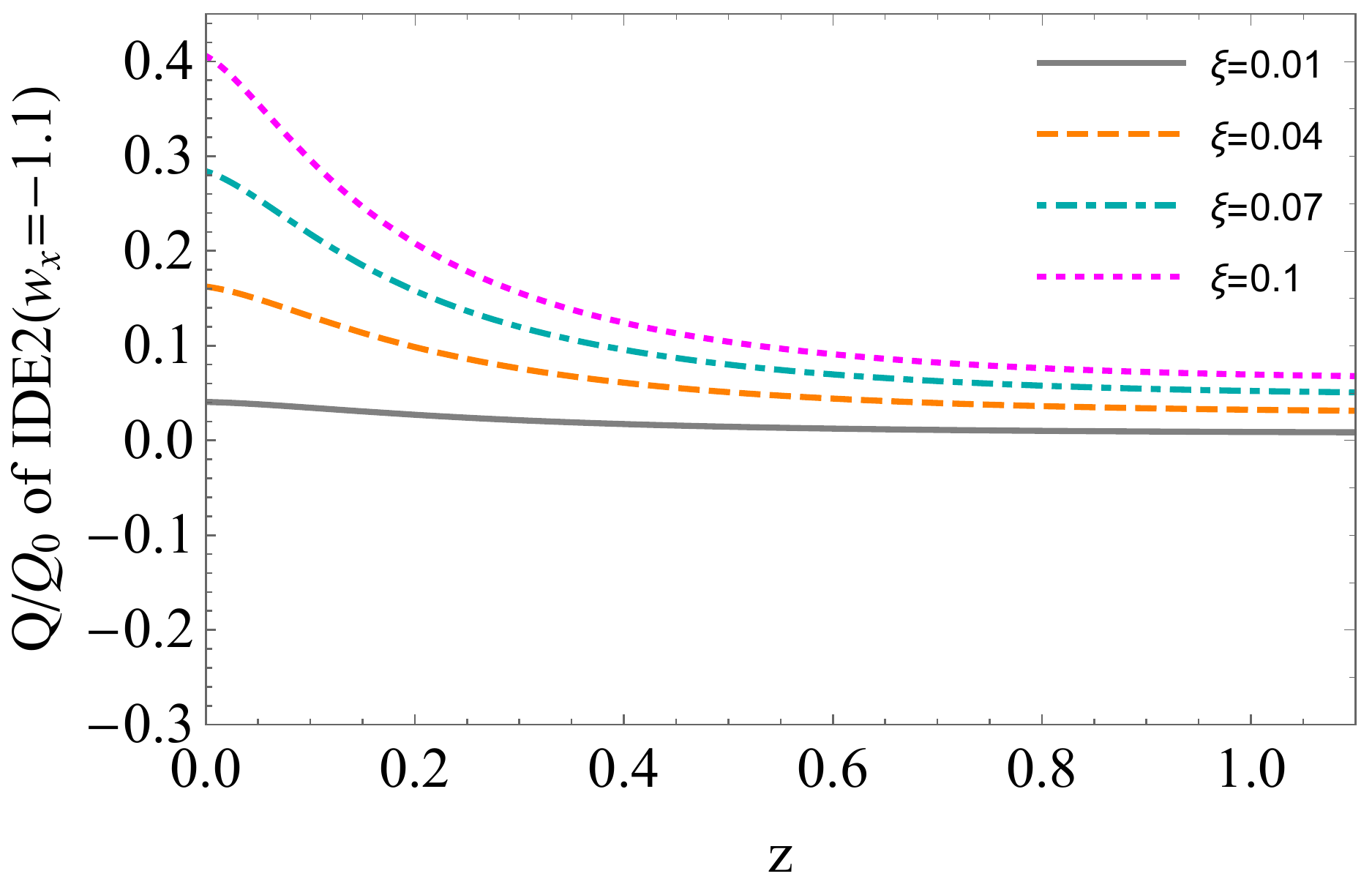}
\caption{The evolution of the interaction rate for different values of the coupling parameter $\protect\xi $ assuming both quintessence
and phantom DE equation of state parameters. The upper panel shows
the behaviour of the model $Q(t)=3H\protect\xi \protect\rho _{x}\sin \left( 
\frac{\protect\rho _{x}}{\protect\rho _{c}}-1\right) $ for the quintessence
(upper left) and phantom DE (upper right) state parameters for some
values of the coupling parameter. The lower panel shows the behaviour of the model $%
Q(t)=3H\protect\xi \protect\rho _{x}\left[ 1+\sin \left( \frac{\protect\rho %
_{x}}{\protect\rho _{c}}-1\right) \right] $ where the lower left stads for $%
w_{x}>-1$ and the lower right stands for $w_{x}<-1$. In both the
plots we have used a parameter $Q_{0}=H_{0}\protect\rho _{0}$, and thus the
quantity $Q/Q_{0}$ becomes dimensionless, where $\protect\rho _{0}=\frac{%
3H_{0}^{2}}{8\protect\pi G}$, is the present value of the total energy
density defined in eqn. (\ref{Hubble}). While drawing the plots we have fixed $\Omega_{c0} = 0.28$, $\Omega_{x0} = 0.68$, $\Omega_{r0} = 0.0001$, and $\Omega_{b0} = 1- \Omega_{r0}-\Omega_{c0}-\Omega_{x0} =  0.0399$. }
\label{fig:Q}
\end{figure*}
\begin{figure*}
\includegraphics[width=0.4\textwidth]{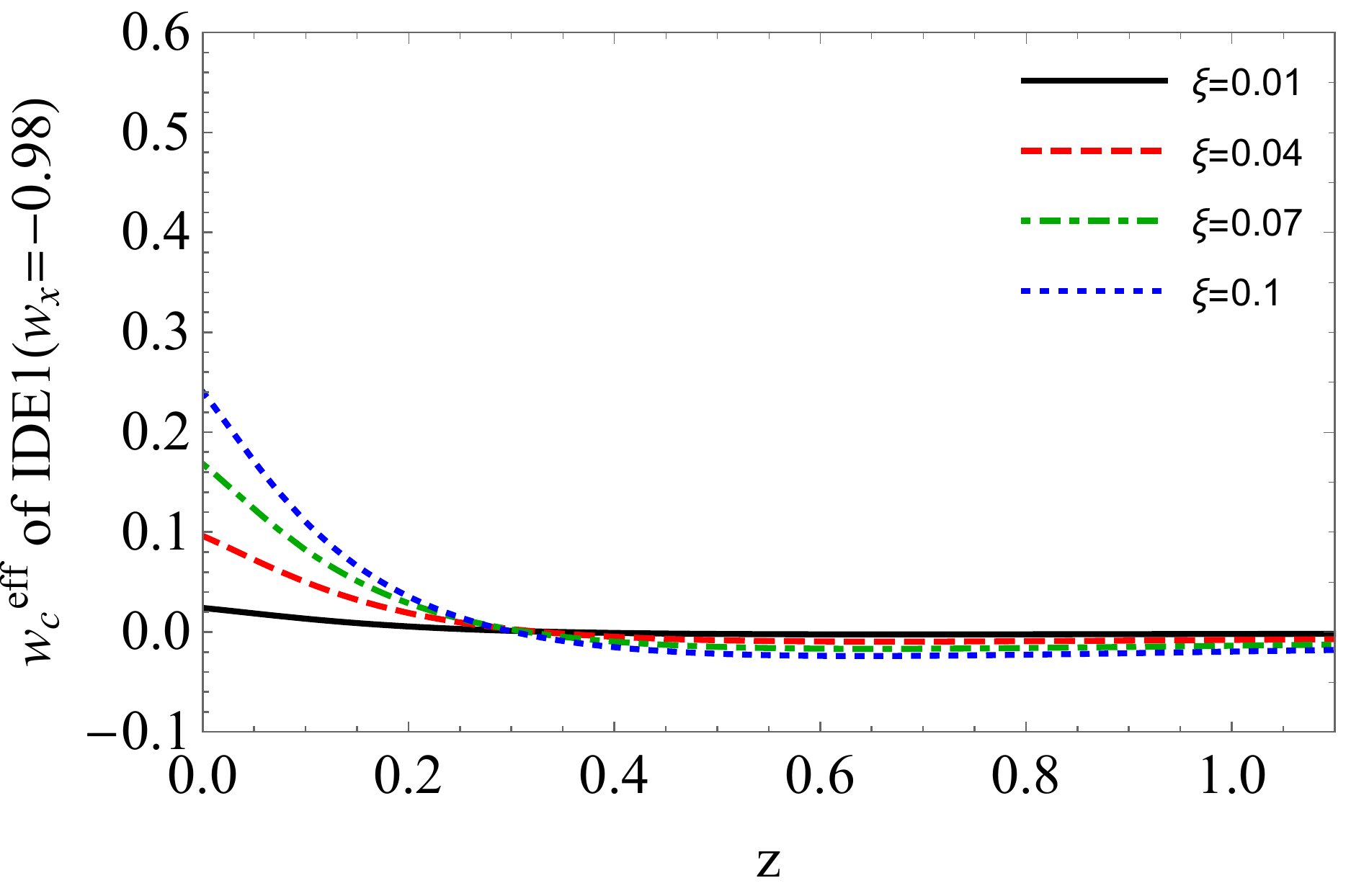} %
\includegraphics[width=0.4\textwidth]{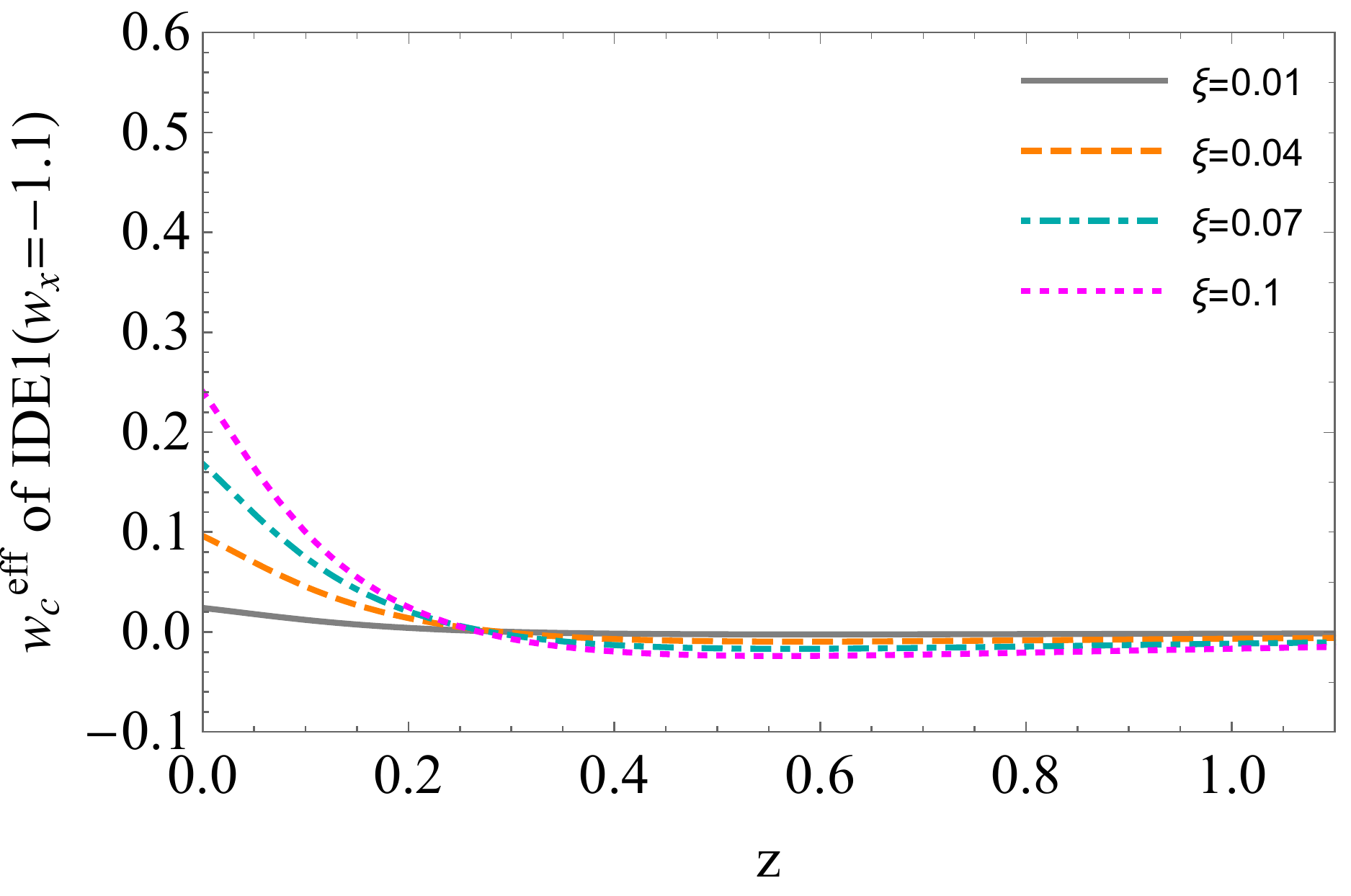}
\includegraphics[width=0.4\textwidth]{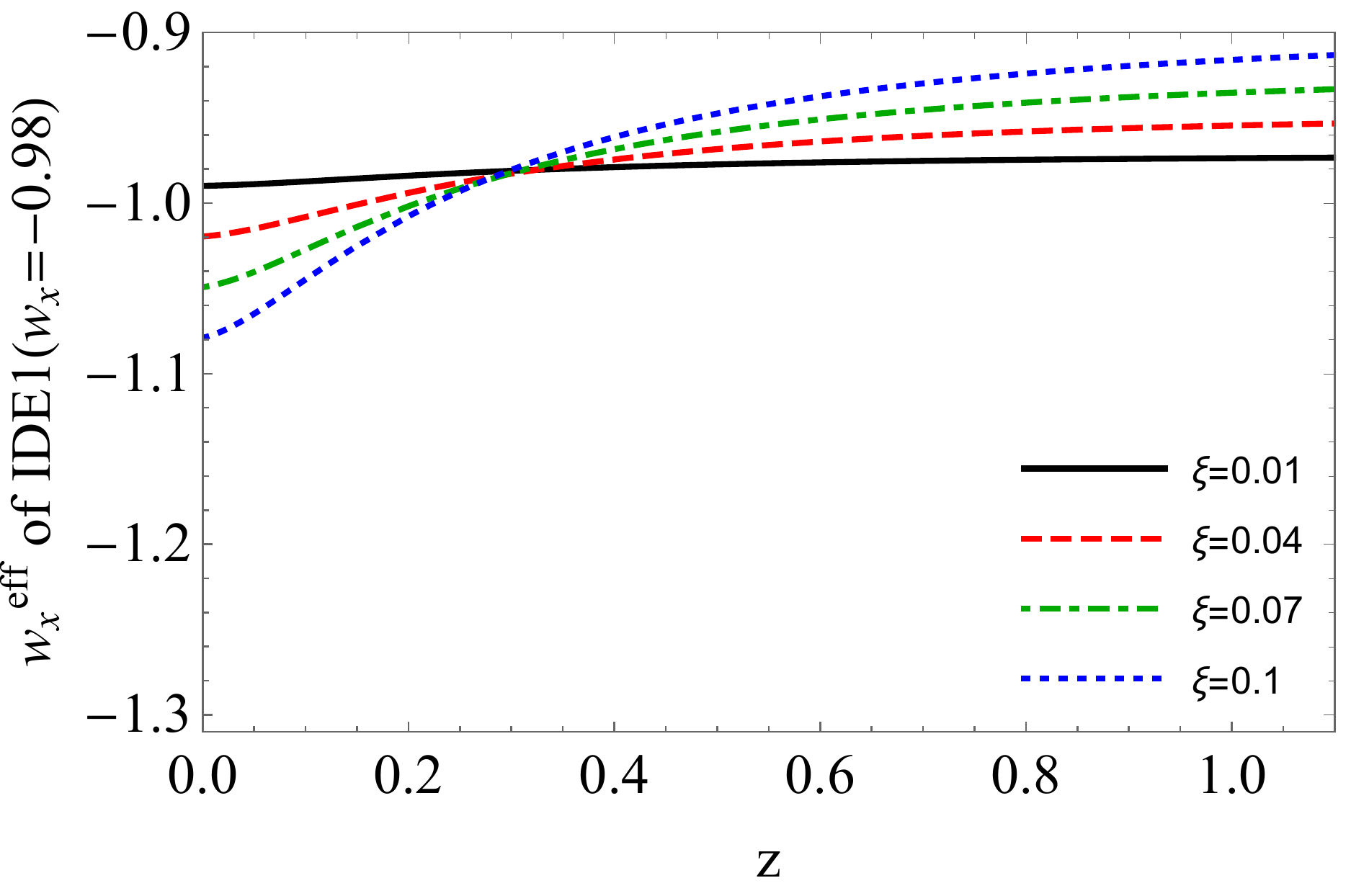} %
\includegraphics[width=0.4\textwidth]{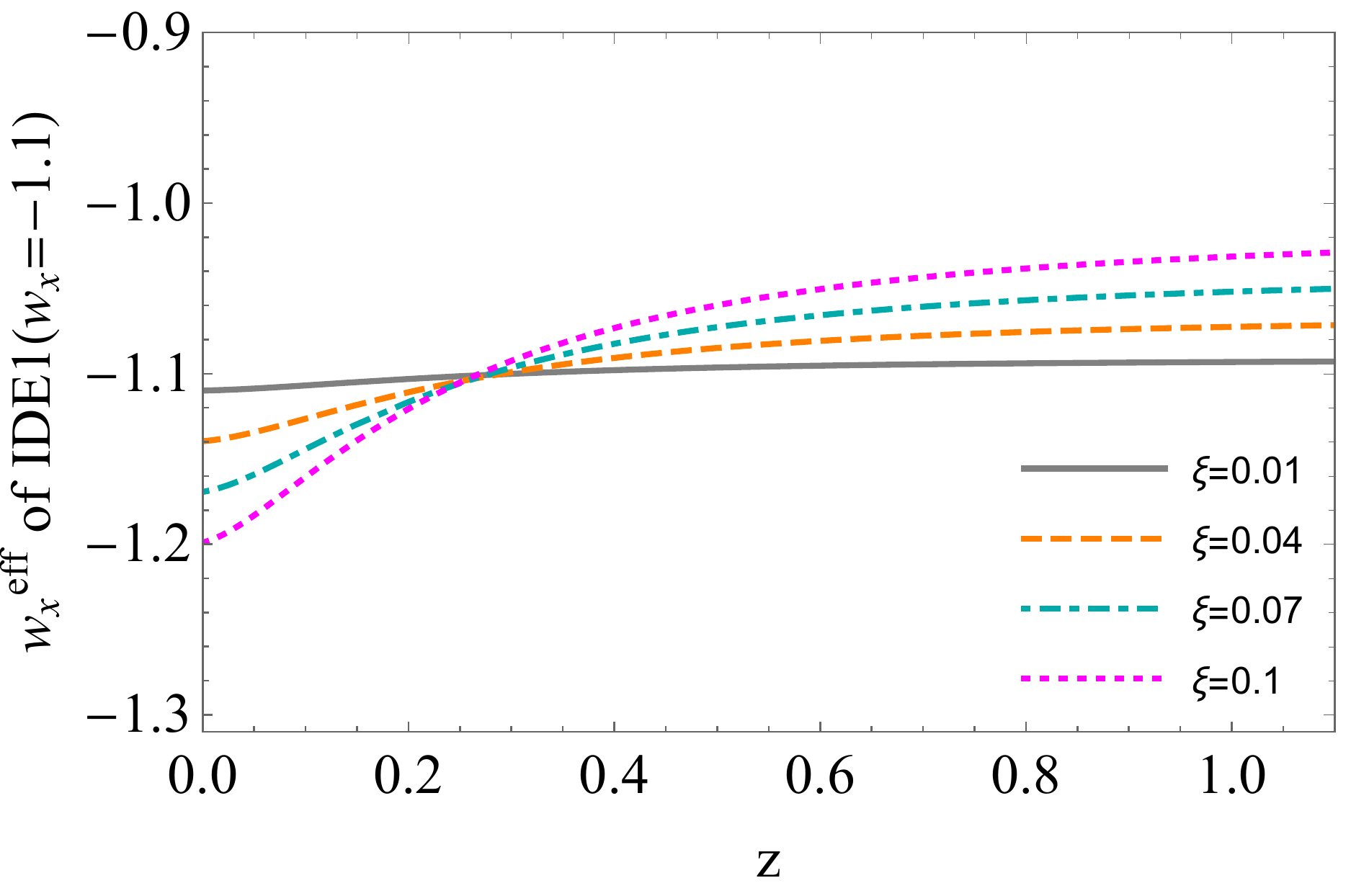}
\caption{We demonstrate the evolution of the effective equation of states
for CDM and DE for the interaction model $Q(t)=3H\protect\xi \protect\rho %
_{x}\sin \left( \frac{\protect\rho _{x}}{\protect\rho _{c}}-1\right) $
considering both the possibilities of the DE equation of state,
i.e., whether $w_{x}$ is in the quintessence or in the phantom regime. We have fixed $\Omega_{c0} = 0.28$, $\Omega_{x0} = 0.68$, $\Omega_{r0} = 0.0001$, and $\Omega_{b0} = 1- \Omega_{r0}-\Omega_{c0}-\Omega_{x0} =  0.0399$. }
\label{fig:effeos-IDE1}
\end{figure*}
\begin{figure*}
\includegraphics[width=0.4\textwidth]{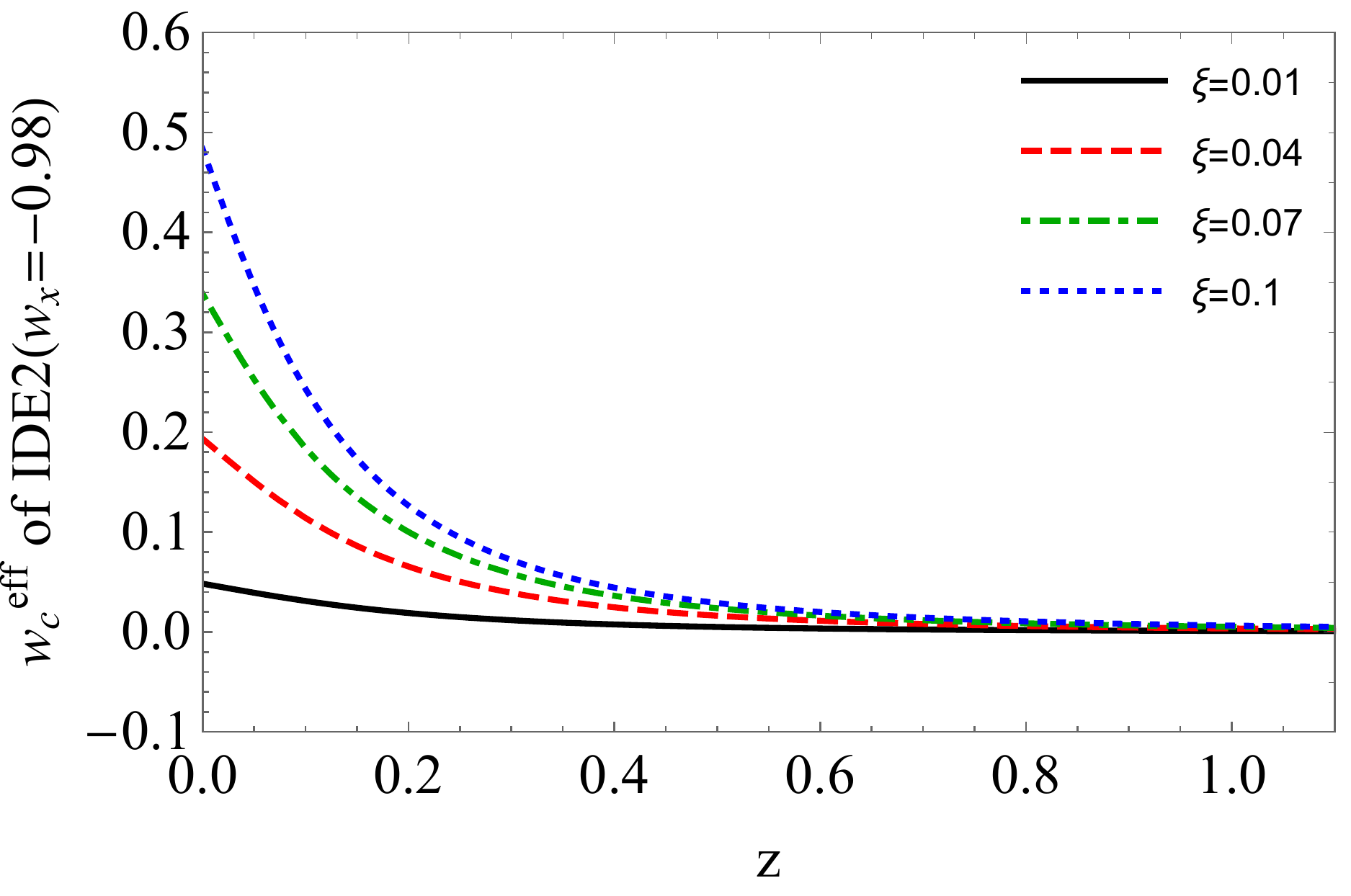}
\includegraphics[width=0.4\textwidth]{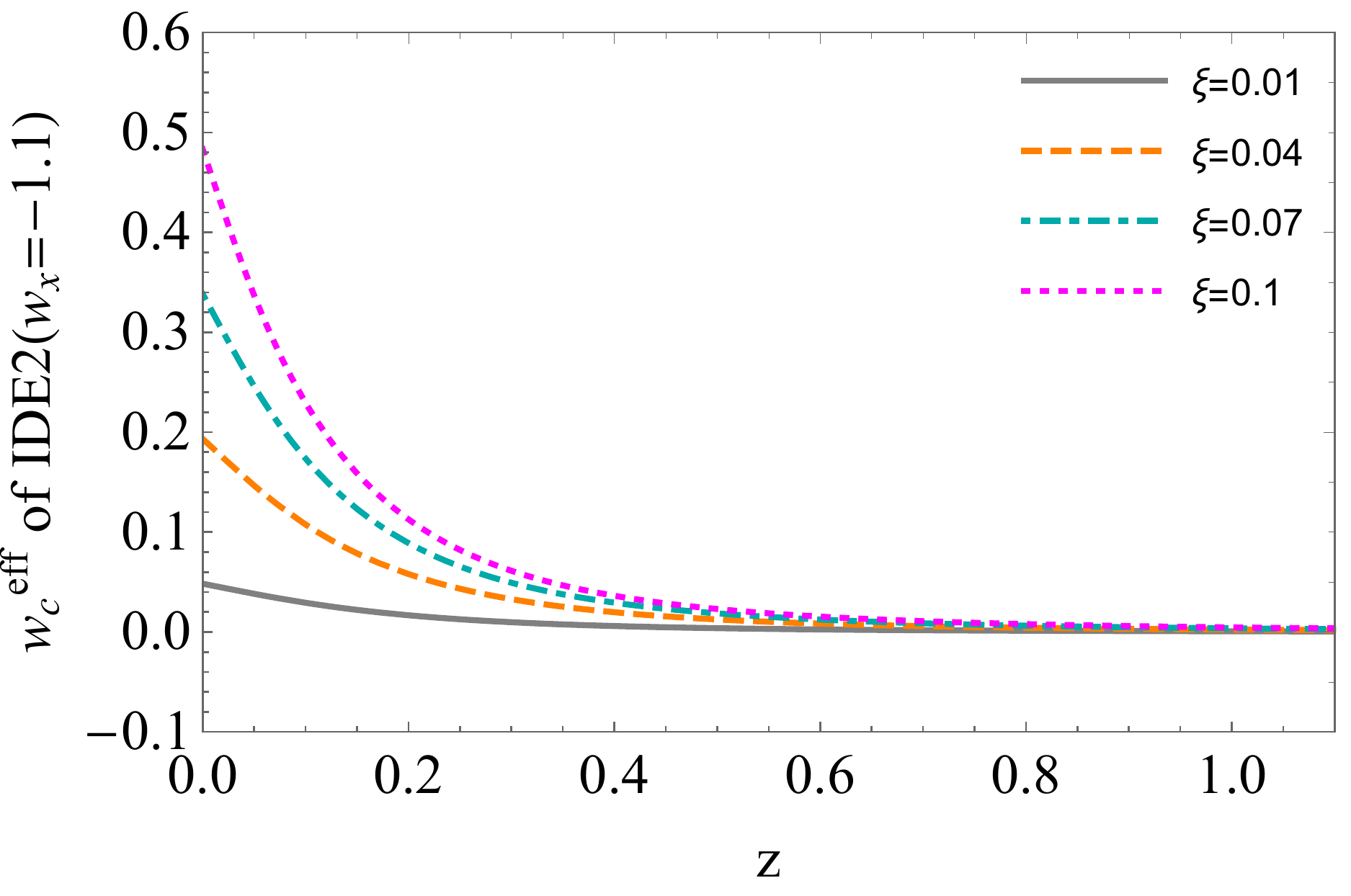}
\includegraphics[width=0.4\textwidth]{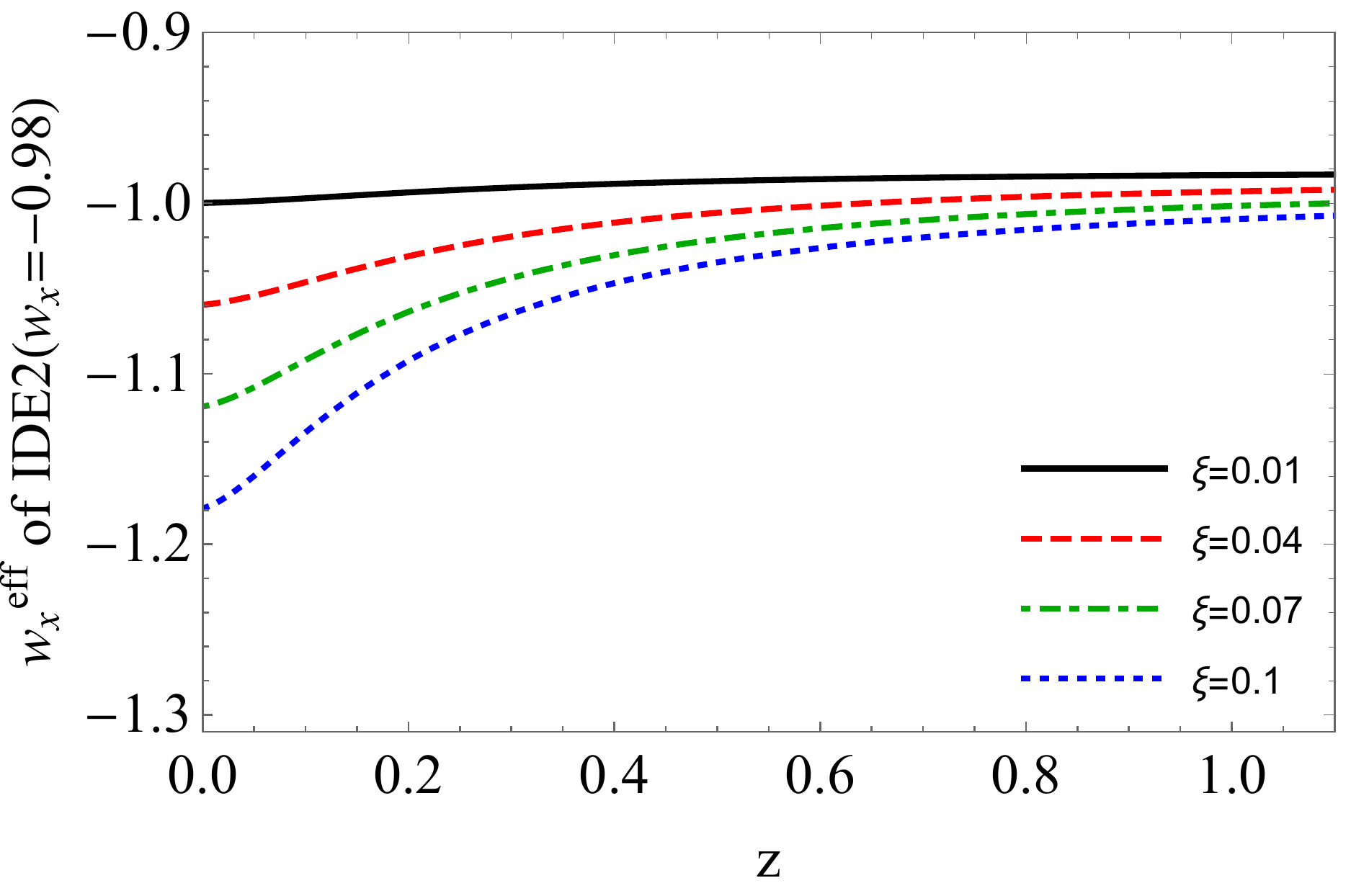}
\includegraphics[width=0.4\textwidth]{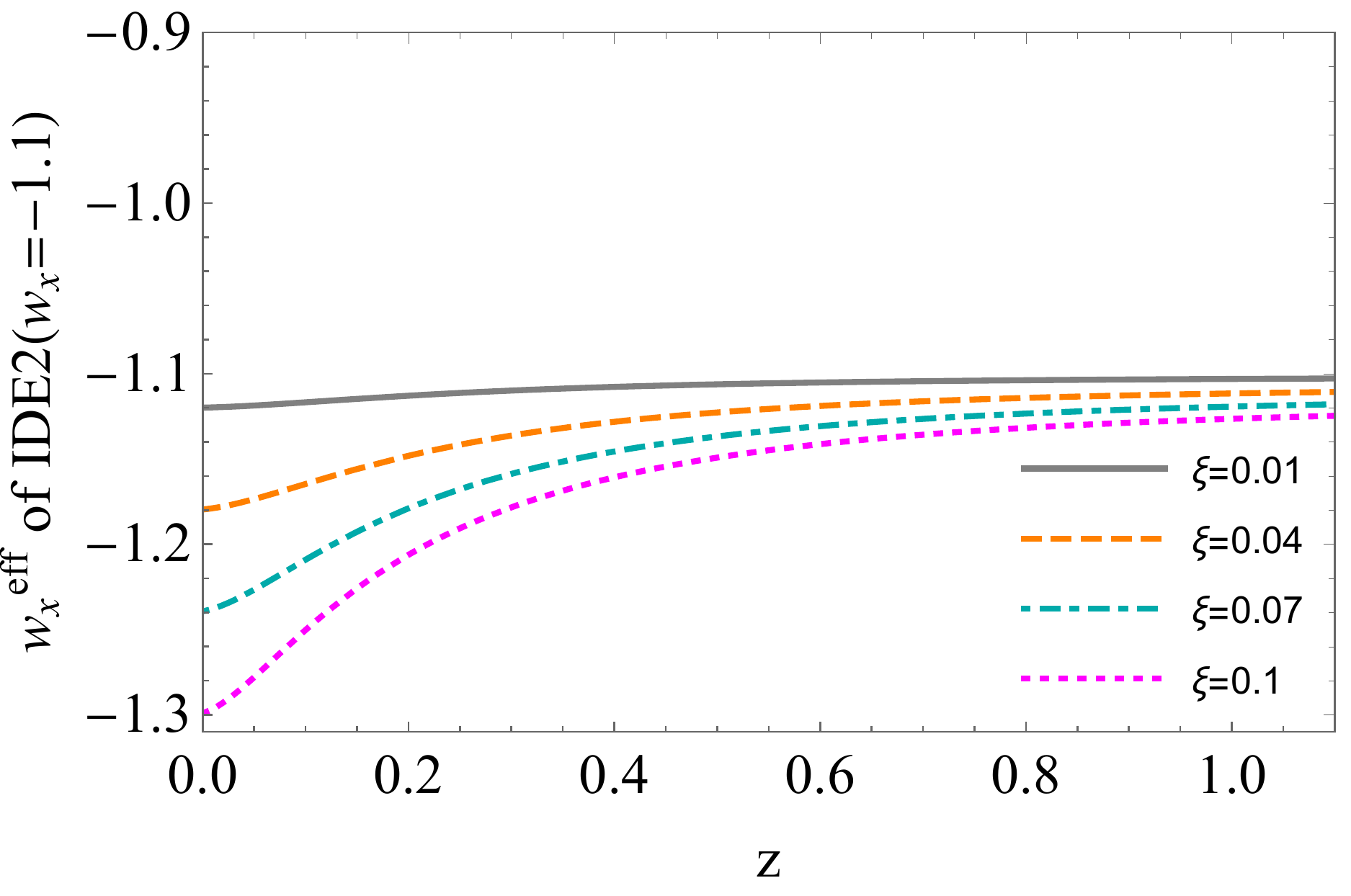}
\caption{We demonstrate the evolution of the effective equations of state
for CDM and DE for the interaction model $Q(t)=3H \xi \rho _{x}\left[ 1+\sin \left( \frac{\rho_{x}}{\rho _{c}}-1\right) \right] $ considering both the possibilities of the DE equation of state, i.e., whether $w_{x}$ is
in the quintessence or in the phantom regime. For all the plots we have fixed $\Omega_{c0} = 0.28$, $\Omega_{x0} = 0.68$, $\Omega_{r0} = 0.0001$, and $\Omega_{b0} = 1- \Omega_{r0}-\Omega_{c0}-\Omega_{x0} =  0.0399$. }
\label{fig:effeos-IDE2}
\end{figure*}

Let us now present the qualitative behaviour of the interaction models (\ref{Q1})
and (\ref{Q2}) as well as their impacts on the universe evolution through
the effective equations of state for CDM and DE respectively given in eqn. (%
\ref{eff-eos-cdm}) and eqn. (\ref{eff-eos-de}).

In Fig. \ref{fig:Q}, we present the evolution of the interaction rates (\ref{Q1}) and (\ref{Q2}) for two different regimes of the DE equation of state, namely
quintessence ($w_{x}>-1$) and phantom ($w_{x}<-1$) considering different
values of the coupling parameter $\xi $, namely, $\xi = 0.01, 0.04, 0.07$ and $\xi =0.1$ aiming to understand how different values of the coupling parameter affect the qualitative features of the interaction models.  The upper panel of Fig. \ref%
{fig:Q} stands for IDE1 of (\ref{Q1}) while the lower panel stands for IDE 2
of (\ref{Q2}).  We note that while showing 
the behaviour of the interaction model, we have
plotted the dimensionless quantity $Q/Q_{0}$ (here $Q_{0}=H_{0}\rho _{0}$, $\rho
_{0}$ being the total energy density of the matter sector defined through
the Hubble equation (\ref{Hubble})). Now, looking at the upper panel of Fig. %
\ref{fig:Q} (IDE1), one can notice that the interaction model allows a fine transition from its negative values ($Q< 0$) to positive values ($Q>0$). For the case with $Q <0$, the energy flow takes place from DE to pressureless DM. As a consequence, as long as the energy flow takes place from DE $\longrightarrow$ pressureless DM, the energy density of the DM sector continues to increase  and the dynamics of the universe is dominated by the DM sector.  We note that for higher values of the coupling parameter, as one can see from the curves representing $\xi = 0.1$ (see both the upper plots of Fig. \ref{fig:Q}), the energy flow is higher in terms of the quantity $|Q/Q_0|$. 
When the energy flow reverses its direction ($Q < 0 \longrightarrow Q >0$), the energy flow takes place from pressureless DM to DE, and hence the energy density of DE becomes higher compared to the energy density of the pressureless DM. As a result, the DE sector dominates the universe's dynamics compared to the pressureless DM and we observe the late accelerating phase of the universe. On the other hand, for the second interaction model (\ref{Q2}), i.e. for IDE2, we do not find any such sign changeable feature, see the lower panel of Fig. \ref{fig:Q}.  Thus, one can see that IDE1 scenario has an appealing characteristic in term of its sign changeable behaviour. We note that only a few IDE models with such special aspect exist in the literature \cite{Wei:2010cs,Sun:2010vz,Guo:2017deu,Arevalo:2019axj,Pan:2019jqh}. The imprints of this sign changeable nature in $Q$ can also be reflected in other parameters, namely the effective equation of state parameters that we have described below.  

In order to have a better understanding on the behaviour of the effective equation of state parameters of the dark fluids, namley, $w_c^{\rm eff}$ and $w_x^{\rm eff}$, we have considered two different DE regimes,
namely, quintessence ($w_{x}>-1$) and phantom ($w_{x}<-1$). Let us note that since the interaction function includes the energy density of DE, thus, the effective equation of state parameter for pressureless DM shown in (\ref{eff-eos-cdm}) is dependent on the equation of state parameter for DE. Thus, while describing the evolution of $w_c^{\rm eff}$, it is necessary to consider a specific value of $w_x$. Following these, in Fig. \ref{fig:effeos-IDE1} and Fig. \ref{fig:effeos-IDE2}, we have described the evolution of $w_c^{\rm eff}$ and $w_x^{\rm eff}$ for two different regimes of $w_x$ as well as for different values of the coupling parameter for the interacting scenarios IDE1 and IDE2, respectively. Let us describe the evolution of $w_c^{\rm eff}$ and $w_x^{\rm eff}$ for each IDE model. 

Fig. \ref{fig:effeos-IDE1} corresponds to IDE1. The upper plots of Fig. \ref{fig:effeos-IDE1} present the evolution of  $w_c^{\rm eff}$ and the lower plots of Fig. \ref{fig:effeos-IDE1} present the evolution of $w_{x}^{\rm eff}$ for different values of the coupling parameters as well as for two different regimes of $w_x$, namely, quintessence and phantom. For IDE1, we have already noticed that a transition from $Q < 0$ (energy flows from DE to CDM) to $Q >0$ (energy flows from CDM to DE) is allowed (see the upper plots of Fig. \ref{fig:Q}). Thus, when the energy flow occurs from DE to CDM (i.e. $Q <0$), the effective nature of CDM becomes exotic (see eqn. (\ref{eff-eos-cdm})) meaning that $w_c^{\rm eff} < 0$; see the evolution of the curves (corresponding to each coupling parameter $\xi$, after $z \gtrsim 0.25$) in the upper plots of Fig. \ref{fig:effeos-IDE1}. On the other hand, concerning the effective nature of the DE sector we find, $w_x^{\rm eff} > w_x$; similarly, see the evolution of the curves (corresponding to each coupling parameter $\xi$, after $z \gtrsim 0.25$) in the lower plots of Fig. \ref{fig:effeos-IDE1}.  When the energy flow reverses its direction ($Q < 0 \longrightarrow Q > 0$), that means when energy flow occurs from CDM to DE (i.e. $Q >0$), then $w_c^{\rm eff}$ becomes positive but more negative nature of the effective DE sector is enhanced, i.e. $w_x^{\rm eff} < w_x$. These features are encoded in the curves of all plots of Fig. \ref{fig:effeos-IDE1} for $z \lesssim 0.25$. In a similar fashion, Fig. \ref{fig:effeos-IDE2} corresponds to the evolution of the effective state parameters for IDE2. Since for IDE2, the interaction rate does not allow any transition in its sign, thus, we have quite straightforward behavior in both $w_c^{\rm eff}$ and $w_{x}^{\rm eff}$ as shown in various plots of Fig. \ref{fig:effeos-IDE2}.

Having presented the evolution equations of the IDE models at the level of background, we are now interested to investigate the models at the level of perturbations. Combining the evolution equations at the level of background and perturbations equations, 
one can fully determine the exact nature of the interaction
models.  Thus, in order to do that, we consider the most general perturbed
FLRW metric \cite{Mukhanov, Ma:1995ey, Malik:2008im} 
\begin{eqnarray}
ds^{2} &=&a^{2}(\tau )\Bigg[-(1+2\phi )d\tau ^{2}+2\partial _{i}Bd\tau dx^{i}
\notag \\
&&+\Bigl((1-2\psi )\delta _{ij}+2\partial _{i}\partial _{j}E\Bigr)%
dx^{i}dx^{j}\Bigg],  \label{perturbed-flrw}
\end{eqnarray}%
in which $\tau $ is the conformal time; $\phi $, $B$, $\psi $, $E$ are the
the gauge-dependent scalar perturbation quantities. Now, using the above
metric (\ref{perturbed-flrw}), one can derive the gravitational equations
following \cite{Majerotto:2009np, Valiviita:2008iv, Clemson:2011an}: $\nabla
_{\nu }T_{A}^{\mu \nu }=Q_{A}^{\mu },~\sum\limits_{\mathrm{A}}{Q_{A}^{\mu }}%
=0,$ where $A=c$ (for pressureless DM or CDM) or $A=x$ (for dark
energy) and $Q_{A}^{\mu }=(Q_{A}+\delta Q_{A})u^{\mu }+a^{-1}(0,\partial
^{i}f_{A}),$ relative to the four-velocity vector $u^{\mu }$. Here, $Q_{A}$
denotes the background energy transfer (i.e., $Q_{A}=Q$) and the quantity $%
f_{A}$ refers to the momentum transfer potential. Following the well
accepted earlier theories \cite{Majerotto:2009np, Valiviita:2008iv,
Clemson:2011an}, we restrict the interaction scenario where the momentum
transfer potential is set to zero in the rest frame of the DM fluid
and consequently, one can find that $k^{2}f_{A}=Q_{A}(\theta -\theta _{c})$,
where $k$ denotes the wave number; $\theta =\theta _{\mu }^{\mu }$, and $%
\theta _{c}$ are respectively the volume expansion scalar of the total fluid
and the volume expansion scalar for the pressureless DM (CDM) fluid.

Now, one can consider either a synchronous gauge or the conformal Newtonian
gauge to describe the perturbations equations. Here, we adopt the
synchronous gauge for which $\phi =B=0$, $\psi =\eta $, and $%
k^{2}E=-h/2-3\eta $ (where $h$ and $\eta$ are the metric perturbations, see 
\cite{Ma:1995ey} for a detailed reading), and moreover, assuming zero
anisotropic stress in the interacting scenario, the density and velocity
perturbations for the dark fluids can be written as

\begin{widetext}
\begin{eqnarray}
\delta _{x}^{\prime } =-(1+w_{x})\left( \theta _{x}+\frac{h^{\prime }}{2}%
\right) -3\mathcal{H}(c_{sx}^{2} -w_{x}) \Bigg[ \delta _{x}+3\mathcal{H}%
(1+w_{x})\frac{\theta _{x}}{k^{2}}\Bigg]  -3\mathcal{H}w_{x}^{\prime }\frac{%
\theta _{x}}{k^{2}} +\frac{aQ}{\rho _{x}}\left[ -\delta _{x}+\frac{\delta Q}{Q}+3\mathcal{H}%
(c_{sx}^{2}-w_{x})\frac{\theta _{x}}{k^{2}}\right],\nonumber
\end{eqnarray}
\begin{eqnarray}
\theta _{x}^{\prime } =-\mathcal{H}(1-3c_{sx}^{2})\theta _{x}+\frac{%
c_{sx}^{2}}{(1+w_{x})}k^{2}\delta _{x}+\frac{aQ}{\rho _{x}}\left[ \frac{%
\theta _{c}-(1+c_{sx}^{2})\theta _{x}}{1+w_{x}}\right],\nonumber 
\end{eqnarray}
\begin{eqnarray}
\delta _{c}^{\prime } = -\left( \theta _{c}+\frac{h^{\prime }}{2}\right) +%
\frac{aQ}{\rho _{c}}\left( \delta _{c}-\frac{\delta Q}{Q}\right),\nonumber 
\end{eqnarray}
\begin{eqnarray}
\theta _{c}^{\prime } =-\mathcal{H}\theta _{c},  \nonumber 
\end{eqnarray}
\end{widetext}
where $\delta _{A}=\delta \rho _{A}/\rho _{A}$ ($A=c$ or $x$,
mentioned above) is the density perturbations; $\mathcal{H}=a^{\prime }/a$,
is the conformal Hubble parameter and $\delta Q/Q$ includes the
perturbations for the Hubble rate $\delta H$ where $\mathcal{H}=aH$. Now
using $\delta H$, one may derive the gauge invariant equations for the
interacting dark fluids, see \cite{Gavela:2009cy} for more details. Thus,
the above set of equations present the general perturbation equations for
any coupling $Q$ between DM and DE. Let us now present the
exact perturbation equations for the two interacting functions. The
perturbation equations of DE and CDM for IDE1 are, 
\begin{widetext}
\begin{eqnarray}
\delta _{x}^{\prime } =-(1+w_{x})\left( \theta _{x}+\frac{h^{\prime }}{2}%
\right) -3\mathcal{H}(c_{sx}^{2}-w_{x})\Bigg[ \delta _{x} +3\mathcal{H}%
(1+w_{x})\frac{\theta _{x}}{k^{2}}\Bigg] -3\mathcal{H}w_{x}^{\prime }\frac{%
\theta _{x}}{k^{2}} \notag\\ + 3\mathcal{H}\xi\sin\left(\frac{\rho_x}{\rho_c}-1\right)
\Bigg[\frac{\cos(1-\rho_x/\rho_c)}{\sin(\rho_x/\rho_c-1)}
\frac{\rho_x}{\rho_c}(\delta_x-\delta_c) \notag\\+\frac{\theta+h'/2}{3\mathcal{H}}
+3\mathcal{H}(c^2_{sx}-w_x)\frac{\theta_x}{k^2}\Bigg], \nonumber 
\end{eqnarray}
\begin{eqnarray}
\theta _{x}^{\prime } =-\mathcal{H}(1-3c_{sx}^{2})\theta _{x}+\frac{%
c_{sx}^{2}}{(1+w_{x})}k^{2}\delta _{x} +3\mathcal{H}\xi\sin\left(\frac{\rho_x}{\rho_c}-1\right)\left[ \frac{%
\theta _{c}-(1+c_{sx}^{2})\theta _{x}}{1+w_{x}}\right],\nonumber 
\end{eqnarray}
\begin{eqnarray}
\delta _{c}^{\prime } =-\left( \theta _{c}+\frac{h^{\prime }}{2}\right) +%
3\mathcal{H}\xi\frac{\rho_x}{\rho_c}\sin\left(\frac{\rho_x}{\rho_c}-1\right)\Bigg[ \delta _{c}-\delta_x  -\frac{\cos(1-\rho_x/\rho_c)}{\sin(\rho_x/\rho_c-1)}\frac{\rho_x}{\rho_c}(\delta_x-\delta_c)-\frac{\theta+h'/2}{3\mathcal{H}}\Bigg],\nonumber 
\end{eqnarray}
\begin{eqnarray}
\theta _{c}^{\prime } =-\mathcal{H}\theta _{c},  \nonumber 
\end{eqnarray}%
\end{widetext}
while on the other hand, the perturbation equations for IDE2 are 
\begin{widetext}
\begin{eqnarray}
\delta _{x}^{\prime } = -(1+w_{x})\left( \theta _{x}+\frac{h^{\prime }}{2}%
\right) -3\mathcal{H}(c_{sx}^{2}-w_{x})\left[ \delta _{x}+3\mathcal{H}%
(1+w_{x})\frac{\theta _{x}}{k^{2}}\right] -3\mathcal{H}w_{x}^{\prime }\frac{%
\theta _{x}}{k^{2}} \notag\\+ 3\mathcal{H}\xi\left[1+\sin\left(\frac{\rho_x}{\rho_c}-1\right)\right]
\Bigg[\frac{\cos(1-\rho_x/\rho_c)}{1+\sin(\rho_x/\rho_c-1)}
\frac{\rho_x}{\rho_c}(\delta_x-\delta_c)  +\frac{\theta+h'/2}{3\mathcal{H}}
+3\mathcal{H}(c^2_{sx}-w_x)\frac{\theta_x}{k^2}\Bigg], \nonumber 
\end{eqnarray}
\begin{eqnarray}
\theta _{x}^{\prime } =-\mathcal{H}(1-3c_{sx}^{2})\theta _{x}+\frac{%
c_{sx}^{2}}{(1+w_{x})}k^{2}\delta _{x} +3\mathcal{H}\xi\left[1+\sin\left(\frac{\rho_x}{\rho_c}-1\right)\right]\left[ \frac{%
\theta _{c}-(1+c_{sx}^{2})\theta _{x}}{1+w_{x}}\right],\nonumber 
\end{eqnarray}
\begin{eqnarray}
\delta _{c}^{\prime } = -\left( \theta _{c}+\frac{h^{\prime }}{2}\right) +%
3\mathcal{H}\xi\frac{\rho_x}{\rho_c}\left[1+\sin\left(\frac{\rho_x}{\rho_c}-1\right)\right]\Bigg[ \delta _{c}-\delta_x  -\frac{\cos(1-\rho_x/\rho_c)}{1+\sin(\rho_x/\rho_c-1)}\frac{\rho_x}{\rho_c}(\delta_x-\delta_c)-\frac{\theta+h'/2}{3\mathcal{H}}\Bigg],\nonumber 
\end{eqnarray}
\begin{eqnarray}
\theta _{c}^{\prime } = -\mathcal{H}\theta _{c}, \nonumber 
\end{eqnarray}%
\end{widetext}

We close this section with a general treatment for the growth rate of matter perturbations valid for any 
interacting DE model as well as some other quantities that are also
affected in presence of the interaction. The general expression describing the
growth rate of matter perturbations in presence of an arbitrary interaction
function $Q$ is, 

\begin{widetext}
\begin{eqnarray}\label{growth-rate}
\delta''_c+\left(1-\frac{Q}{H\rho_c}\right)\mathcal{H}\delta'_c
=\frac{3}{2}\mathcal{H}^2\Omega_b\delta_b+\frac{3}{2}\mathcal{H}^2\Omega_c\delta_c
\Bigg\{ 1 +\frac{2}{3}\frac{\rho_t}{\rho_c}\frac{Q}{H\rho_c}
\Bigg[ \frac{\mathcal{H}'}{\mathcal{H}^2} +1-3w_x+\frac{w'_x}{\mathcal{H}(1+w_x)}
+\frac{Q}{H\rho_c}\left(1+\frac{\rho_x}{\rho_c}\right) \Bigg] \Bigg\},\nonumber 
\end{eqnarray}
\end{widetext}
from which it is evident that in presence of no interaction (i.e. $Q = 0$), 
the standard equation of growth rate of matter perturbations for the
non-interacting cosmologies can be easily recovered. In (\ref{growth-rate}), 
$\mathcal{H}$ is the conformal Hubble factor that has already been mentioned
earlier and the prime stands for the derivative with respect to the
conformal time. The growth rate of pressureless DM, $f_c$, which measures 
the direct effects of interaction on the matter perturbations is, 
$f_c  = \frac{d}{da} (\ln \delta_c)$. Thus, plugging the interaction rate 
$Q$ considered in this work, into (\ref{growth-rate}), one could find the 
growth rate of pressureless DM.  We have numerically solved the 
equation (\ref{growth-rate}) for both interaction functions considering the initial condition, $\delta_c[0.001]=\delta_b[0.001]=0.001, \delta_c^{\prime}[0.001]= \delta_b^{\prime}[0.001]=1$.  

\begin{figure*}
\includegraphics[width=0.4\textwidth]{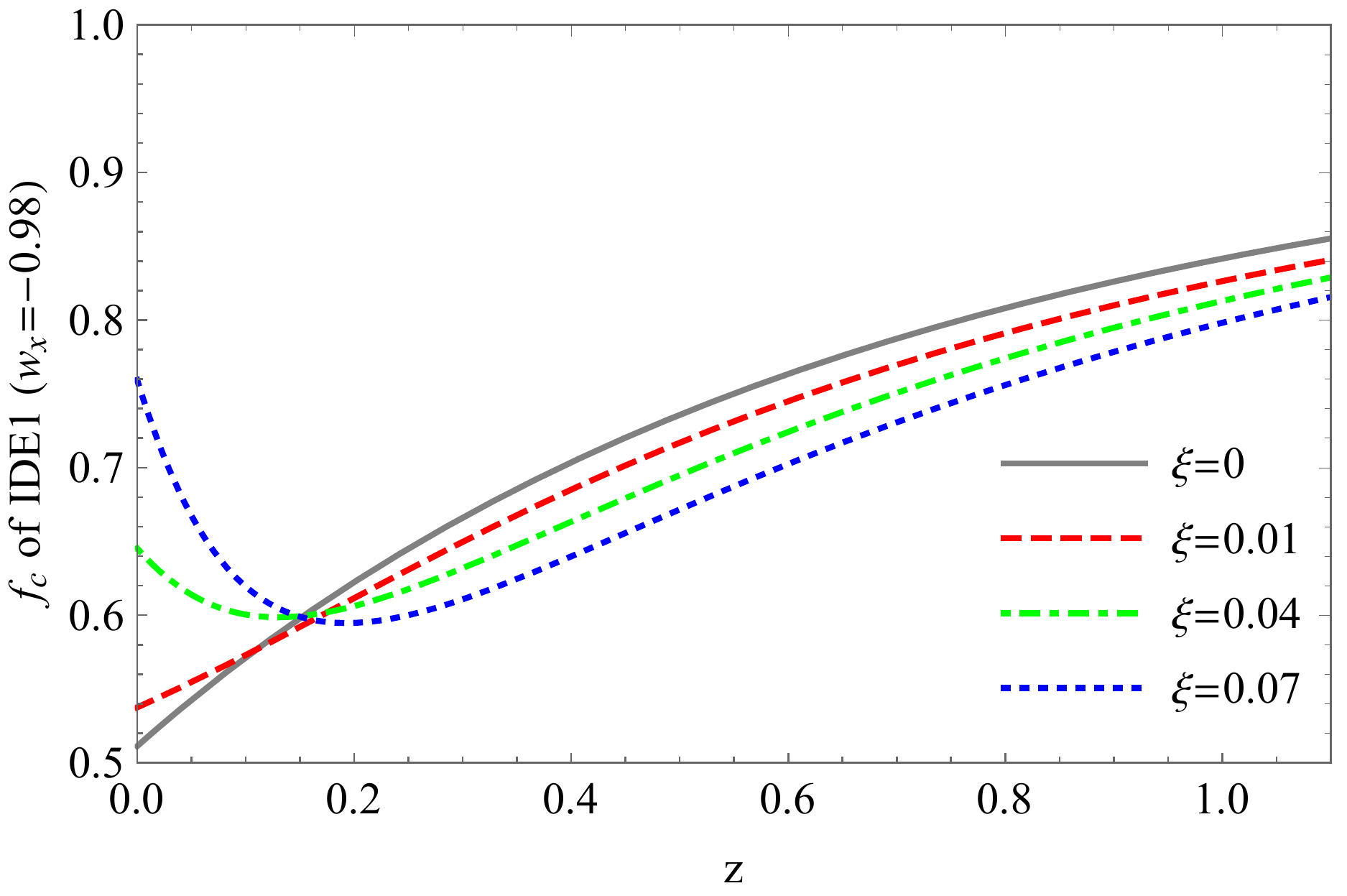} %
\includegraphics[width=0.4\textwidth]{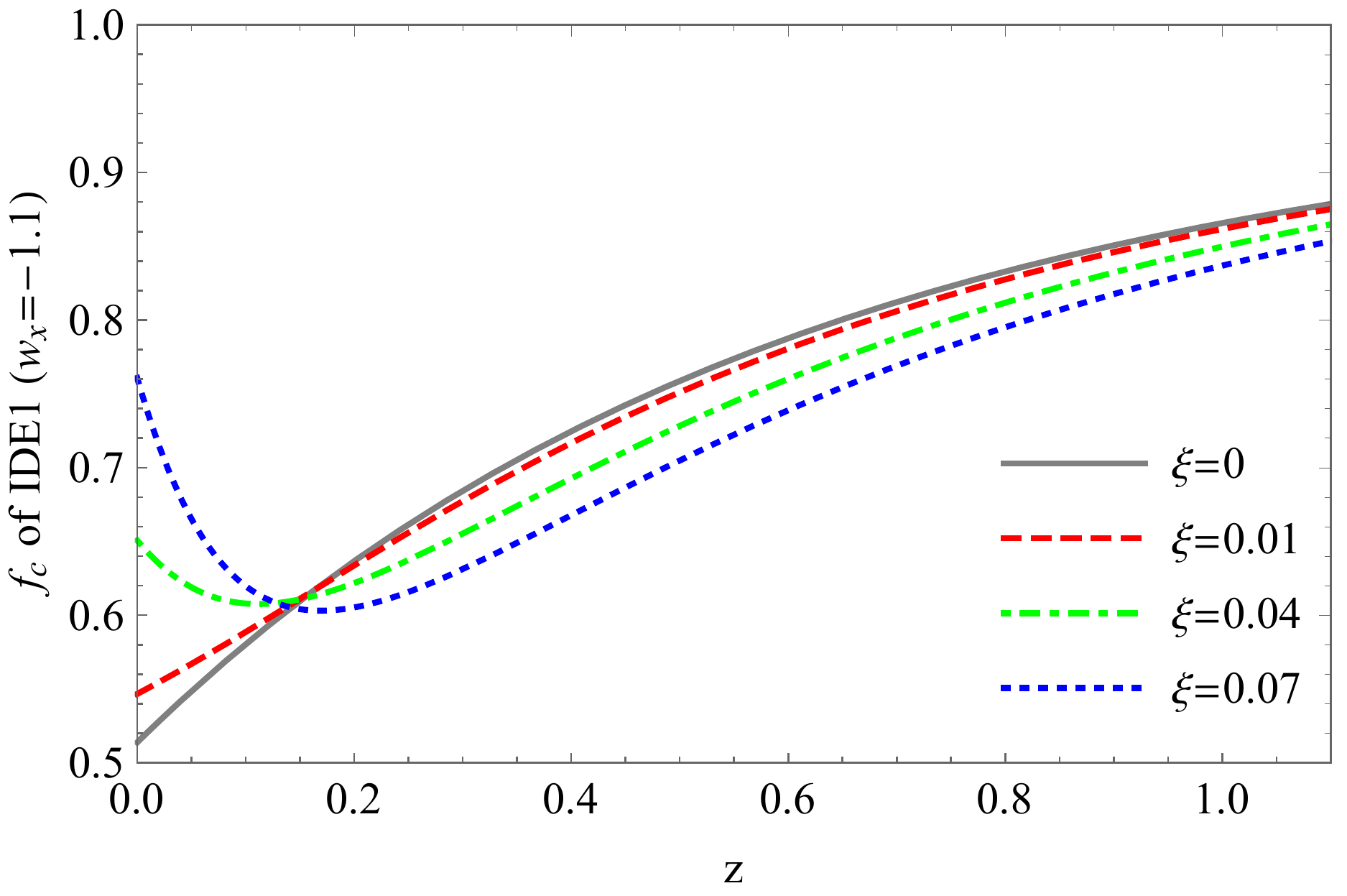}
\caption{Growth rate of CDM, $f_c$, for IDE1 corresponding to the
interaction function (\protect\ref{Q1}) has been shown for different
values of the coupling parameter. In the left plot we assume that DE has
quintessence behaviour where we set $w_x =-0.98$ as a typical value whilst
for the right plot we fix phantom DE with a typical value $w_x =-1.1
$. For both the plots we set $\Omega_{c0} = 0.28$, $\Omega_{x0} = 0.68$, $\Omega_{r0} = 0.0001$, and $\Omega_{b0} = 1- \Omega_{r0}-\Omega_{c0}-\Omega_{x0} =  0.0399$.}
\label{fig-fc-ide1}
\end{figure*}

\begin{figure*}
\includegraphics[width=0.4\textwidth]{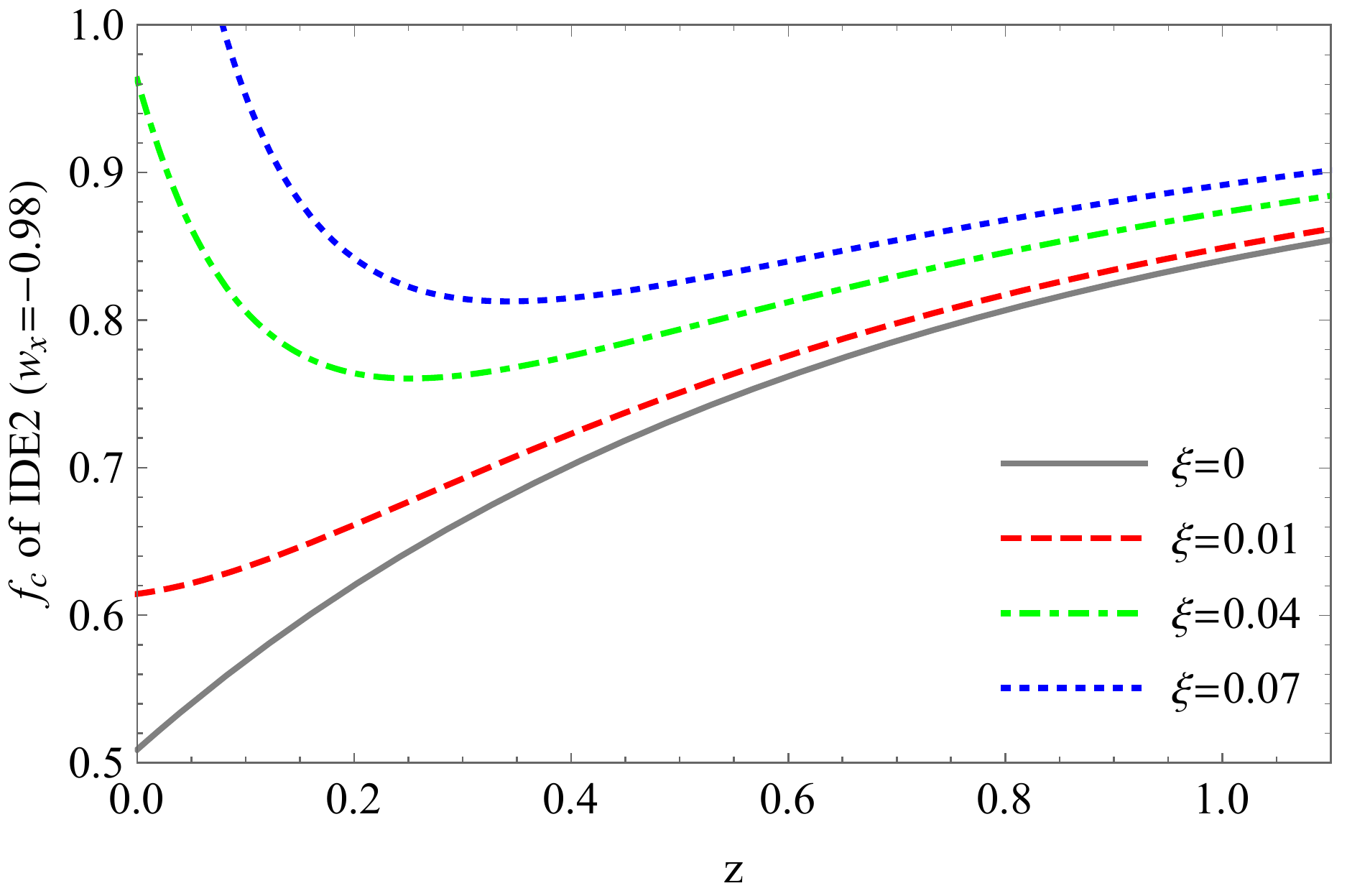} %
\includegraphics[width=0.4\textwidth]{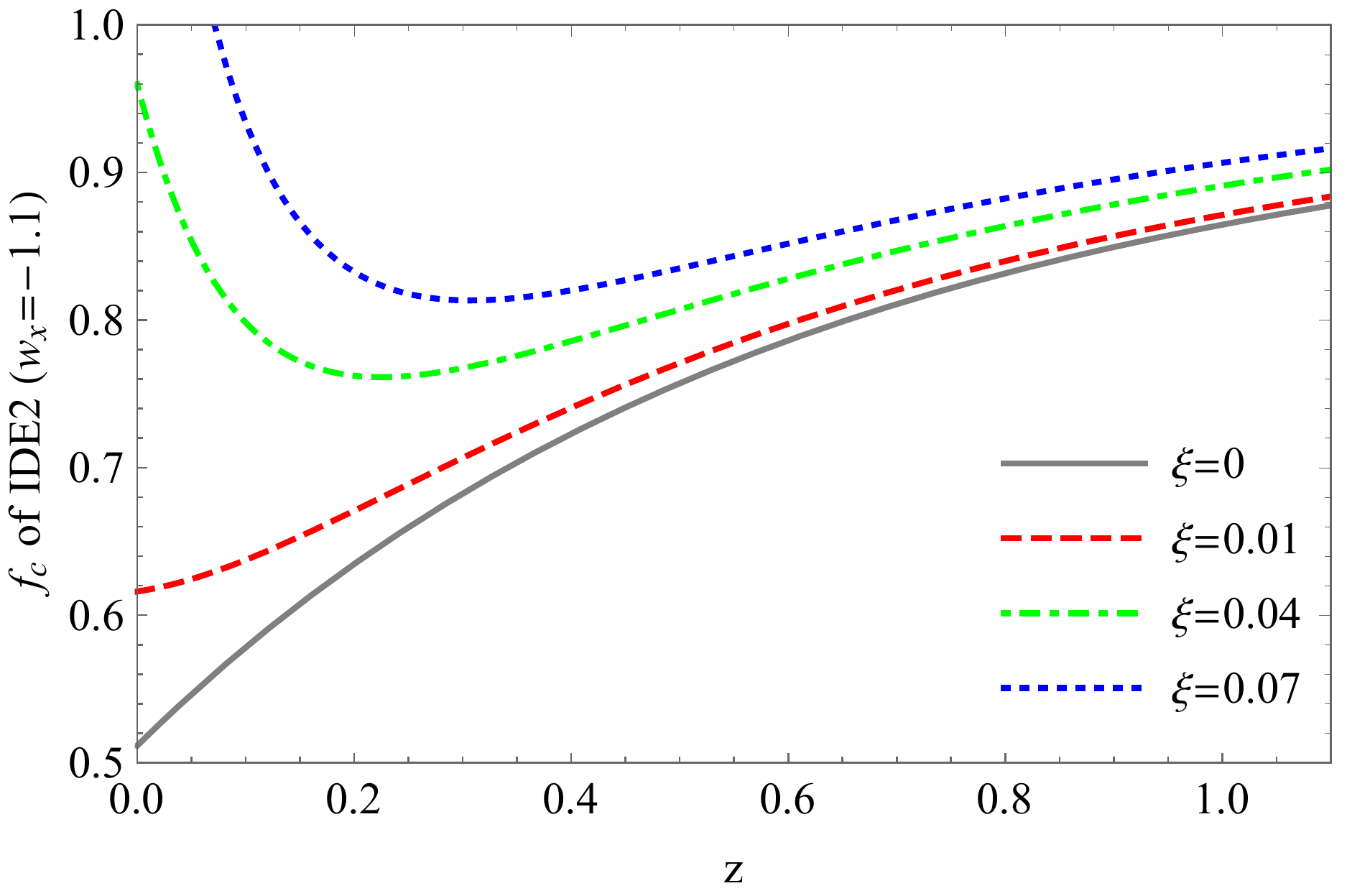}
\caption{Growth rate of CDM, $f_c$, for IDE2 corresponding to the 
interaction function (\protect\ref{Q2}) has been shown for different
values of the coupling parameter. In the left plot we assume DE has
quintessence behaviour where we set $w_x =-0.98$ as a typical value whilst
for the right plot we fix phantom DE with a typical value $w_x =-1.1
$. For both the plots we set $\Omega_{c0} = 0.28$, $\Omega_{x0} = 0.68$, $\Omega_{r0} = 0.0001$, and $\Omega_{b0} = 1- \Omega_{r0}-\Omega_{c0}-\Omega_{x0} =  0.0399$. }
\label{fig-fc-ide2}
\end{figure*}

In Fig. \ref{fig-fc-ide1} we show the
growth rate of the CDM, $f_{c}$, for IDE1 considering various values of the coupling parameter, $\xi$. In particular, we checked the evolution history of $f_c$ for both quintessence DE, i.e. $w_x> -1$ (left plot of Fig. \ref%
{fig-fc-ide1} ) and phantom DE, i.e. $w_x <-1$ (right plot of Fig. \ref{fig-fc-ide1}) DE.  From both the plots in Fig. \ref{fig-fc-ide1}, one could clearly see that the coupling parameter significantly affects the entire growth history of the universe, and in particular, at late times the growth rate for CDM presents large differences. This actually implies that the growth history of CDM  is much sensitive to the coupling parameter. We note that for small coupling parameter, for instance $\xi  =0.01$, the changes in the growth rate is not pronounced, while the changes are prominent as long as the coupling parameter increases (see the curves in Fig. \ref{fig-fc-ide1} corresponding to $\xi = 0.04$ and $0.07$).
We also note that the evolution of $f_c$, does not significantly depend on the DE state parameters.  Similarly, in Fig. \ref{fig-fc-ide2}, we present the same quantity, namely, $f_c$ but for IDE2 using different values of the coupling parameter, which again clearly shows that the growth history of the universe is significantly affected in presence of a non-zero coupling parameter. In both Figs. \ref{fig-fc-ide1} and \ref{fig-fc-ide2}, the violent changes in the evolution of $f_c$ appear mainly in the late times.    
Moreover, we also notice that for IDE2, the evolution of $f_c$ is slightly different than IDE1 and this differences mainly appear due to their different evolution, see Fig. \ref{fig:Q}. In fact, from Fig. \ref{fig-fc-ide1}, one can see that all the curves representing $f_c$ for $\xi \neq 0$ cross the curve $f_c$ with $\xi  =0$. This actually happens since the interaction function  (\ref{Q1}) has a sign changeable property. In summary, we observe that the growth history of the universe is highly dependent on the interaction in the dark sector, which is natural because in this case the evolution of CDM changes which directly affects the growth rate of CDM. For more discussions on this issue, we refer to \cite{Yang:2014gza,yang:2014vza}.

\begin{figure*}
\includegraphics[width=0.4\textwidth]{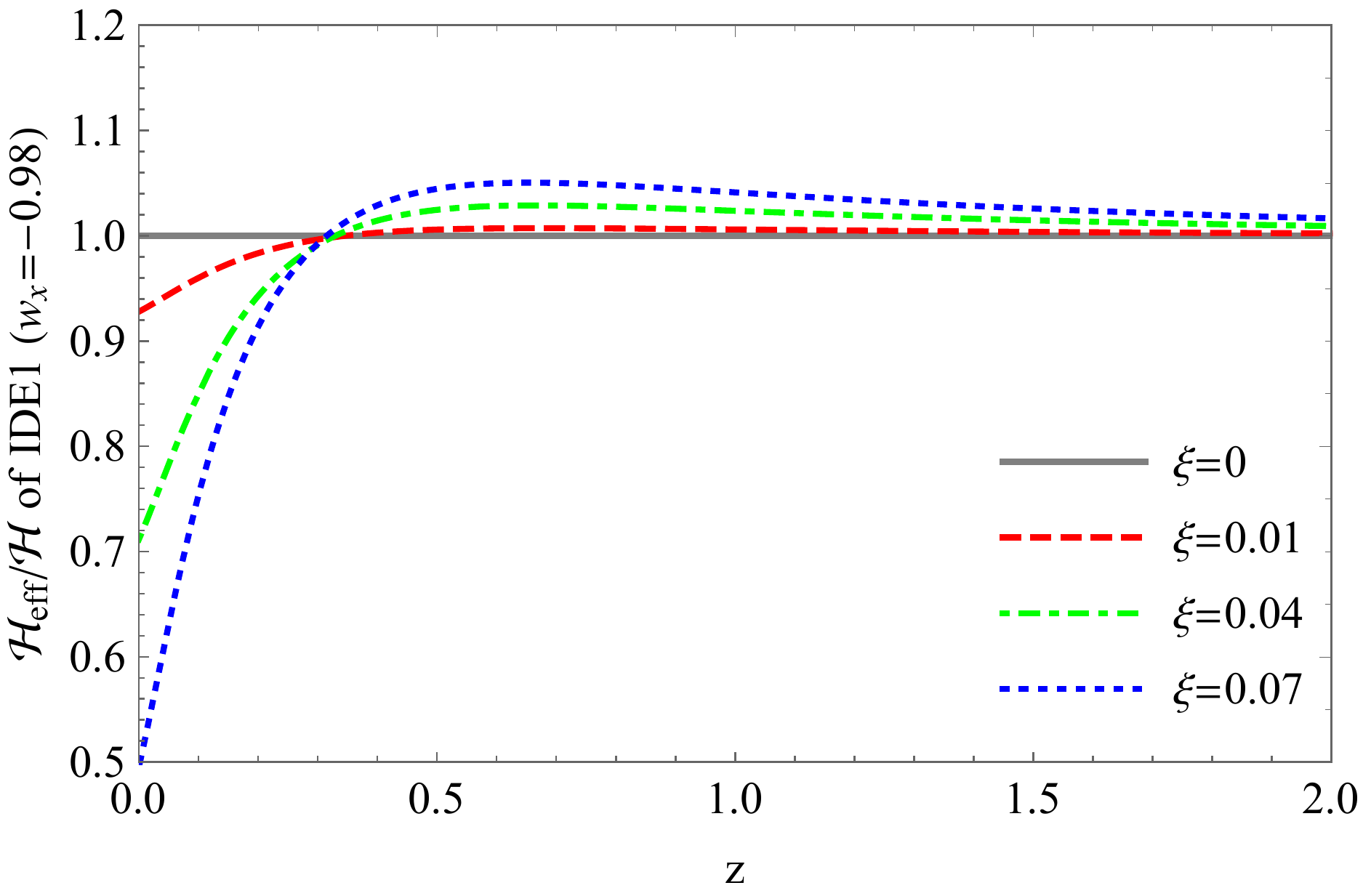} %
\includegraphics[width=0.4\textwidth]{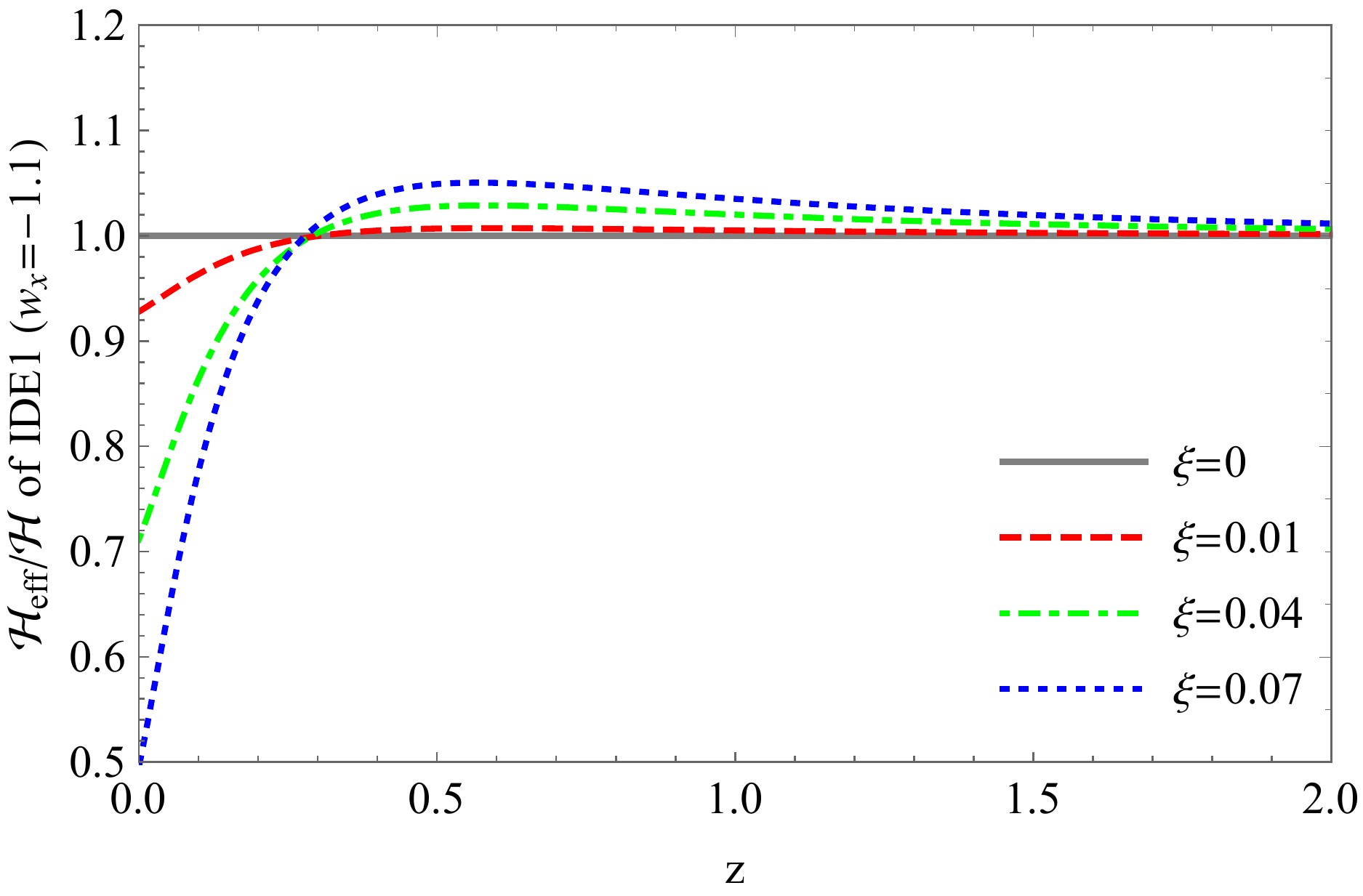} 
\caption{We show the evolution of the effective expansion history $%
\mathcal{H}_{\mathrm{eff}}$ for the IDE1 scenario. The left panel stands for
the quintessence DE state parameter ($w_x > -1$) whilst the right
panel stands for phantom DE state parameter ($w_x< -1$). Similar to earlier plots, here too, we fix $\Omega_{c0} = 0.28$, $\Omega_{x0} = 0.68$, $\Omega_{r0} = 0.0001$, and $\Omega_{b0} = 1- \Omega_{r0}-\Omega_{c0}-\Omega_{x0} =  0.0399$. }
\label{fig-H-ide1}
\end{figure*}

\begin{figure*}
\includegraphics[width=0.4\textwidth]{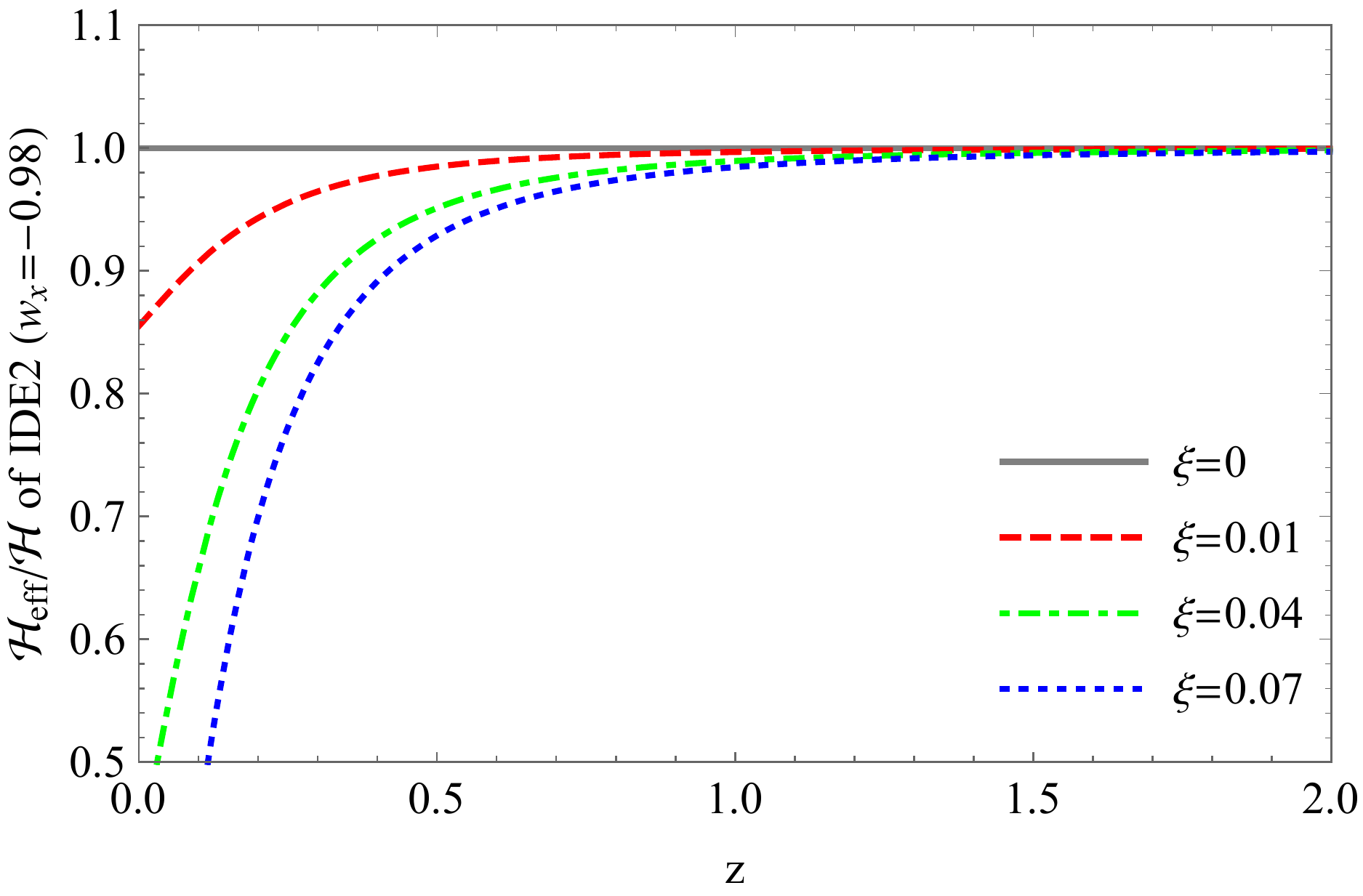} %
\includegraphics[width=0.4\textwidth]{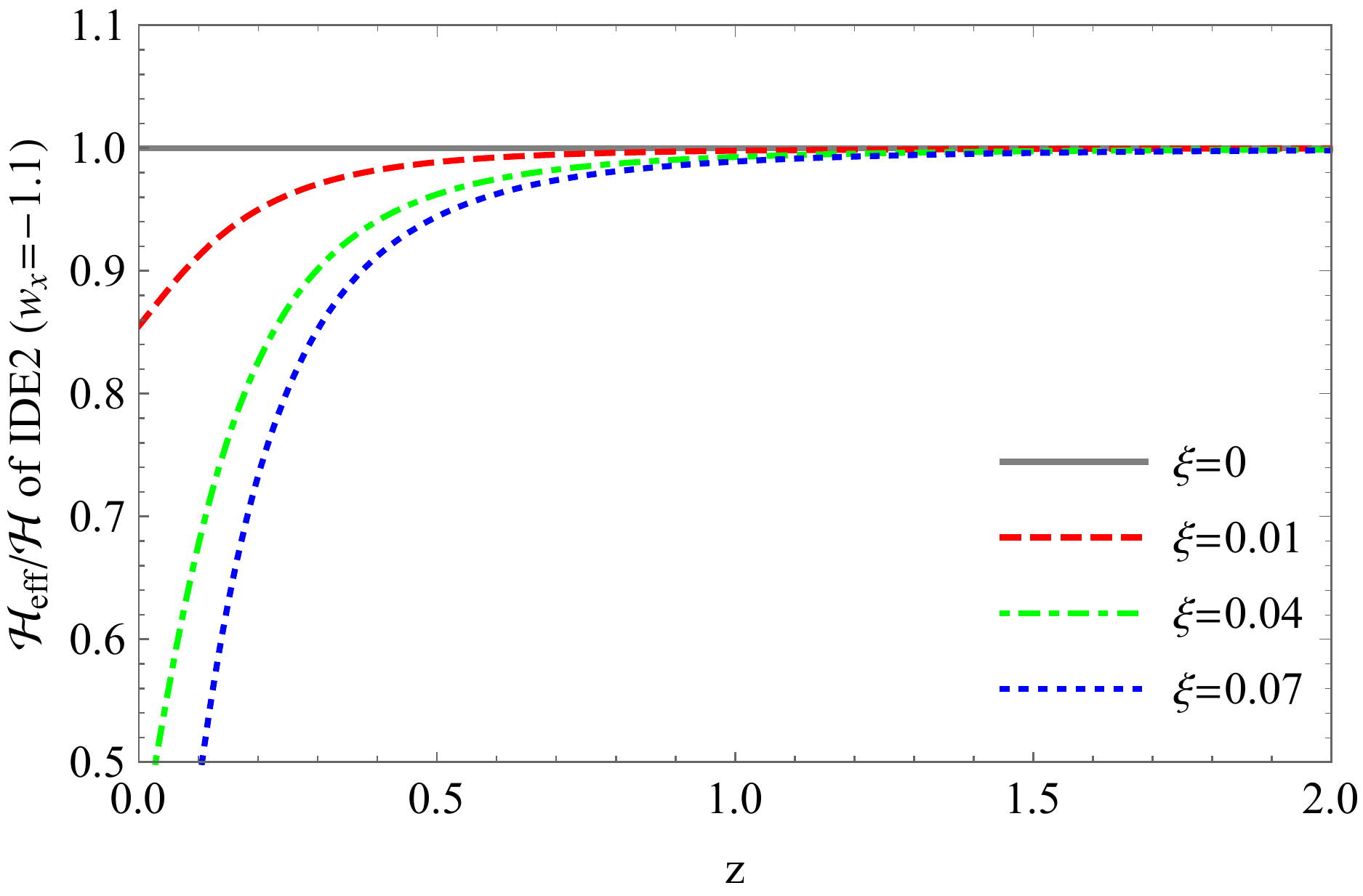} 
\caption{We show the evolution of the effective expansion history $%
\mathcal{H}_{\mathrm{eff}}$ for the IDE2 scenario. The left panel stands for
the quintessence DE state parameter ($w_x > -1$) whilst the right
panel stands for phantom DE state parameter ($w_x< -1$). The fixed values that we set while drawing the plots are, $\Omega_{c0} = 0.28$, $\Omega_{x0} = 0.68$, $\Omega_{r0} = 0.0001$, and $\Omega_{b0} = 1- \Omega_{r0}-\Omega_{c0}-\Omega_{x0} =  0.0399$. }
\label{fig-H-ide2}
\end{figure*}

Finally, we introduce the effective expansion history of the
universe, $\mathcal{H}_{\mathrm{eff}}$, which is the expansion history of
the universe in presence of the interaction function $Q$. The effective 
expansion history of the universe has the following expression

\begin{eqnarray}
\frac{\mathcal{H}_{\mathrm{eff}}}{\mathcal{H}}=\left(1-\frac{Q}{H\rho_c}%
\right).
\end{eqnarray}
which for $Q = 0$, recovers the non-interacting case $\mathcal{H}_{\mathrm{eff}} = \mathcal{H}$. We have graphically shown the behavior of $\mathcal{H}_{\mathrm{eff}}$ for both the IDE models. In Figs. \ref{fig-H-ide1} (for IDE1) and \ref{fig-H-ide2} (for IDE2) we show the effective expansion history of the universe for different values of the coupling parameter considering both quintessence and phantom dark energy state parameter. The left plot of Fig. \ref{fig-H-ide1} and \ref{fig-H-ide2} stands for $w_x> -1$ regime and the right plot of Fig. \ref{fig-H-ide1} and \ref{fig-H-ide2} stands for $w_x< -1$ regime. From all the  plots, one can clearly notice  that as long as the coupling parameter increases, the deviation of $\mathcal{H}_{\mathrm{eff}}$  becomes evident from its corresponding non-interacting expression, i.e. $\mathcal{H}$, and interestingly such effect remains independent of the dark
energy equation of state.


\section{The data and methodology}

\label{sec-data}

In this section we describe the observational datasets aiming to use for
analyzing the present interacting DE models. In what follows we
describe each dataset with their corresponding references.

\begin{enumerate}
\item \textbf{Cosmic microwave background data:} We consider the cosmic microwave background (CMB) temperature and polarization anisotropies together with their cross correlations from the latest Planck 2018 release \cite{Aghanim:2018oex,Aghanim:2019ame}. This dataset is known as the Planck TT,TE,EE+lowE. In this work  we refer to this dataset as Planck 2018.

\item \textbf{Baryon acoustic oscillations distance measurements:} 
We use the baryon acoustic oscillations (BAO) distance measurements from the experiments 6dFGS \cite{Beutler:2011hx}, SDDS MGS \cite{Ross:2014qpa}, and BOSS DR12 \cite{Alam:2016hwk}. One can quickly look at the Table 1 of \cite{Cheng:2019bkh} where the measured values of the observables related to these experiments are displayed.  

\item \textbf{Dark energy survey:}  We consider the measurements of the first-year Dark Energy Survey measurements~\cite{Krause:2017ekm, Troxel:2017xyo, Abbott:2017wau}.

\item \textbf{Hubble Space Telescope (HST):} Finally, we consider the very latest measurement of the Hubble constant from the Hubble Space Telescope. The measurement of the Hubble constant yields $H_0 = 74.03 \pm 1.42$ km/s/Mpc (at $68\%$ CL)~\cite{Riess:2019cxk}. Let us note that $H_0$ from   \cite{Riess:2019cxk} is in tension ($4.4 \sigma$) with the Planck's report within the $\Lambda$CDM cosmology. We refer to this data as R19 in this work.

\end{enumerate}

Now, using the above observational datasets, we have fitted both the
interaction scenarios where the algorithm to extract the cosmological
constraints is the well known Markov Chain Monte Carlo (MCMC) 
package \texttt{\small COSMOMC} \cite{Lewis:1999bs, Lewis:2002ah} in which a convergence diagnostic, namely the
Gelman and Rubin statistic is already equipped. Additionally the package
includes the support for the Planck 2018 Likelihood Code 
\cite{Aghanim:2019ame} (see \url{http://cosmologist.info/cosmomc/}). For both the
interaction scenarios, since the DE has a constant equation of
state, the parameters space of our interest is the following $\mathcal{P}%
\equiv \Bigl\{\Omega _{b}h^{2},\Omega _{c}h^{2},100\theta _{MC},\tau
,n_{s},log[10^{10}A_{S}],w_{x},\xi \Bigr\}$, where the symbols in $\mathcal{P%
}$ have the following meanings: $\Omega _{b}h^{2}$ is the baryons density; $%
\Omega _{c}h^{2}$ is the cold DM density; $100\theta _{MC}$ is the
ratio of sound horizon to the angular diameter distance; $\tau $ is the
optical depth; $n_{s}$, is the scalar spectral index and $A_{S}$ is the
amplitude of the initial power spectrum; $w_{x}$ is the constant equation of
state parameter for DE; $\xi $ is the coupling strength. In Table %
\ref{tab:priors} we describe the priors on the model parameters of the
interacting scenarios that we have taken during the statistical analysis.

\begin{table}
\begin{center}
\begin{tabular}{|c|c|}
\hline
Parameter & Prior \\ \hline
$\Omega_{b} h^2$ & $[0.005,0.1]$ \\ 
$\Omega_{c} h^2$ & $[0.01,0.99]$ \\ 
$\tau$ & $[0.01,0.8]$ \\ 
$n_s$ & $[0.5, 1.5]$ \\ 
$\log[10^{10}A_{s}]$ & $[2.4,4]$ \\ 
$100\theta_{MC}$ & $[0.5,10]$ \\ 
$\xi$ & $[0, 1]$ \\ 
$w_x$ & $[-2, 0]$ \\ \hline\hline
\end{tabular}%
\end{center}
\caption{Flat priors on various free parameters of the interacting scenarios used 
during the statistical analyses. }
\label{tab:priors}
\end{table}

\begingroup                                                                                                                     
\begin{center}                                                                                                                  
\begin{table*}
\scalebox{0.9}
{                                                                                                                   
\begin{tabular}{cccccccccccccc}                                                                                                            
\hline\hline                                                                                                                    
Parameters & Planck 2018 & Planck 2018+BAO & Planck 2018+DES & Planck 2018+R19 & Planck 2018+BAO+R19 \\ \hline
$\Omega_c h^2$ & $    0.1162_{-    0.0042-    0.0079}^{+    0.0038+    0.0074}$ & $    0.1149_{-    0.0033-    0.0052}^{+    0.0028+    0.0057}$ & $    0.1128_{-    0.0034-    0.0051}^{+    0.0025+    0.0056}$ &  $    0.1139_{-    0.0044-    0.0063}^{+    0.0033+    0.0068}$  & $    0.1138_{-    0.0041-    0.0056}^{+    0.0026+    0.0064}$  \\

$\Omega_b h^2$ & $    0.02232_{-    0.00015-    0.00029}^{+    0.00015+    0.00029}$ & $    0.02235_{-    0.00014-    0.00027}^{+    0.00014+    0.00027}$ & $    0.02240_{-    0.00013-    0.00027}^{+    0.00014+    0.00027}$ & $    0.02233_{-    0.00015-    0.00028}^{+    0.00014+    0.00029}$  & $    0.02233_{-    0.00015-    0.00028}^{+    0.00014+    0.00029}$ \\

$100\theta_{MC}$ & $    1.04085_{-    0.00037-    0.00073}^{+    0.00038+    0.00070}$ & $    1.04096_{-    0.00033-    0.00067}^{+    0.00035+    0.00064}$  & $    1.04110_{-    0.00032-    0.00068}^{+    0.00034+    0.00068}$ & $    1.04099_{-    0.00036-    0.00072}^{+    0.00037+    0.00069}$ & $    1.04100_{-    0.00035-    0.00068}^{+    0.00036+    0.00069}$ \\

$\tau$ & $    0.054_{-    0.0088-    0.015}^{+    0.0077+    0.017}$ & $    0.054_{-    0.0076-    0.016}^{+    0.0076+    0.016}$ & $    0.055_{-    0.0080-    0.016}^{+    0.0078+    0.016}$  & $    0.054_{-    0.0076-    0.015}^{+    0.0076+    0.016}$   &  $    0.054_{-    0.0082-    0.015}^{+    0.0075+    0.016}$  \\

$n_s$ & $    0.9726_{-    0.0046-    0.0088}^{+    0.0043+    0.0089}$ & $    0.9734_{-    0.0040-    0.0078}^{+    0.0041+    0.0078}$ & $    0.9745_{-    0.0039-    0.0076}^{+    0.0040+    0.0078}$  & $    0.9726_{-    0.0042-    0.0080}^{+    0.0042+    0.0082}$  & $    0.9723_{-    0.0043-    0.0074}^{+    0.0038+    0.0077}$ \\

${\rm{ln}}(10^{10} A_s)$ & $    3.054_{-    0.018-    0.031}^{+    0.016+    0.034}$ & $ 3.054_{-    0.015-    0.032}^{+    0.015+    0.033}$ & $    3.054_{-    0.016-    0.031}^{+    0.016+    0.032}$  & $    3.054_{-    0.016-    0.031}^{+    0.015+    0.032}$  & $    3.054_{-    0.015-    0.031}^{+    0.016+    0.031}$   \\

$w_x$ & $   -1.204_{-    0.178-    0.616}^{+    0.241+    0.456}$ & $   -1.096_{-    0.064-    0.120}^{+    0.062+    0.121}$ & $   -1.307_{-    0.136-    0.279}^{+    0.175+    0.266}$  & $   -1.253_{-    0.051-    0.101}^{+    0.054+    0.099}$  &  $   -1.187_{-    0.046-    0.095}^{+    0.046+    0.089}$  \\

$\xi$ & $    0.063_{-    0.053-    0.063}^{+    0.046+    0.052}$ & $    0.072_{-    0.027-    0.069}^{+    0.050+    0.054}$ & $    0.069_{-    0.026-    0.060}^{+    0.040+    0.052}$  & $    0.063_{-    0.030-    0.063}^{+    0.048+    0.053}$  & $    0.074_{-    0.023-    0.066}^{+    0.048+    0.053}$ \\

$\Omega_{m0}$ & $    0.272_{-    0.063-    0.118}^{+    0.047+    0.119}$ & $    0.289_{-    0.015-    0.030}^{+    0.015+    0.030}$ & $    0.235_{-    0.034-    0.054}^{+    0.026+    0.058}$  & $    0.250_{-    0.013-    0.023}^{+    0.011+    0.024}$  & $    0.266_{-    0.012-    0.021}^{+    0.0099+    0.021}$  \\

$\sigma_8$ & $    0.857_{-    0.070-    0.130}^{+    0.055+    0.157}$ & $    0.824_{-    0.021-    0.040}^{+    0.021+    0.040}$ & $    0.884_{-    0.049-    0.080}^{+    0.042+    0.081}$  & $    0.872_{-    0.017-    0.033}^{+    0.017+    0.034}$  & $    0.852_{-    0.017-    0.030}^{+    0.016+    0.031}$  \\

$H_0$ & $   72.67_{-    8.26-   15.21}^{+    5.43+   16.94}$ & $   69.17_{-    1.71-    2.78}^{+    1.53+    3.02}$ &  $   76.46_{-    6.28-    9.09}^{+    4.05+    9.93}$ & $   74.03_{-    1.40-    2.82}^{+    1.42+    2.78}$ & $   71.77_{-    1.17-    2.05}^{+    1.05+    2.18}$   \\

\hline\hline
                                                                                                                    
\end{tabular}
}                                                                                                                   
\caption{Observational constraints at 68\% and 95\% CL on
the first interaction scenario (IDE1) corresponding to the interaction rate
$Q (t)=3H\protect\xi\protect\rho_x\sin\left(\frac{\protect\rho_x}{%
\protect\rho_c}-1\right)$. Here, $\Omega_{m0}$ is the value of $\Omega_m =
\Omega_c +\Omega_b$ calculated at $z= 0$ (i.e., present epoch), and $H_0$ is in the
units of km/Mpc/sec. }
\label{tab:IDE1}                                                                                                   
\end{table*}                                                                                                                     
\end{center}                                                                                                                    
\endgroup     

\begin{figure*}
\includegraphics[width=0.32\textwidth]{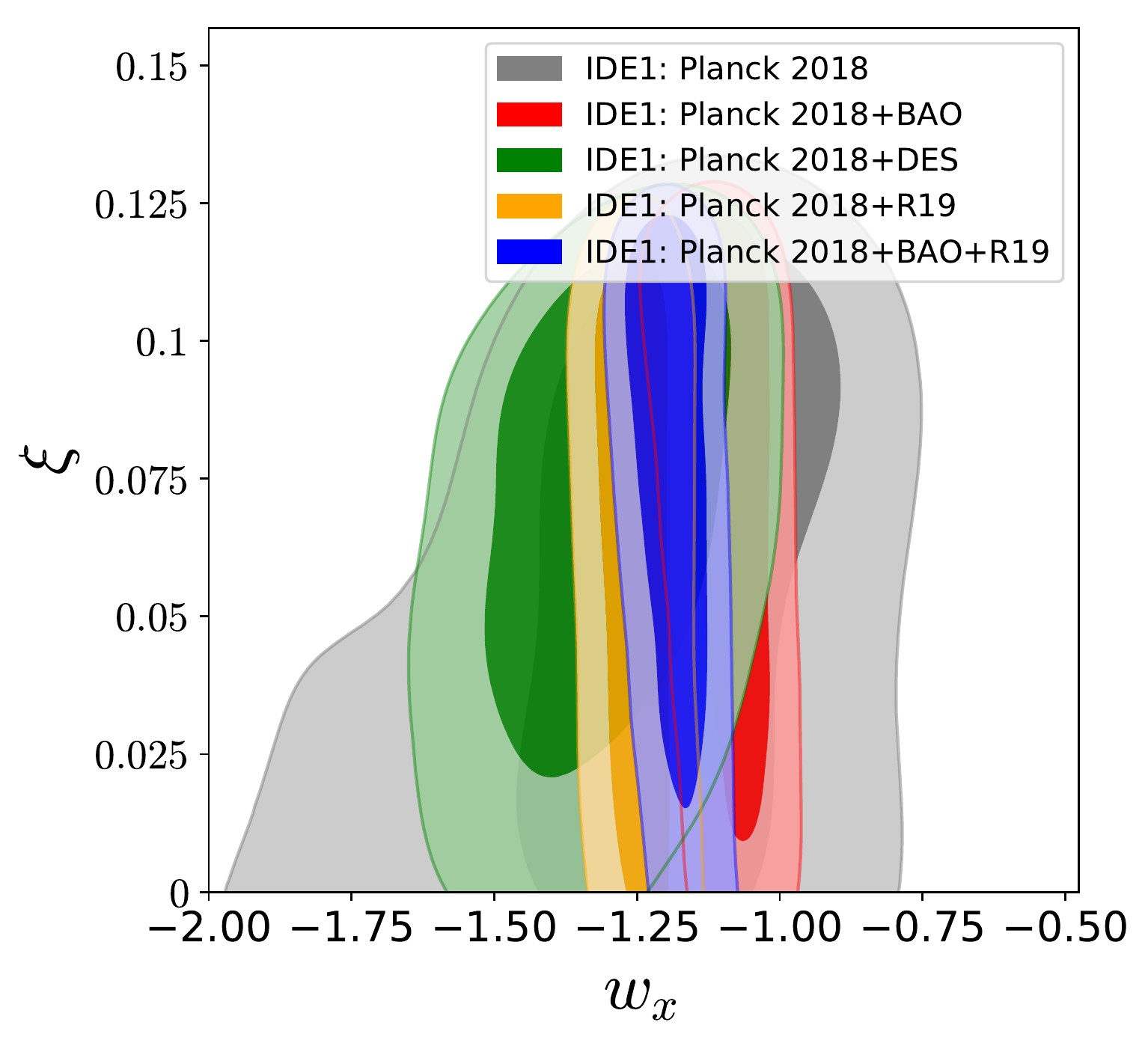}
\includegraphics[width=0.3\textwidth]{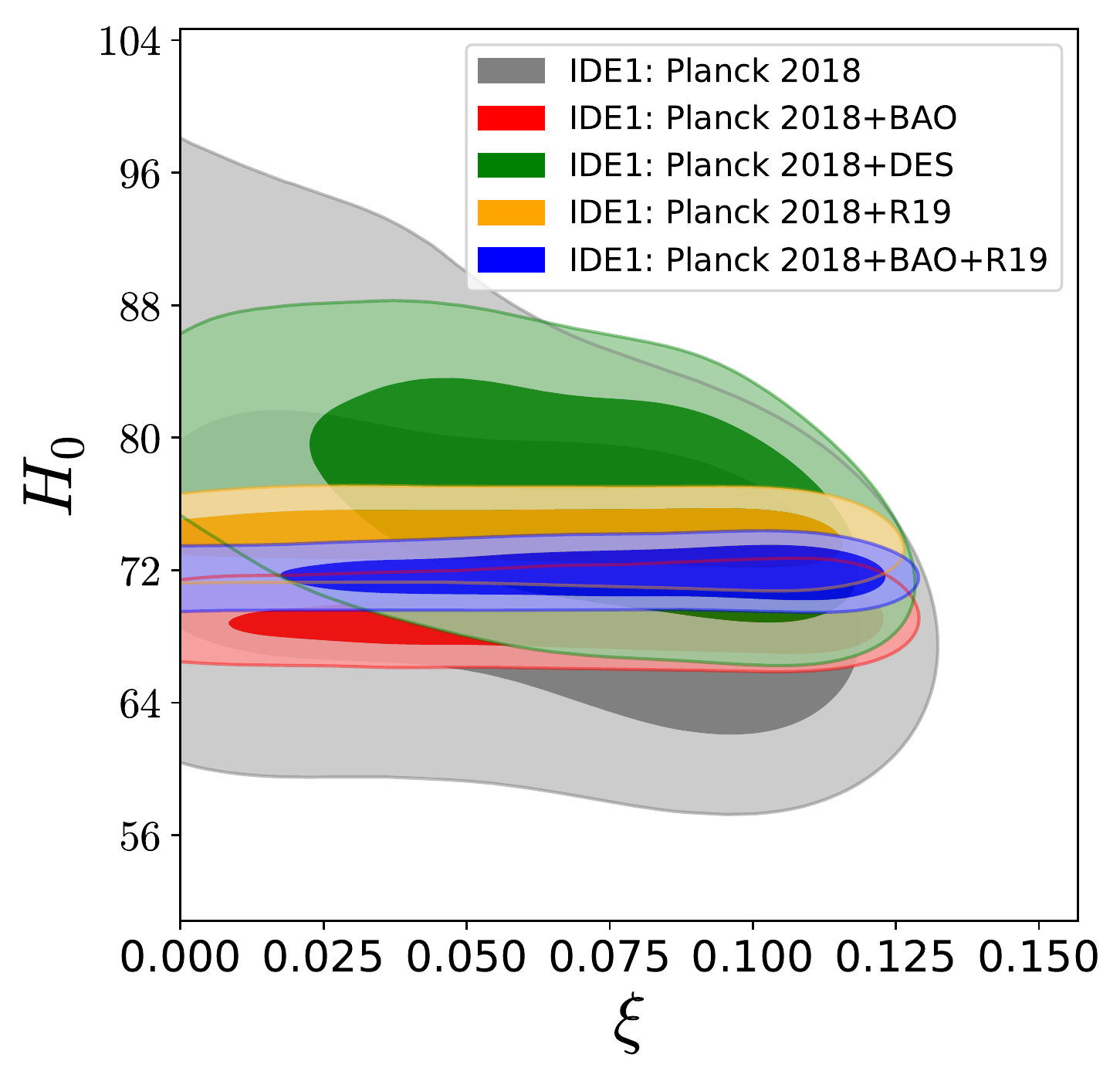}
\includegraphics[width=0.3\textwidth]{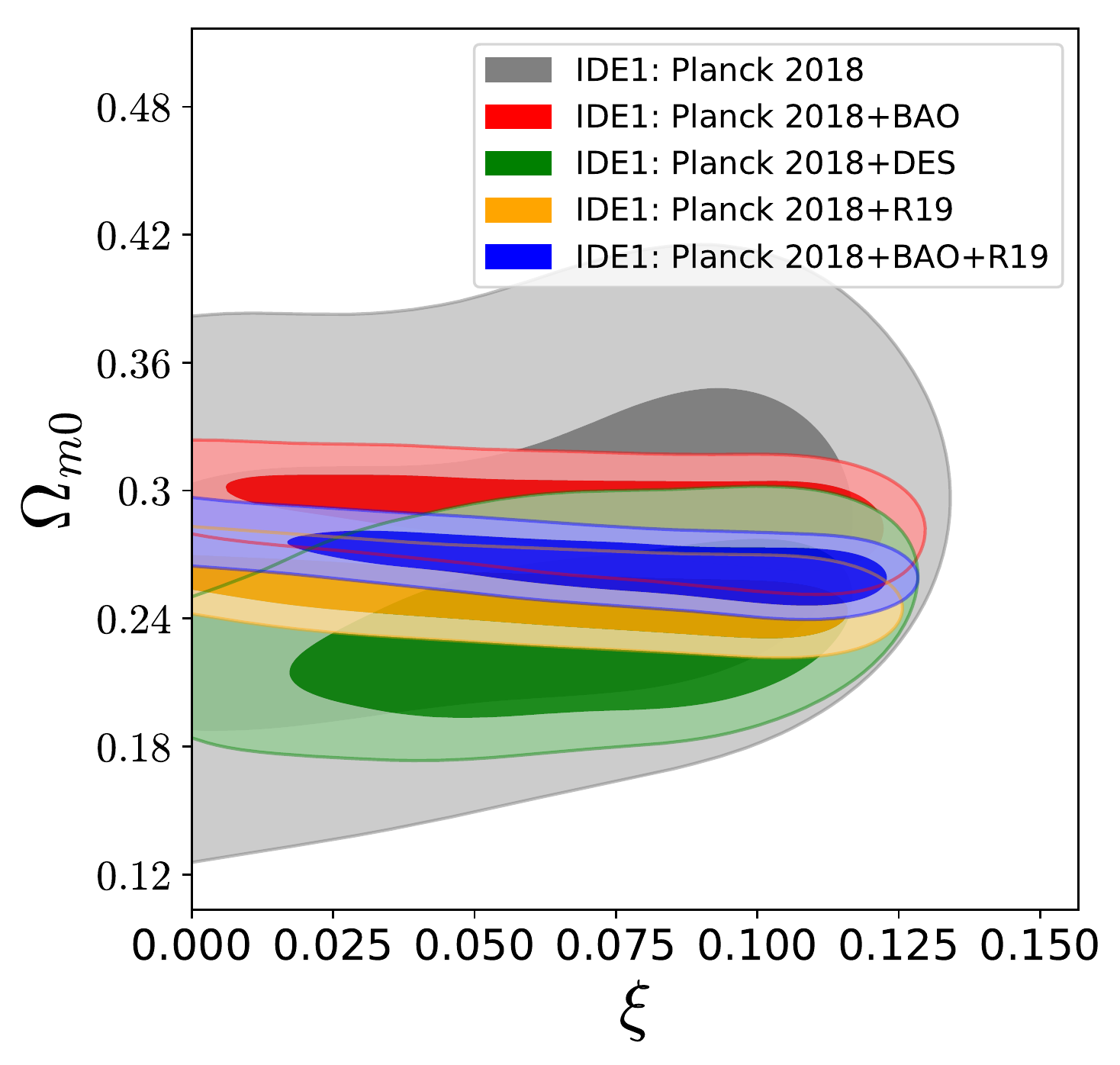}
\caption{68\% and 95\% confidence-level contour plots for the IDE1 scenario showing the effects of the coupling parameter on other observables using different observational datasets.  }
\label{contour-ide1}
\end{figure*}
\begin{figure*}
\includegraphics[width=0.6\textwidth]{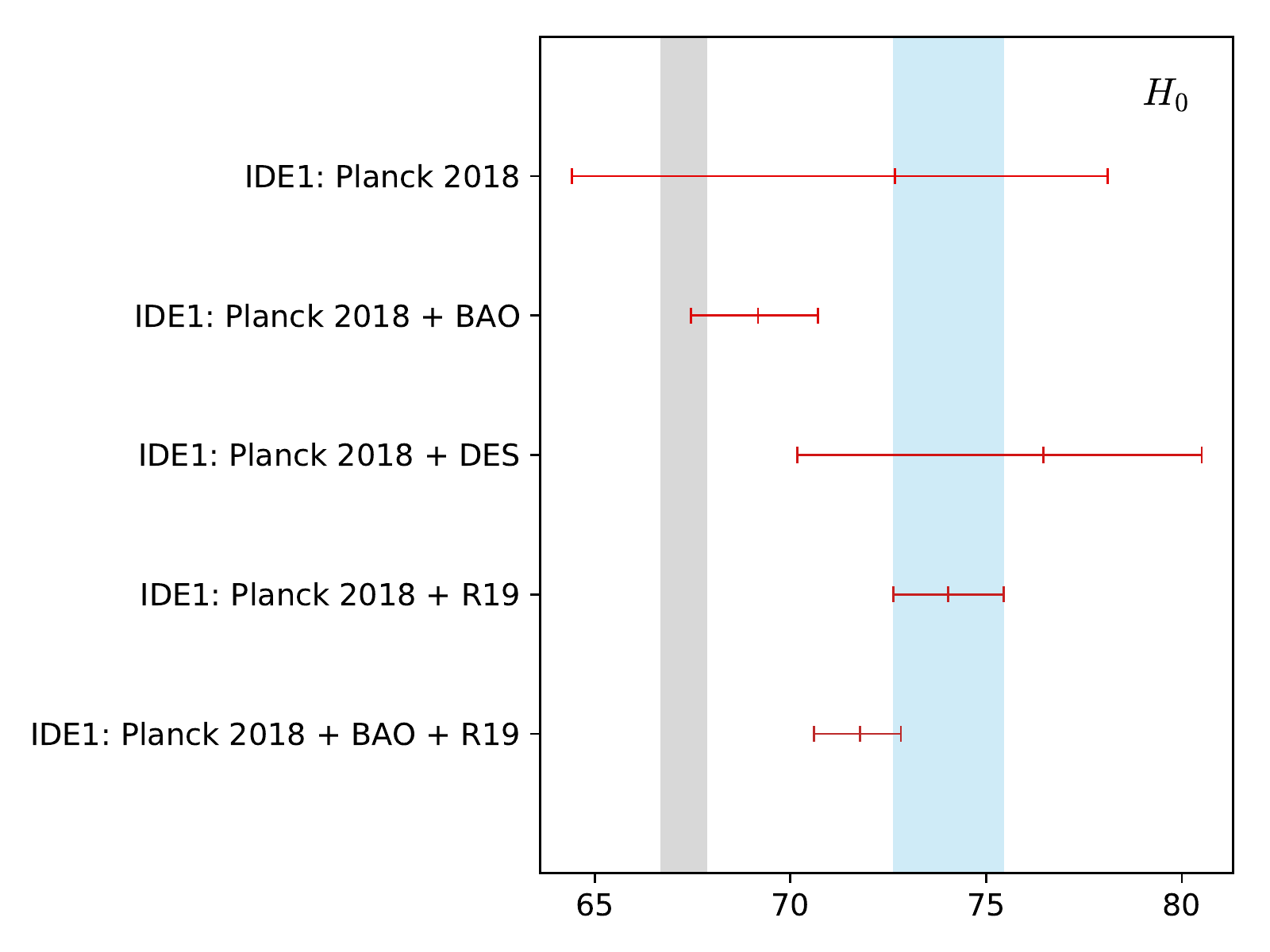}
\caption{Whisker plot with 68\% CL constraints on $H_0$ for the IDE1 scenario considering all the cosmological datasets employed in this work. The grey vertical corresponds to $H_0$ measurement from Planck 2018 release and the bluish vertical band stands for $H_0$ measurement (R19). }
\label{fig:whiskerH0-IDE1}
\end{figure*}
\begin{figure*}
\includegraphics[width=0.45\textwidth]{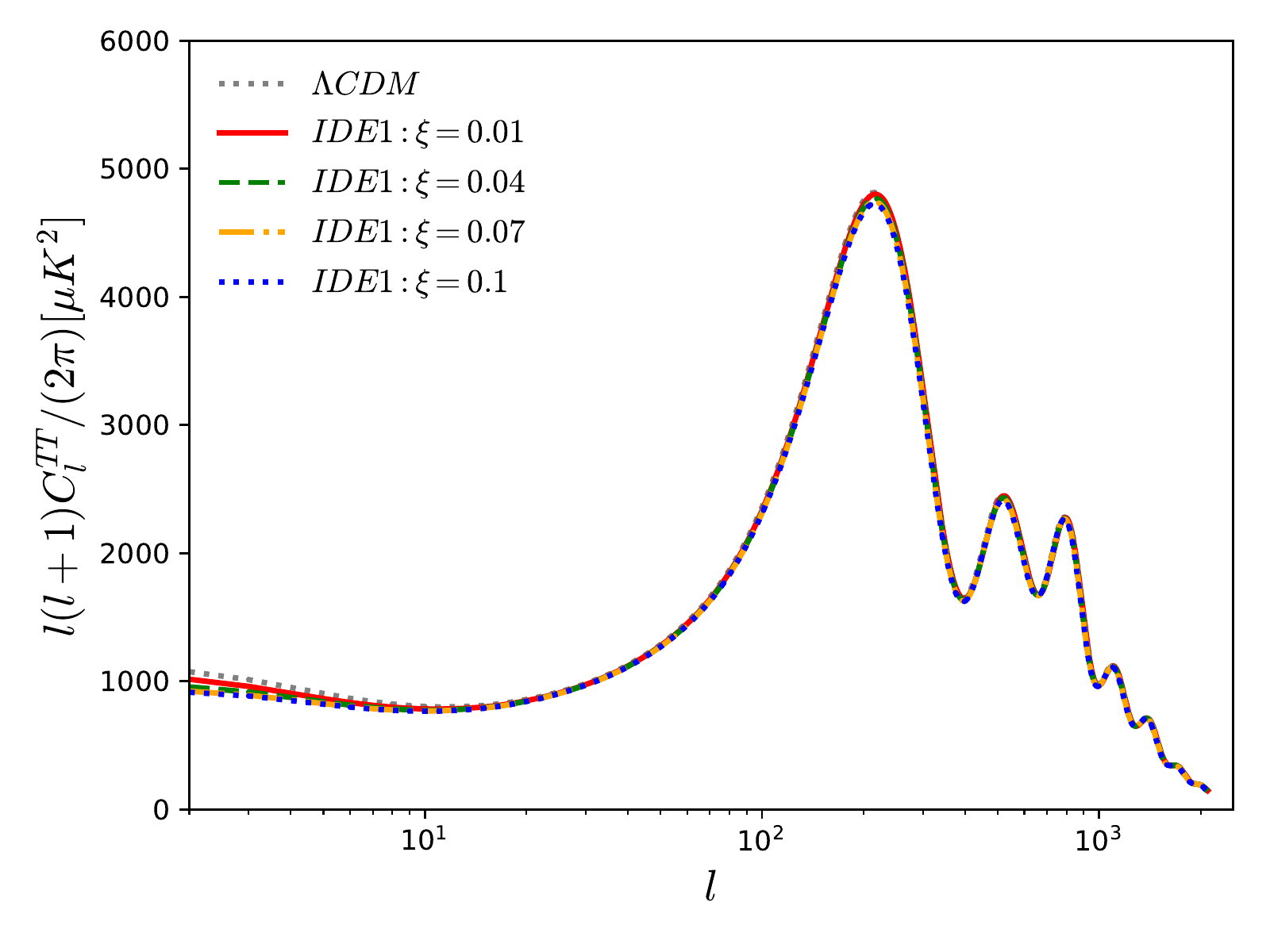}
\includegraphics[width=0.45\textwidth]{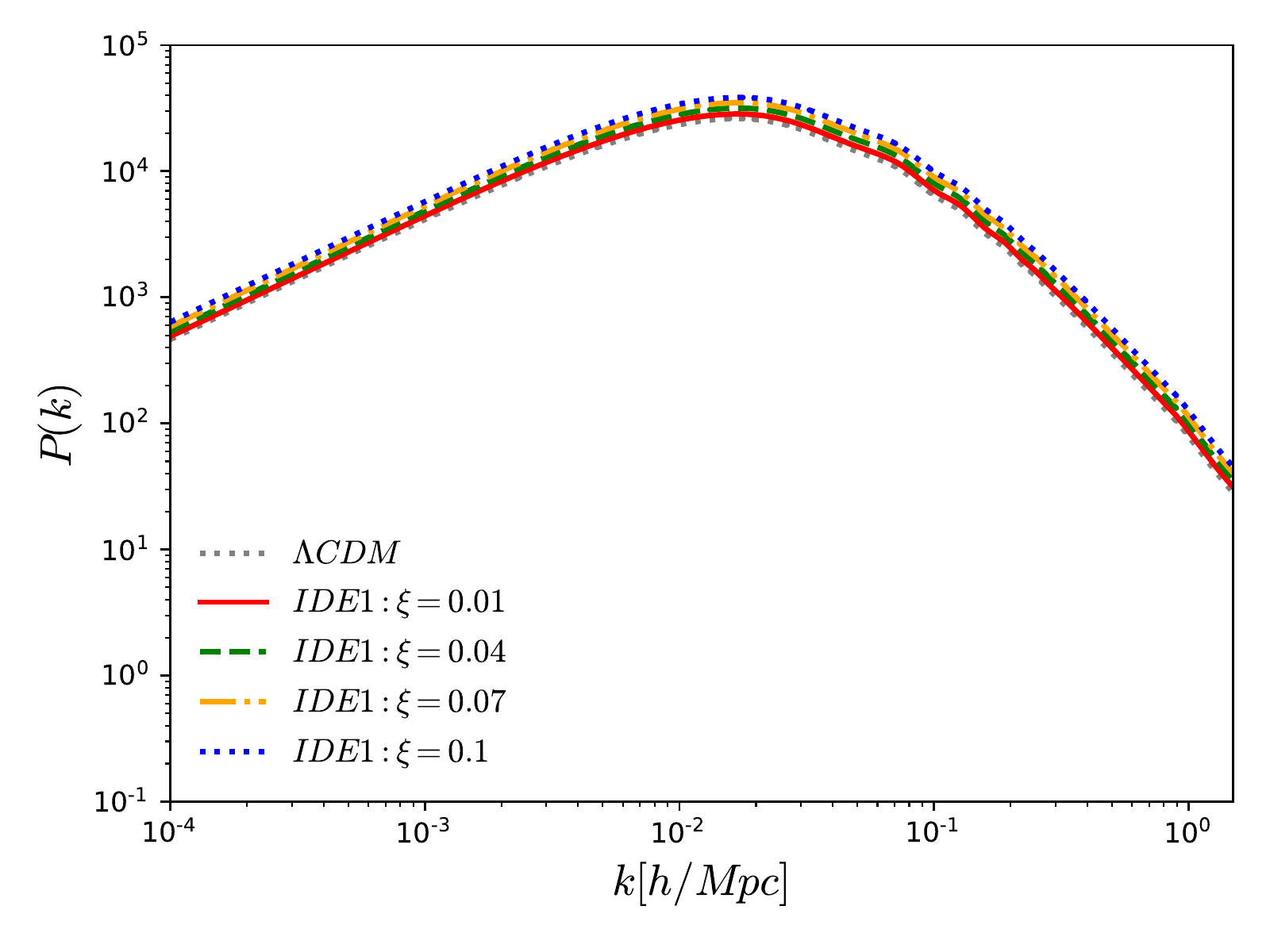}
\caption{CMB TT (left plot), matter power spectra (right plot) for IDE1  corresponding to the interaction function (\ref{Q1}) have been shown for different values of the coupling parameter $\xi$. Let us note that for all the plots we fix the parameters, $\Omega_{c0} = 0.28$, $\Omega_{x0} = 0.68$, $\Omega_{r0} = 0.0001$, and $\Omega_{b0} = 1- \Omega_{r0}-\Omega_{c0}-\Omega_{x0} =  0.0399$. }
\label{fig-ide1-TT-mower}
\end{figure*}
\begin{figure}
\includegraphics[width=0.45\textwidth]{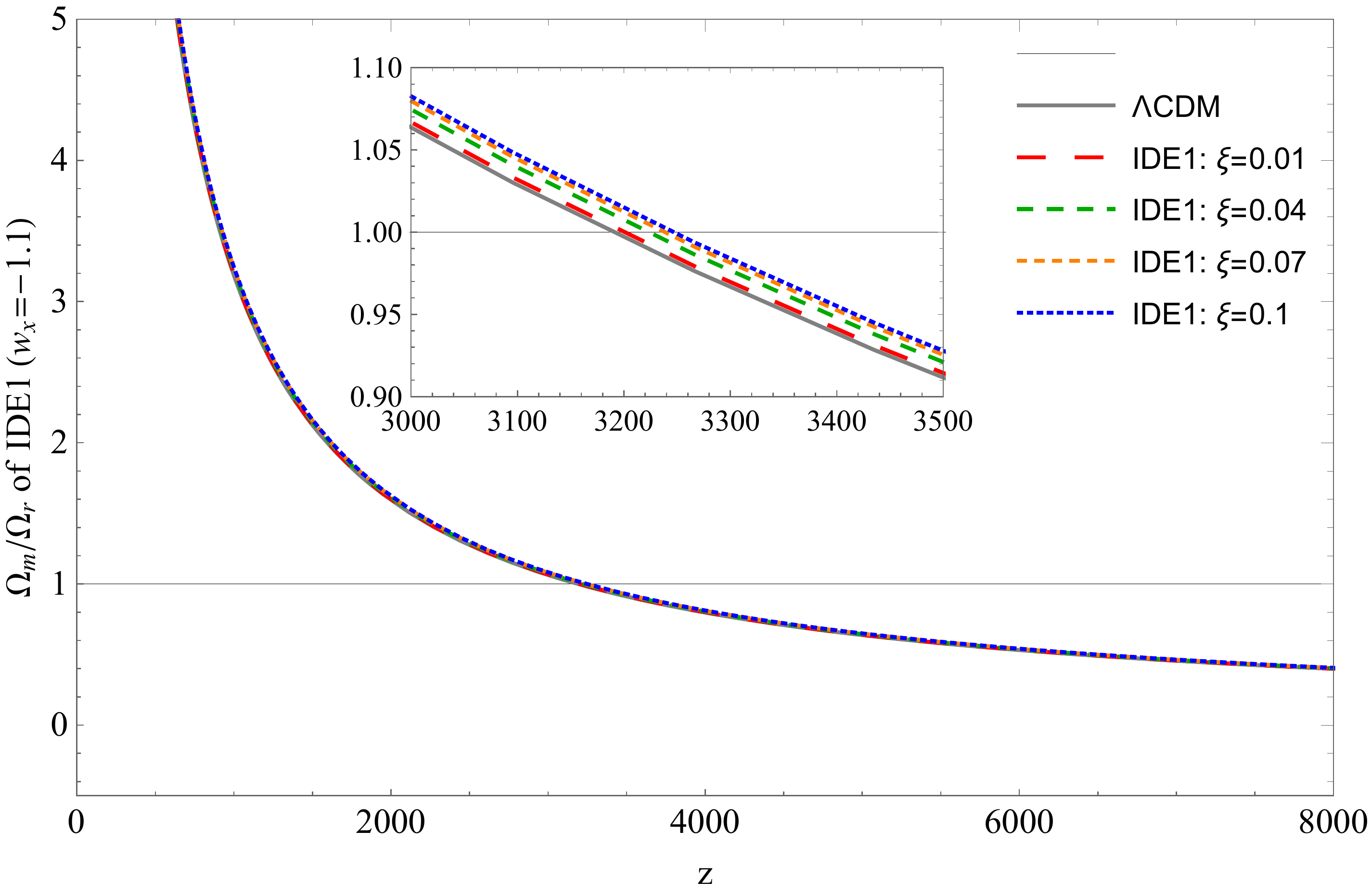}
\caption{The evolution of the quantity $\Omega_m/\Omega_r$ for the IDE1 scenario (corresponding to the interaction function in eqn. (\ref{Q1}))  has been shown for various values of the coupling parameter $\xi$. We note that here $\Omega_m = \Omega_c + \Omega_b$. The horizaontal line corresponds to the case, $\Omega_m = \Omega_r$ implying the matter-radiation equality. We see that the curves corresponding to different values of $\xi$ cannot be distinguished from one another, but actually they are different as shown in the subplot inside the main plot. The parameters that we fix to depict the plots are, $\Omega_{c0} = 0.28$, $\Omega_{x0} = 0.68$, $\Omega_{r0} = 0.0001$, and $\Omega_{b0} = 1- \Omega_{r0}-\Omega_{c0}-\Omega_{x0} =  0.0399$.}
\label{fig:ratio1}
\end{figure}

\section{Results and analysis}

\label{sec-results}

In this section we present the observational analysis for the present
interaction scenarios using CMB measurements from Planck 2018, BAO, DES and a local measurement of $H_0$ from HST. We first constrain the models using CMB data only from Planck 2018 and then added other datasets considering the tensions between them so that the results are not biased due to the tensions between the datasets. In what follows we describe the main observational consequences of the interacting scenarios and check how the coupling parameter.  

\subsection{IDE1: $Q(t)=3H\xi \rho _{x}\sin \left( \frac{\rho _{x}}{\rho _{c}}-1\right) $}

\label{sec-ide1}

The observational constraints  on this nonlinear interaction model at 68\% and 95\% confidence-level (CL) have been shown in Table \ref{tab:IDE1} for various cosmological datasets, namely, CMB alone from Planck 2018, Planck 2018+BAO, Planck 2018+DES, Planck 2018+R19 and Planck 2018+BAO+R19. And in Fig. \ref{contour-ide1} we show the $(w_x, \xi)$, $(\xi, H_0)$ and $(\xi, \Omega_{m0})$ planes at 68\% and 95\% CL. Fig. \ref{contour-ide1} is very interesting because as one can see, the coupling parameter is found to be uncorrelated with $w_x$, $H_0$ and $\Omega_{m0}$. In particular, the joint contour $(w_x, \xi)$ is vertical for all the datasets showing no correlation between them. The same is true for other two cases.  
We note that within this interaction scenario one could safely combine R19 data with Planck 2018 and Planck 2018+R19.  
Because as we shall show below, the Planck 2018 data for this interaction scenario do not show any tension with the local estimation of $H_0$. Similarly, for this model consideration the tension between the datasets, Planck 2018+BAO and R19 are less than $2\sigma$, thus one can safely add R19 with Planck 2018+BAO. Let us now describe how these datasets could constrain a possible coupling in the dark sector driven by this nonlinear interaction as well as we try to explore other consequences.  

In the second column of Table \ref{tab:IDE1} we show the constraints on this model for Planck 2018 alone. We find that the coupling parameter is non-null and within 68\% CL this gives its nonzero estimation  leading to $\xi = 0.063_{-    0.053}^{+    0.046}$ (68\% CL). The dark energy equation of state, $w_x$, although favors its phantom nature ($w_x =  -1.204_{-0.178}^{+    0.241}$ at 68\% CL), but as one can see, within 68\% CL, its quintessence nature is equally allowed. The most interesting result of this anlysis is that the Hubble constant is almost same with its local determination. We find that within this interaction scenario Planck 2018 data alone constrain $H_0 = 72.67_{-    8.26}^{+    5.43}$ km/s/Mpc (at 68\% CL) and the local determination of $H_0$ by Riess et al. $H_0 = 74.03 \pm 1.42$ km/s/Mpc (at $68\%$ CL)~\cite{Riess:2019cxk}. This clearly shows that the $H_0$ tension is resolved in presence of a non-zero interaction. Thus, we find that within this model consideration, a mild non-zero interaction  is able to successfully solve the $H_0$ tension. 

We now consider the analysis with Planck 2018+BAO. The inclusion of BAO to Planck 2018 is purely motivated to break the statistical degeneracies in the parameter space that may appear during the analysis with Planck 2018 data only. We find that 
when BAO data are added to Planck 2018,  the evidence for a non-zero coupling becomes stronger. Our results show that at more than 95\% CL, the coupling parameter remains nonzero ($\xi =  0.072_{- 0.069}^{+ 0.054}$, Planck 2018+BAO), and thus, it signals for a strong interaction in the dark sector.  We also find that $H_0$ is slightly goes down ($H_ 0= 69.17_{- 1.71}^{+    1.53}$ km/s/Mpc at 68\% CL) compared to Planck 2018 only ($H_0 = 72.67_{-    8.26}^{+    5.43}$ km/s/Mpc at 68\% CL) but due to its higher error bars, the tension is reduced and within 95\% CL, it is reconciled. We also notice that the dark energy equation has a phantom character at more than 68\% CL ($w_x= -1.096_{-    0.064}^{+    0.062}$, 68\% CL). For Planck 2018+DES, we find similar conclusion on $\xi$ as observe with Planck 2018+BAO. We find that for Planck 2018+DES, the evidence of a nonzero coupling also increases with $\xi \neq 0$ at more than 95\% CL together with a high value of the Hubble constant which is slightly higher than the local measurements but consistent to it within 68\% CL. Thus, Planck 2018+DES is in favor of alleviating the $H_0$ tension together with a nonzero coupling in the dark sector as well as the phantom nature of the dark energy equation of state at more than 95\% CL. 

We now come to the final two analyses, namely Planck 2018+R19 and Planck 2018+BAO+R19. We see that the constraint on $H_0$ from Planck 2018 is not in tension with its local measurements \cite{Riess:2019cxk}. Thus, one could safely consider the combined analysis Planck 2018+R19. On the other hand, concerning the constraints on $H_0$ since the tension between the datasets Planck 2018+BAO and R19 are less $3\sigma$, thus, the combination Planck 2018+BAO+R19 can also be considered safely. The results of Planck 2018+R19 and Planck 2018+BAO+R19 are shown in the last two columns of Table \ref{tab:IDE1}, respectively. Our analyses show that both the combinations show an evidence for a nonzero coupling  in the dark sector. However, for Planck 2018+R19, $\xi$ allows its zero values within 95\% CL, while for Planck 2018+BAO+R19, $\xi \neq 0$ at more than 95\% CL.   For both the datasets, we find an evidence of a phantom DE at more than 95\% CL. Additionally, we find that the estimations of $H_0$ are influenced by R19.  
As one can see that $H_ 0 =  74.03_{-    1.40}^{+    1.42}$ km/Mpc/sec (68\% CL, Planck 2018+R19) and  $H_0 =    71.77_{-    1.17}^{+    1.05}$ km/Mpc/sec (68\% CL, Planck 2018+BAO+R19). Thus, for Planck 2018+R19, $H_0$ tension is perfectly solved. For a better understanding on how $H_0$ tension is solved/alleviated within this interaction scenario we show the whisker plot for $H_0$ at 68\% CL in Fig. \ref{fig:whiskerH0-IDE1}
considering all the observational datasets.

Finally, we focus on the cosmological implications of the interaction in the large scale of our Universe via CMB temperature power spectra (CMB TT) and matter power spectra. To understand how the coupling parameter influences the cosmological dynamics,   
in Fig. \ref{fig-ide1-TT-mower}, we present the CMB TT (left graph of Fig. \ref{fig-ide1-TT-mower}) and matter power spectra (right graph of Fig. \ref{fig-ide1-TT-mower}) for different values of the coupling parameter $\xi$. We note that in Fig. \ref{fig-ide1-TT-mower} we have also included the non-interacting $\Lambda$CDM case ($\xi  = 0$) in order to show how the model with $\xi \neq 0$ behaves differently with $\xi  = 0$ case (here $\Lambda$CDM). 
From the left panel of Fig. \ref{fig-ide1-TT-mower}, we find that with the increase of the coupling parameter, the height of the first acoustic peak in the CMB TT mildly decreases relative to the height of the first acoustic peak corresponding to the non-interacting case (here $\Lambda$CDM model). The reason of such changes can be realized as follows. In  presence of an interaction, the energy density of CDM does not follow the usual evolution law ($\rho_c \propto a^{-3}$), and hence the evolution of the entire matter sector, $\Omega_m \; (= \Omega_c +\Omega_b)$ changes from its usual law and as a consequence the matter-radiation quality may alter. In Fig. \ref{fig:ratio1} we show the evolution of $\Omega_{m}/\Omega_r$ for different values of the coupling parameter where the epoch at which the equality $\Omega_m = \Omega_r$ holds, is known as the matter radiation equality. From Fig. \ref{fig:ratio1} (see specifically the subplot in Fig. Fig. \ref{fig:ratio1}), we find that due to an interaction in the dark sector, 
the matter-radiation equality occurs earlier compared to the non-interacting case ($\xi =0$). In case of this scenario, the shift of matter-radiation equality is very mild. However, due to earlier matter-radiation equality, the sound horizon is decreased, and as a consequence, the first peak of CMB TT is decreased. 
Moreover, our investigations also reveal that in the lower multipole region some changes in the CMB TT spectra are visible. Precisely, we find that, as $\xi$ increases, the amplitude of the corresponding CMB TT spectrum decreases, relative to the non-interacting case (left panel of Fig. \ref{fig-ide1-TT-mower}). 
As the presence of interaction changes the evolution of CDM from its usual law (without interaction),  hence this directly affects the CMB TT spectra in the lower multipole region via integrated Sachs-Wolfe (ISW) effect due to the gravitational potential.  Let us now discuss the effects of interaction in the matter power spectra. From the matter power spectra for this model (right plot of \ref{fig-ide1-TT-mower}),  
one can notice that the evolution in matter power spectra is completely opposite to the CMB TT spectra. The amplitude of the matter power spectra increases with the increase of the coupling parameter $\xi$. The reason for such enhancement is due to the earlier matter-radiation equality as shown in Fig. \ref{fig:ratio1}.

\begingroup                                                                                                                     
\begin{center}                                                                                                                  
\begin{table*}
\scalebox{0.9}
{                                                                                                                   
\begin{tabular}{cccccccccccccccc}                                                                                                            
\hline\hline                                                                                                                    
Parameters & Planck 2018 & Planck 2018+BAO & Planck 2018+DES & Planck 2018+R19 & Planck 2018+BAO+R19 \\ \hline

$\Omega_c h^2$ & $    0.0821_{-    0.0167-    0.0553}^{+    0.0380+    0.0444}$ & $    0.1028_{-    0.0089-    0.0235}^{+    0.0164+    0.0196}$  & $    0.1081_{-    0.0044-    0.0091}^{+    0.0047+    0.0093}$   & $    0.1034_{-    0.0079-    0.0237}^{+    0.0158+    0.0200}$  & $    0.1138_{-    0.0040-    0.0055}^{+    0.0025+    0.0066}$   \\

$\Omega_b h^2$ & $    0.02230_{-    0.00015-    0.00028}^{+    0.00015+    0.00030}$ & $    0.02233_{-    0.00014-    0.00027}^{+    0.00014+    0.00027}$ & $    0.02236_{-    0.00015-    0.00029}^{+    0.00015+    0.00029}$  & $    0.02230_{-    0.00015-    0.00028}^{+    0.00015+    0.00030}$ & $    0.02232_{-    0.00015-    0.00029}^{+    0.00015+    0.00028}$ \\

$100\theta_{MC}$ & $    1.04300_{-    0.00250-    0.00306}^{+    0.00101+    0.00397}$ & $    1.04166_{-    0.00102-    0.00135}^{+    0.00064+    0.00155}$  & $    1.04135_{-    0.00037-    0.00074}^{+    0.00037+    0.00076}$  & $    1.04159_{-    0.00098-    0.00140}^{+    0.00057+    0.00164}$  & $    1.04101_{-    0.00036-    0.00069}^{+    0.00038+    0.00069}$  \\

$\tau$ & $    0.054_{-    0.0082-    0.015}^{+    0.0078+    0.017}$ & $    0.055_{-    0.0082-    0.015}^{+    0.0071+    0.016}$  & $    0.055_{-    0.0075-    0.015}^{+    0.0074+    0.016}$  & $    0.054_{-    0.0077-    0.015}^{+    0.0077+    0.016}$  & $    0.054_{-    0.0076-    0.015}^{+    0.0075+    0.014}$  \\

$n_s$ & $    0.9720_{-    0.0045-    0.0083}^{+    0.0042+    0.0085}$ & $    0.9731_{-    0.0041-    0.0079}^{+    0.0042+    0.0080}$ & $    0.9728_{-    0.0043-    0.0089}^{+    0.0049+    0.0085}$  & $    0.9722_{-    0.0042-    0.0078}^{+    0.0040+    0.0080}$  & $    0.9724_{-    0.0040-    0.0081}^{+    0.0040+    0.0076}$  \\

${\rm{ln}}(10^{10} A_s)$ & $    3.055_{-    0.018-    0.030}^{+    0.016+    0.033}$ & $    3.056_{-    0.016-    0.031}^{+    0.015+    0.033}$  & $    3.056_{-    0.016-    0.031}^{+    0.015+    0.032}$   & $    3.054_{-    0.017-    0.031}^{+    0.016+    0.032}$  & $    3.054_{-    0.016-    0.029}^{+    0.015+    0.029}$  \\

$w_x$ & $   -1.460_{-    0.418-    0.498}^{+    0.201+    0.611}$ & $   -1.038_{-    0.057-    0.118}^{+    0.058+    0.122}$  & $   -1.299_{-    0.147-    0.283}^{+    0.149+    0.297}$  & $   -1.211_{-    0.054-    0.106}^{+    0.054+    0.107}$  & $   -1.189_{-    0.046-    0.091}^{+    0.047+    0.090}$  \\

$\xi$ & $    0.106_{-    0.106-    0.106}^{+    0.027+    0.166}$ & $    0.042_{-    0.042-    0.042}^{+    0.013+    0.055}$ & $    0.027_{-    0.015-    0.025}^{+    0.012+    0.025}$  & $    0.042_{-    0.042-    0.042}^{+    0.0090+    0.061}$  & $    0.074_{-    0.022-    0.066}^{+    0.046+    0.052}$  \\

$\Omega_{m0}$ & $    0.162_{-    0.100-    0.131}^{+    0.046+    0.171}$ & $    0.264_{-    0.028-    0.056}^{+    0.036+    0.051}$ & $    0.223_{-    0.032-    0.052}^{+    0.027+    0.057}$  & $    0.230_{-    0.021-    0.048}^{+    0.027+    0.044}$  & $    0.265_{-    0.012-    0.020}^{+    0.0096+    0.021}$  \\

$\sigma_8$ & $    0.974_{-    0.060-    0.197}^{+    0.128+    0.151}$ &  $    0.835_{-    0.023-    0.040}^{+    0.019+    0.043}$ & $    0.902_{-    0.042-    0.089}^{+    0.040+    0.083}$  & $    0.884_{-    0.019-    0.035}^{+    0.019+    0.036}$  & $    0.853_{-    0.016-    0.031}^{+    0.016+    0.030}$   \\

$H_0$ & $   84.33_{-    7.39-   21.95}^{+   15.46+   17.91}$ & $   69.06_{-    1.61-    2.82}^{+    1.36+    2.98}$ & $   77.11_{-    4.79-   10.13}^{+    4.76+    9.55}$  & $   74.15_{-    1.37-    2.71}^{+    1.36+    2.71}$  & $   71.81_{-    1.08-    2.11}^{+    1.07+    2.07}$  \\

\hline\hline                                                                                                                    
\end{tabular}
}                                                                                                                   
\caption{Observational constraints at 68\% and 95\% CL on
the second interaction scenario (IDE2) corresponding to the interaction rate
$Q(t)=3H\xi \rho _{x}\left[ 1+\sin \left( \frac{\rho _{x}}{\rho _{c}}-1\right) \right]$. Here, $\Omega_{m0}$ is the value of $\Omega_m =
\Omega_c +\Omega_b$ calculated at $z= 0$ (i.e., present epoch), and $H_0$ is in the
units of km/Mpc/sec. }
\label{tab:IDE2}                                                                                                   
\end{table*}                                                                                                                     
\end{center}                                                                                                                    
\endgroup                                                                                                                       
\begin{figure*}
\includegraphics[width=0.32\textwidth]{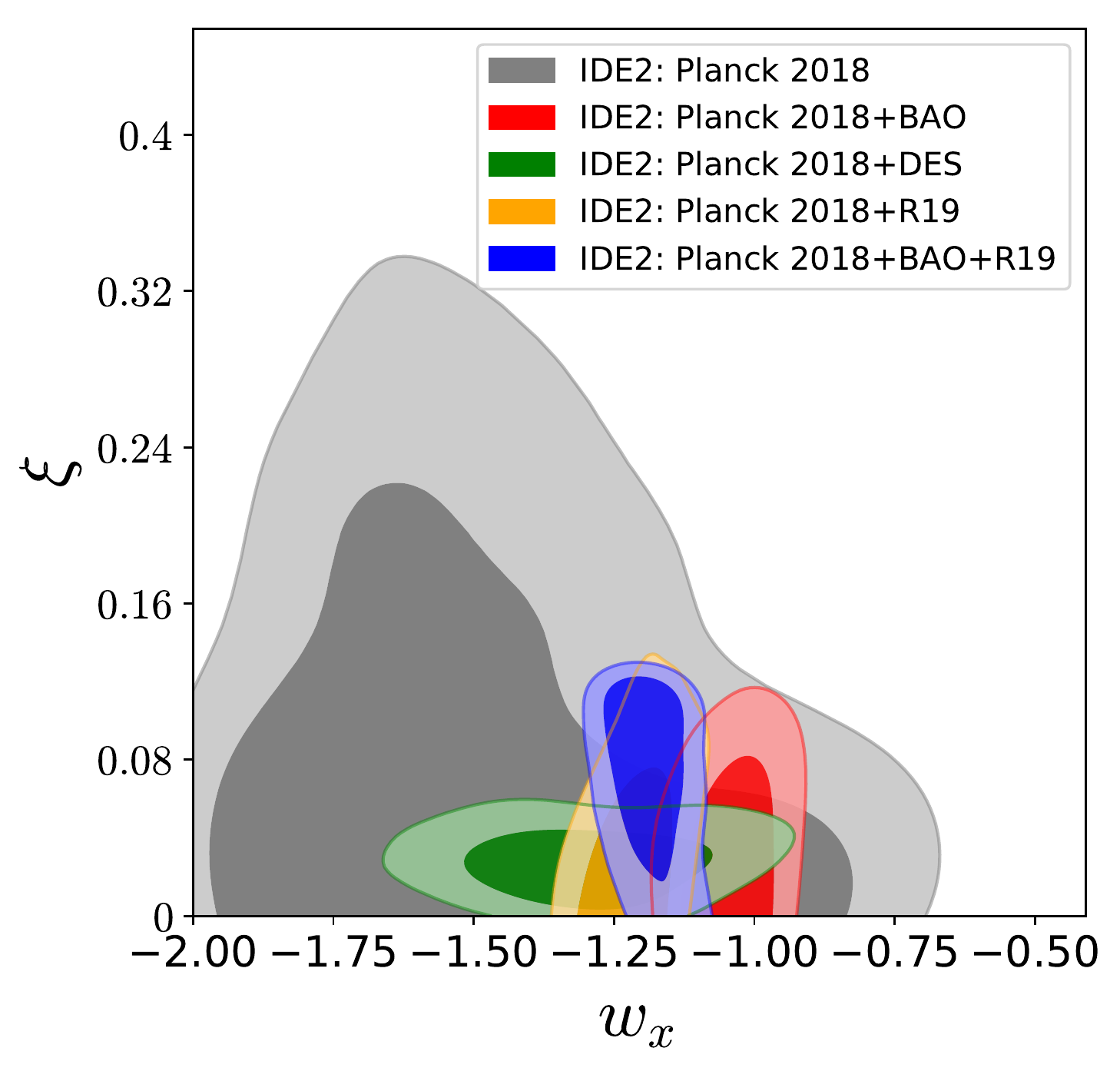}
\includegraphics[width=0.3\textwidth]{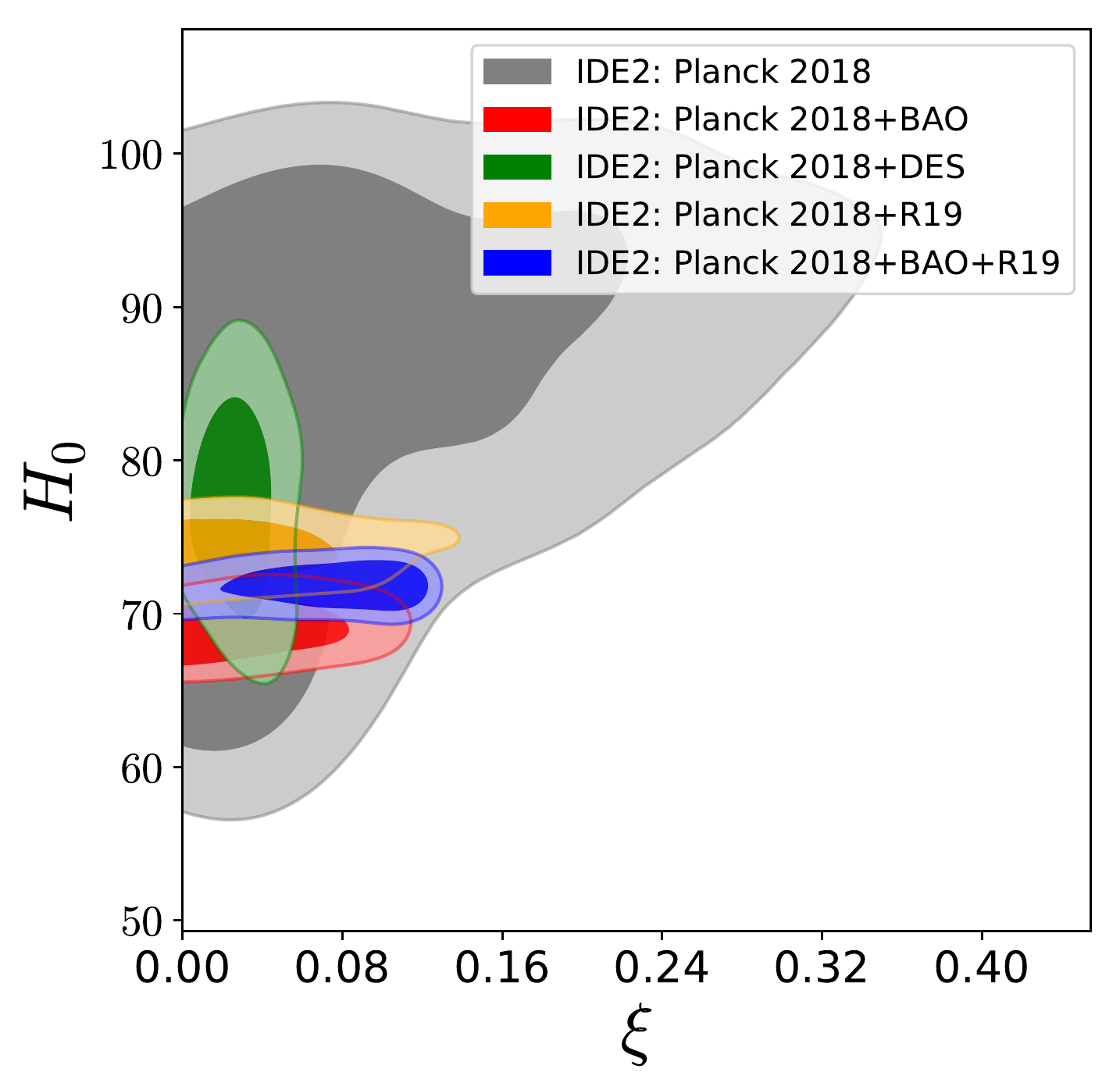}
\includegraphics[width=0.3\textwidth]{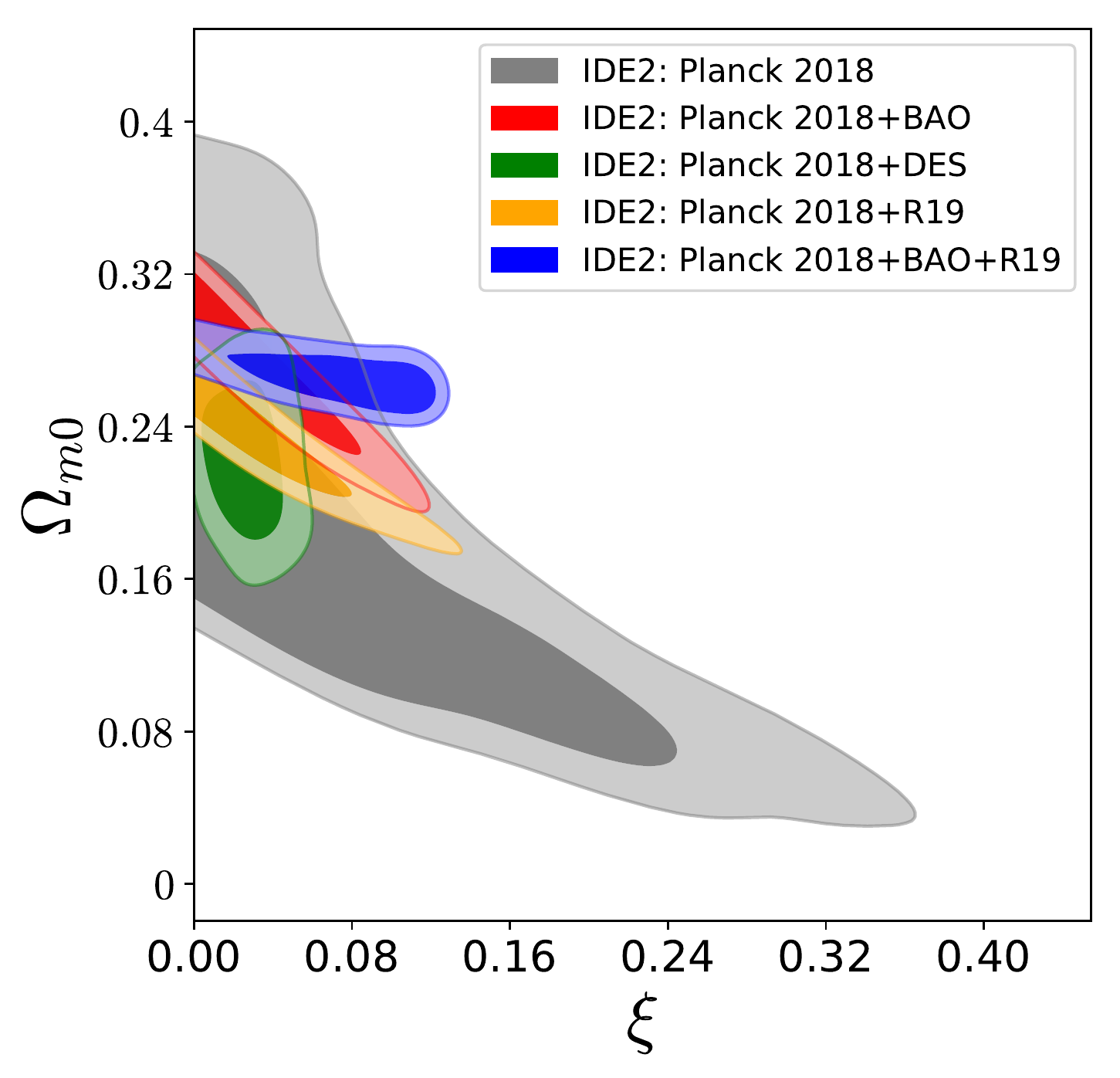}
\caption{68\% and 95\% confidence-level contour plots between several
combinations of the model parameters for the IDE2 scenario using diferent
observational datasets. }
\label{contour-ide2}
\end{figure*}
\begin{figure*}
\includegraphics[width=0.6\textwidth]{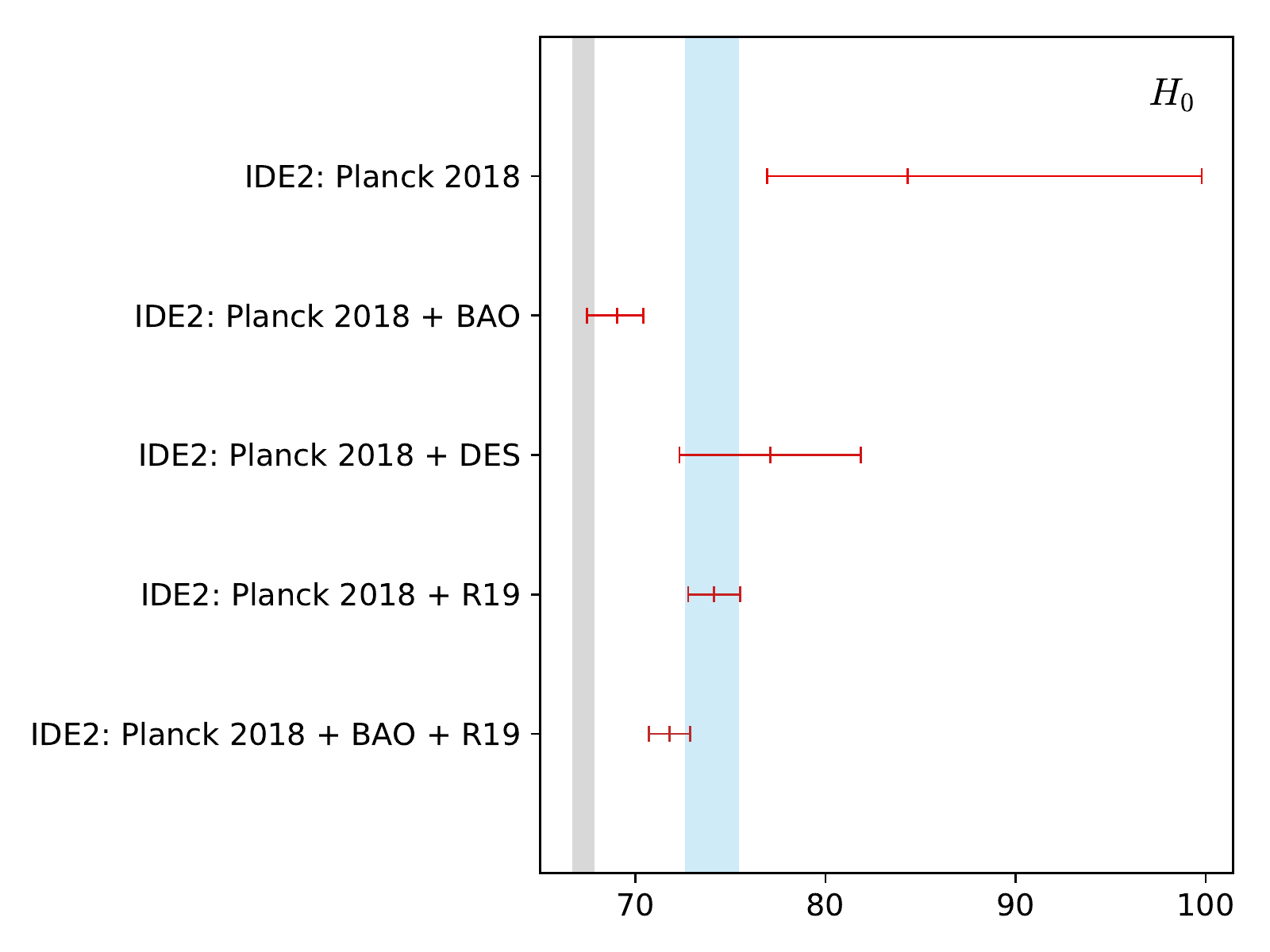}
\caption{Whisker plot with 68\% CL constraints on $H_0$ for the IDE2 scenario considering all the cosmological datasets employed in this work. The grey vertical corresponds to $H_0$ measurement from Planck 2018 release and the bluish vertical band stands for $H_0$ measurement (R19). }
\label{fig:whiskerH0-IDE2}
\end{figure*}
\begin{figure*}
\includegraphics[width=0.45\textwidth]{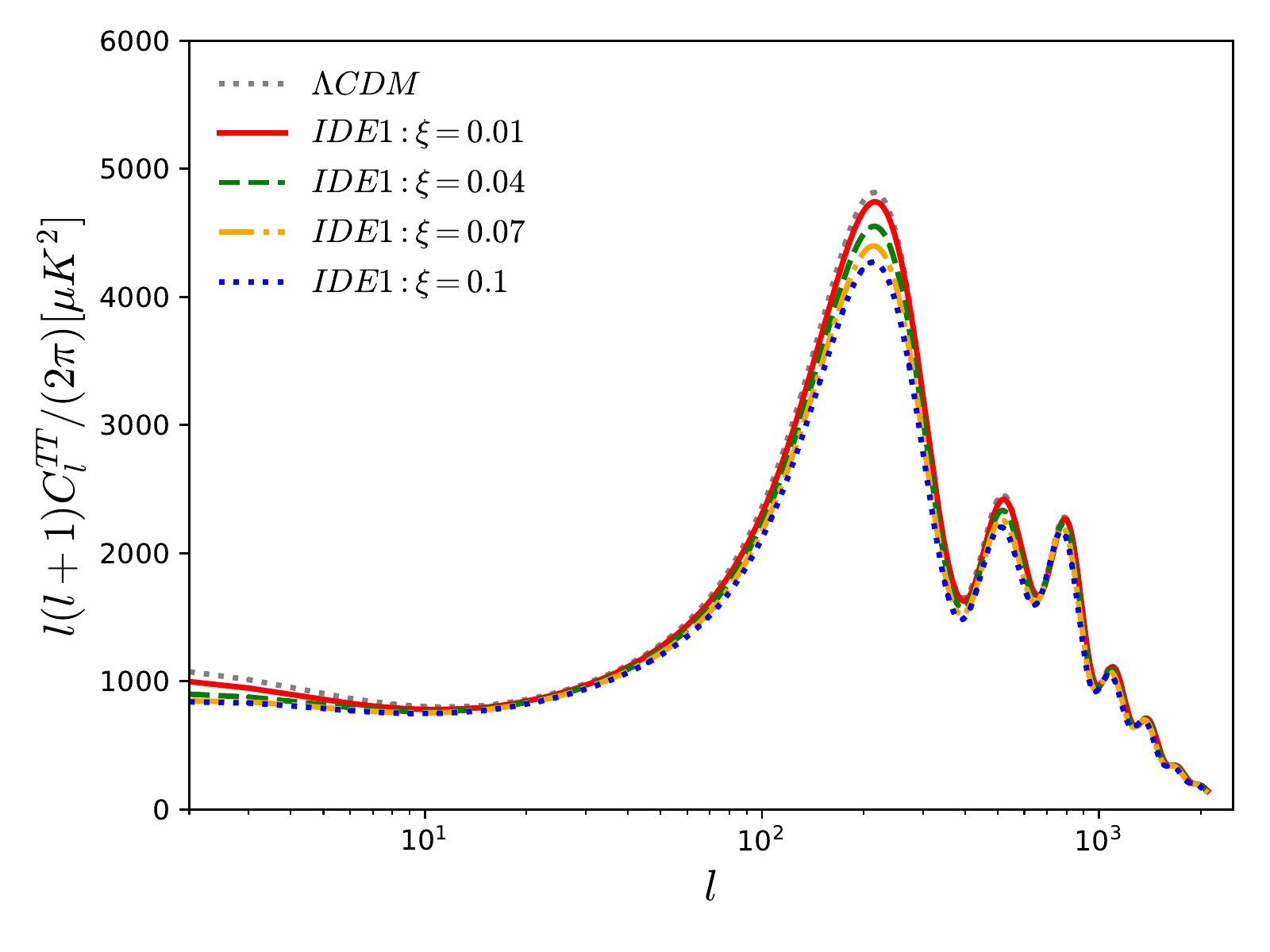} %
\includegraphics[width=0.45\textwidth]{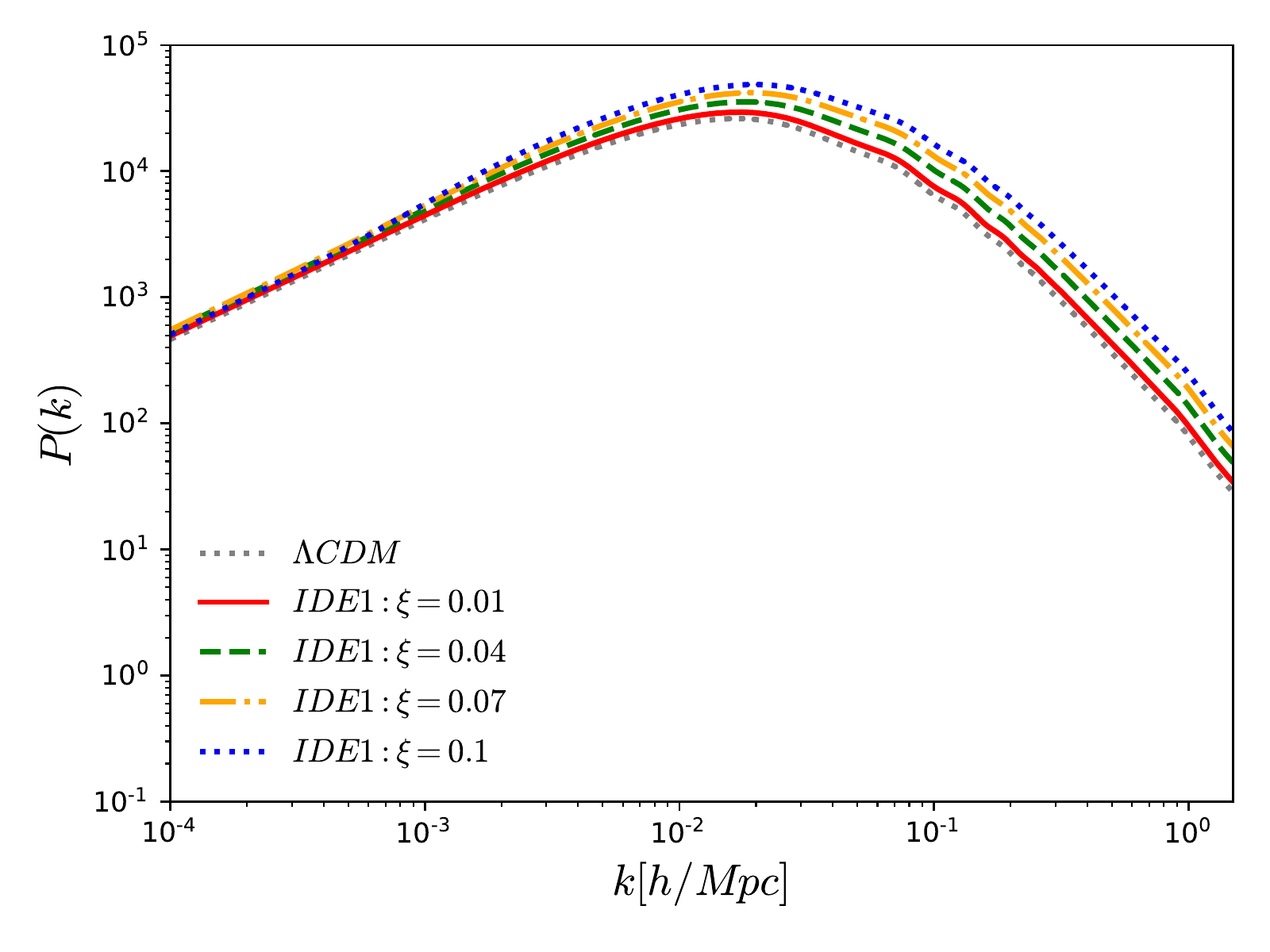}
\includegraphics[width=0.45\textwidth]{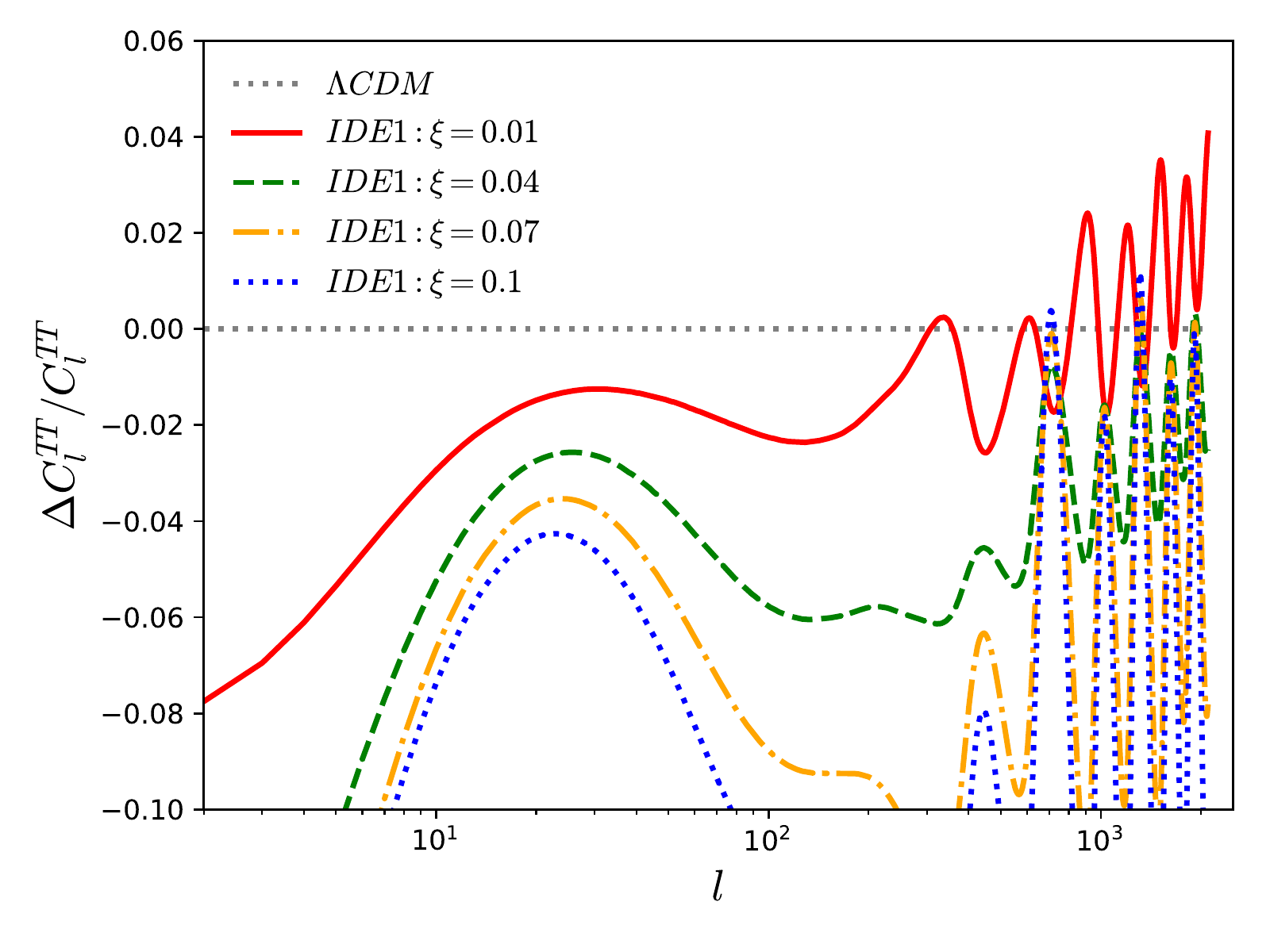}
\includegraphics[width=0.45\textwidth]{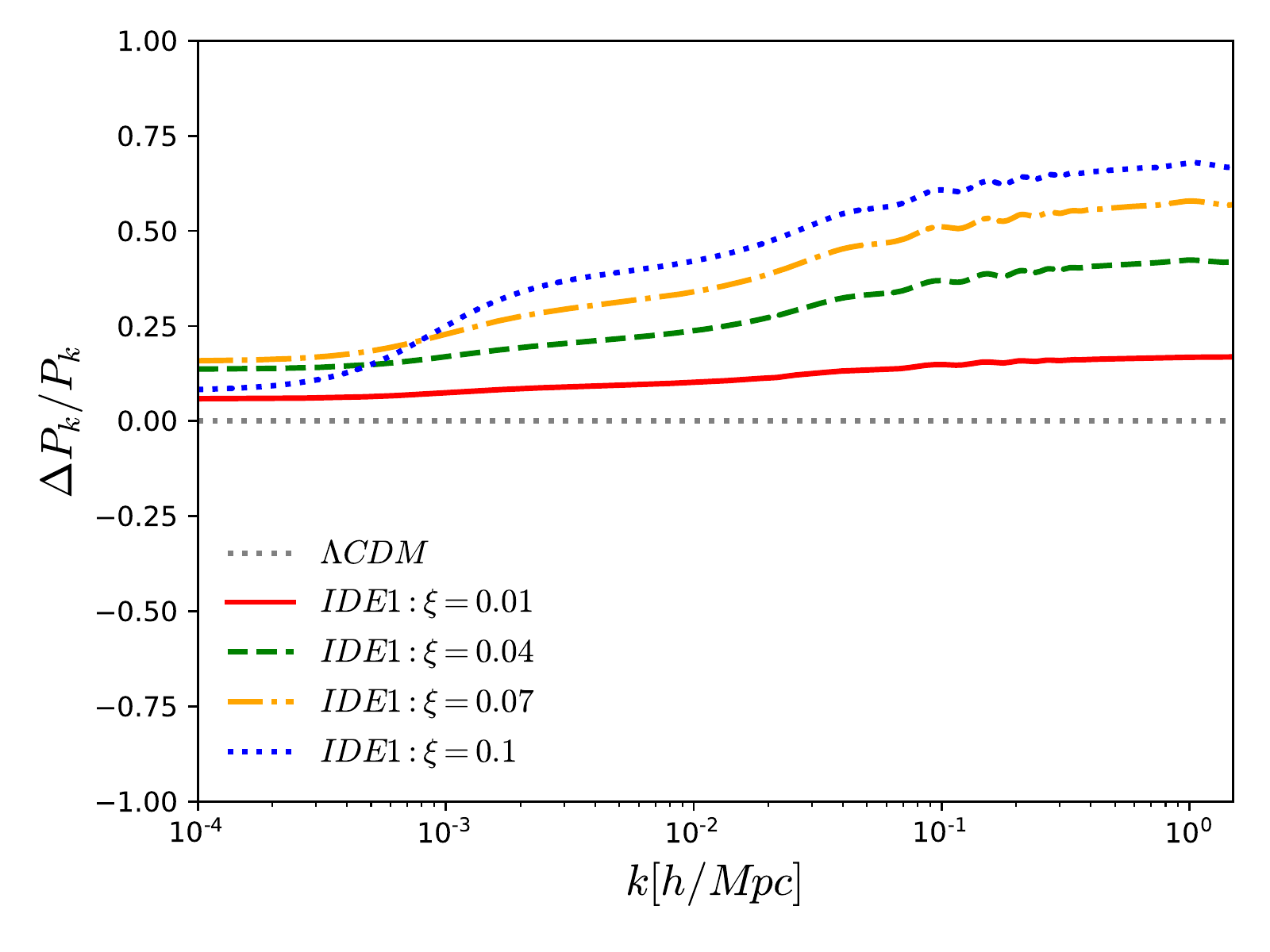}
\caption{CMB TT (upper left), matter power spectra (upper right) for IDE2 corresponding to the interaction function (\ref{Q2}) have been shown for different values of the coupling parameter $\xi$. As usual the fixed parameters for all the plots are,  $\Omega_{c0} = 0.28$, $\Omega_{x0} = 0.68$, $\Omega_{r0} = 0.0001$, and $\Omega_{b0} = 1- \Omega_{r0}-\Omega_{c0}-\Omega_{x0} =  0.0399$. }
\label{fig-ide2-TT-mower}
\end{figure*}
\begin{figure}
\includegraphics[width=0.45\textwidth]{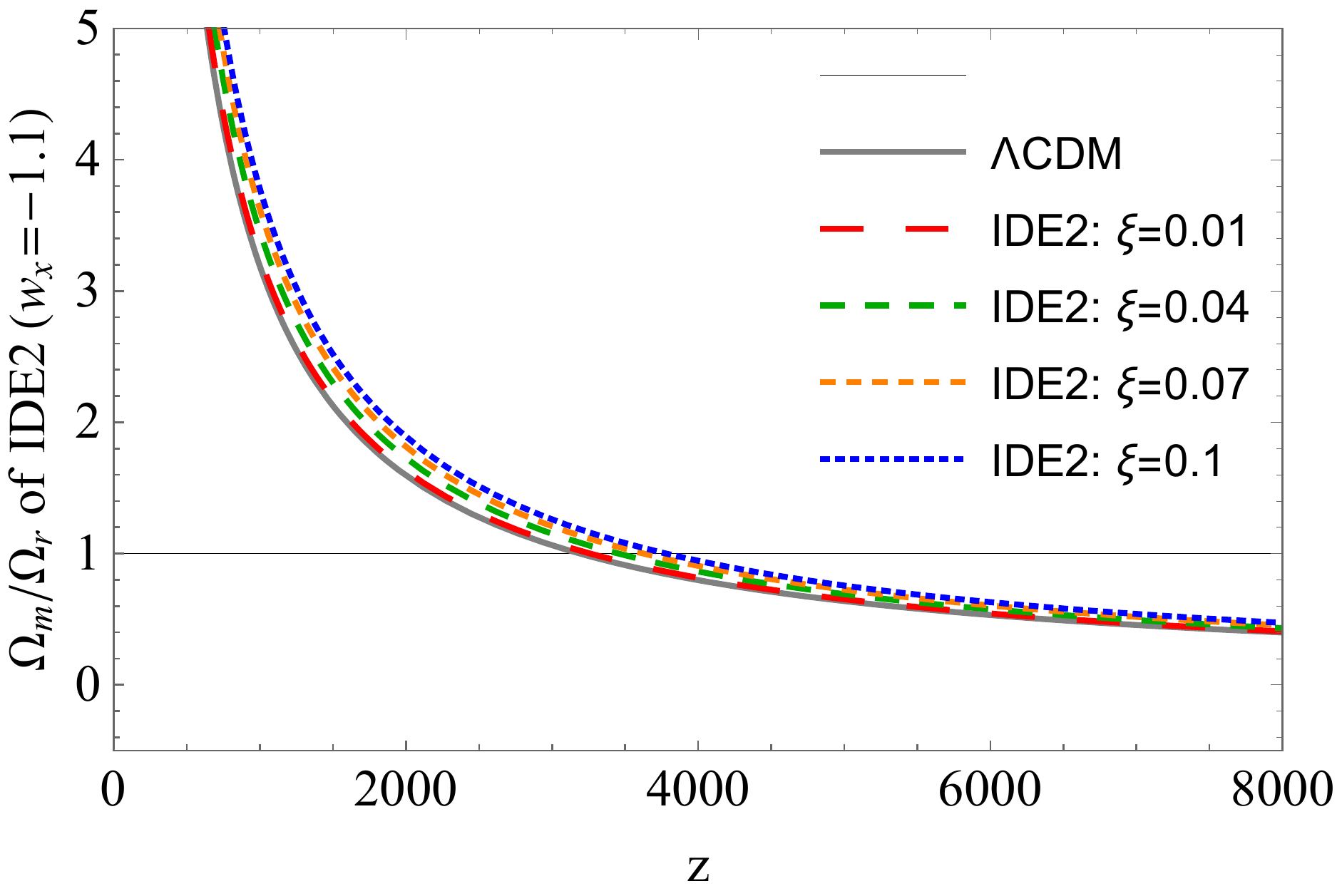}
\caption{The evolution of the quantity $\Omega_m/\Omega_r$ for the IDE2 scenario (corresponding to the interaction function in eqn. (\ref{Q2}))  has been shown for various values of the coupling parameter $\xi$. We note that here $\Omega_m = \Omega_c + \Omega_b$. The horizontal line corresponds to the case $\Omega_m = \Omega_r$ implying the matter-radiation equality. As usual the fixed parameters for the plot are,  $\Omega_{c0} = 0.28$, $\Omega_{x0} = 0.68$, $\Omega_{r0} = 0.0001$, and $\Omega_{b0} = 1- \Omega_{r0}-\Omega_{c0}-\Omega_{x0} =  0.0399$. }
\label{fig:ratio2}
\end{figure}

\subsection{IDE2:  $Q(t)=3H\xi \rho _{x}\left[ 1+\sin \left( \frac{\rho _{x}}{\rho _{c}}-1\right) \right]$}

\label{sec-ide2}

The constraints  on this interaction model have been displayed in Table \ref{tab:IDE2} at 68\% and 95\% CL for the datasets namely, CMB Planck 2018, Planck 2018+BAO, Planck 2018+DES, Planck 2018+R19 and Planck 2018+BAO+R19. In Fig. \ref{contour-ide2} we present the joint contours in the planes $(w_x, \xi)$, $(\xi, H_0)$ and $(\xi, \Omega_{m0})$  at 68\% and 95\% CL for all the observational datasets. From 
Fig. \ref{contour-ide2}, we also observe that $\xi$ is uncorrelated with both $w_x$ and $H_0$, and this is true irrespective of the cosmological datasets. Most of the joint contours representing the planes either $(w_x, xi)$ or $(\xi, H_0)$ are either vertical or horizontal. However, the correction between $\xi$ and $\Omega_{m0}$ is found for some datasets, for example, both Planck 2018+BAO and Planck 2018+R19 exhibit the strong anticorrelation between them.  Let us note that, as explained earlier,  
for this model too we could safely combine R19 with Planck 2018 and Planck 2018+R19. In the following we shall describe the constraints on this model scenario in detail. 

In the second column of Table \ref{tab:IDE2} we show the constraints from CMB alone. We find that the Hubble constant takes a very large value ($H_0= 84.33_{- 7.39}^{+   15.46}$ km/Mpc/sec, at 68\% CL) which is even greater than its local measurement, and thus, the $H_0$ tension persists. On the other hand we see that the coupling parameter is constrained to be small and it  is consistent to $\xi = 0$. Additionally, we find that a phantom dark energy equation of state is suggested 
at more than 68\% CL. 

Similar to the previous model, here too we add BAO to CMB (from Planck 2018) aiming to break the degeneracies in the parameters while analyzing the underlying model with Planck 2018 data only.  
When BAO data are added to  Planck 2018, we see that the mean value of $H_0$ is substantially reduced with reduced error bars ($H_0 =   69.06_{-    1.61}^{+    1.36}$ km/Mpc/sec, at 68\% CL). However, this estimation is higher compared to Planck's recent measurements \cite{Aghanim:2018eyx} and the tension on $H_0$ is indeed reduced a bit but not solved completely. The coupling parameter $\xi$ assumes a mild value but within 68\% CL, it is  consistent to $\xi = 0$.   
Concerning the dark energy equation of state, although the mean value is phantom but within 68\% CL, its quintessence nature is also allowed. Interestingly, the inclusion of DES  to Planck 2018 gives similar constraints as we already found with IDE1. That means for this model scenario, an evidence of a non-zero coupling in the dark sector ($\xi \neq 0$ at more than 95\% CL) is strongly suggested by Planck 2018+DES, and moreover, a phantom $w_x$ is also suggested at more than 95\% CL together with an increased Hubble constant. The measurement of Hubble constant, $H_0$ for Planck 2018+DES is perfectly consistent (within 68\% CL) with its local estimation \cite{Riess:2019cxk}.

We then concentrate on the observational constraints for Planck 2018+R19 and Planck 2018+BAO+R19. If we consider the constraint on $H_0$ from Planck 2018, we see that due to high error bars in it, within 95\% CL, this is not in tension with its local measurements \cite{Riess:2019cxk}, so the addition of R19 with Planck 2018 does not bias the results. Similarly, looking at the constraint on $H_0$ from Planck 2018+BAO, we see that  the tension between the datasets Planck 2018+BAO and R19 are less $3\sigma$, thus, the combination Planck 2018+BAO+R19 can also be taken into consideration. From both the analyses, we find that the Hubble constant increases and becomes close to its local measurement and thus alleviating the tension. In particular, for Planck 2018+R19, $H_0$ tension is solved perfectly within 68\% CL and for Planck 2018+BAO+R19, this is solved within 95\% CL. In Fig. \ref{fig:whiskerH0-IDE2} we show the whisker plot for $H_0$ taking its 68\% CL constraints considering all the datasets.  This gives a clear idea on how the $H_0$ tension is solved/alleviated by different observational data within this model scenario. 
The dark energy equation of state remains phantom at more than 95\% CL. However, concerning the coupling parameter, for Planck 2018+R19, we do not find any strong evidence for it ($\xi$ is mild and consistent to $0$ at 68\% CL), but for Planck 2018+BAO+R19, an evidence for $\xi \neq 0$ is strongly suggested. We see that for the latter combination, that means for Planck 2018+BAO+R19, $\xi = 0.074_{- 0.066}^{+ 0.052}$, at 95\% CL.

In a similar fashion, let us now discuss the effects of  interaction in the CMB temperature power spectra (CMB TT) and matter power spectra. In Fig. \ref{fig-ide2-TT-mower}, we present the CMB TT (left graph of Fig. \ref{fig-ide2-TT-mower}) and matter power spectra (right graph of Fig. \ref{fig-ide2-TT-mower}) for different values of the coupling parameter $\xi$. For comparison, we include the non-interacting $\Lambda$CDM case ($\xi  = 0$) in the left graph of Fig. \ref{fig-ide2-TT-mower} with $\xi \neq 0$ cases. 
From the CMB TT spectra (left plot of Fig. \ref{fig-ide2-TT-mower}), one can clearly visualize that as long as the coupling parameter increases, the height of the first acoustic peak in the CMB TT decreases compared to the non-interacting case. In fact, the reduction in the CMB TT spectra is prominent compared to what we observed in the CMB TT spectra for IDE1 (we refer to the left plot of Fig. \ref{fig-ide1-TT-mower}). 
As already commented earlier, due to an interaction in the dark sector, 
the matter-radiation equality occurs earlier compared to the non-interacting case, see Fog. \ref{fig:ratio2}. And due to earlier matter-radiation equality, the sound horizon is decreased and hence the first acoustic peak of CMB TT is also decreased. Compared to IDE1, in this case, the effects on CMB TT spectra are prominent. 
Similar to IDE1, here too we find that, in the lower multipole region, as the coupling parameter $\xi$ increases, the amplitude of the CMB TT spectrum decreases relative  to the non-interacting case. Since the presence of interaction effectively changes the usual evolution of CDM ($ \rho_c \propto a^{-3}$), as a consequence, this affects the CMB TT spectra in the lower multipole region via integrated Sachs-Wolfe (ISW) effect due to the gravitational potential.  Finally, from the matter power spectra, shown in the right plot of \ref{fig-ide2-TT-mower},  one finds that amplitude of the matter power spectra increases with the increase of the coupling parameter $\xi$. This enhancement occurs due to the earlier matter-radiation equality (see Fig. \ref{fig:ratio2}). 
\begin{figure}
\includegraphics[width=0.4\textwidth]{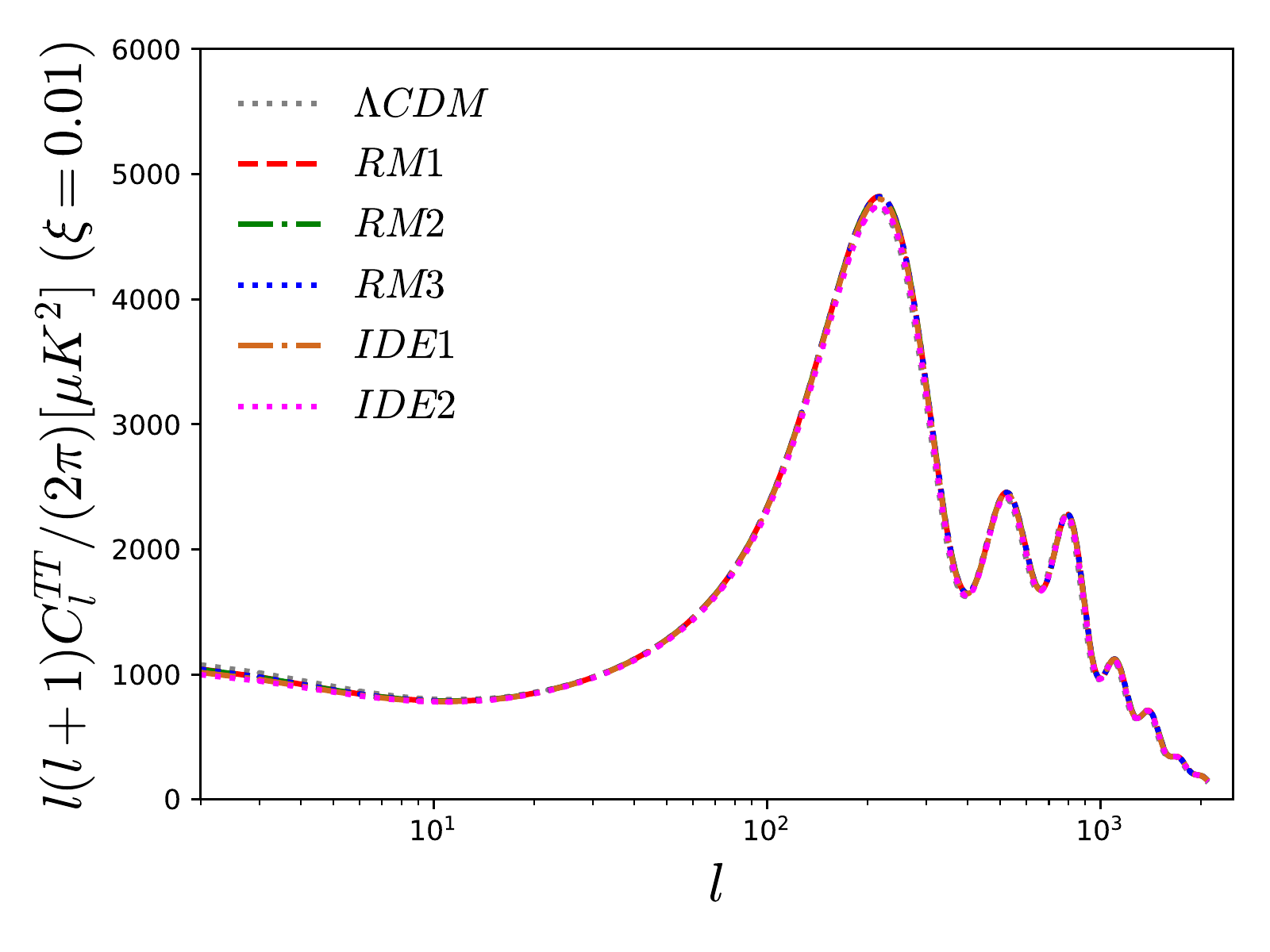} %
\includegraphics[width=0.4\textwidth]{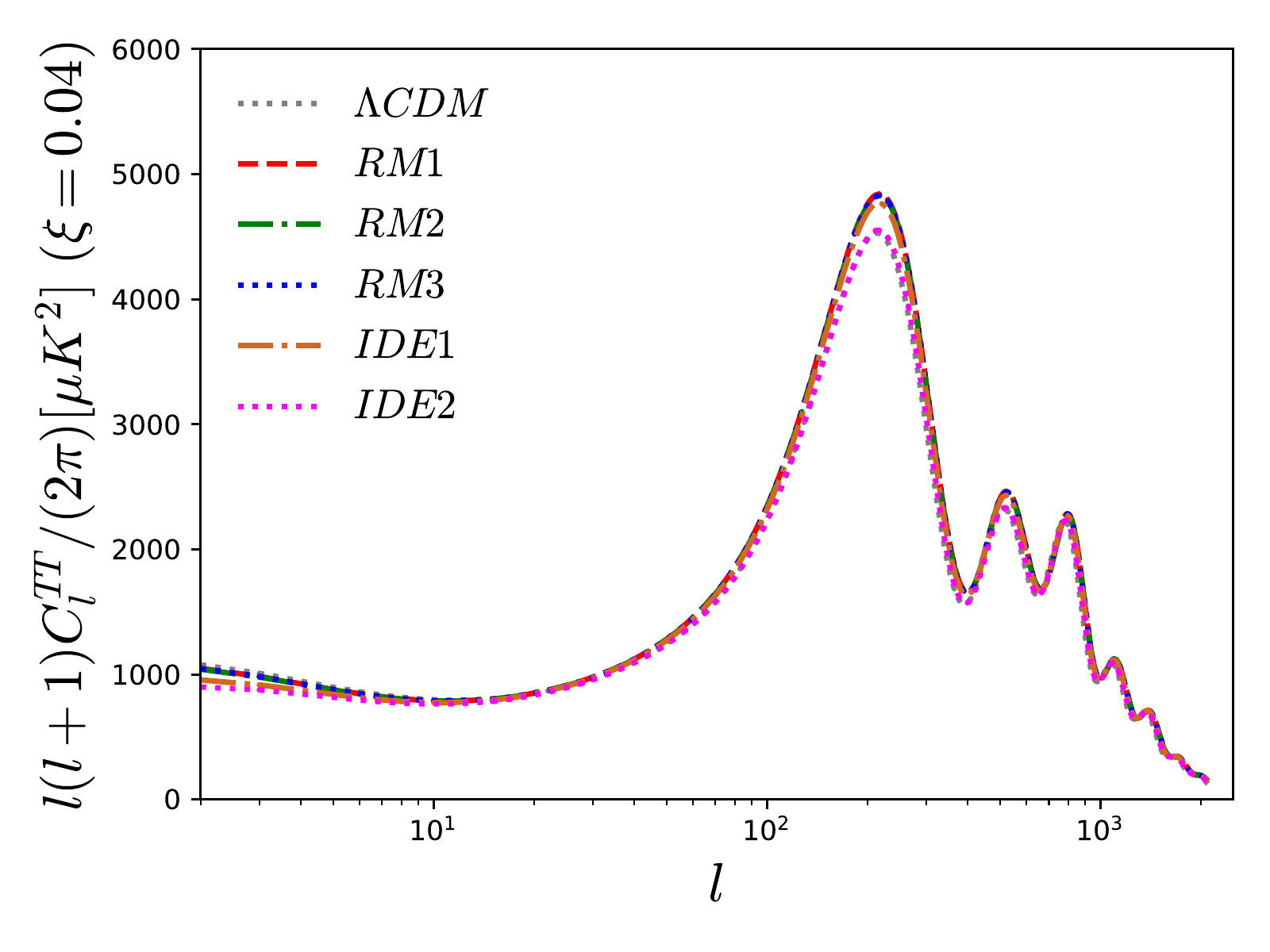} %
\includegraphics[width=0.4\textwidth]{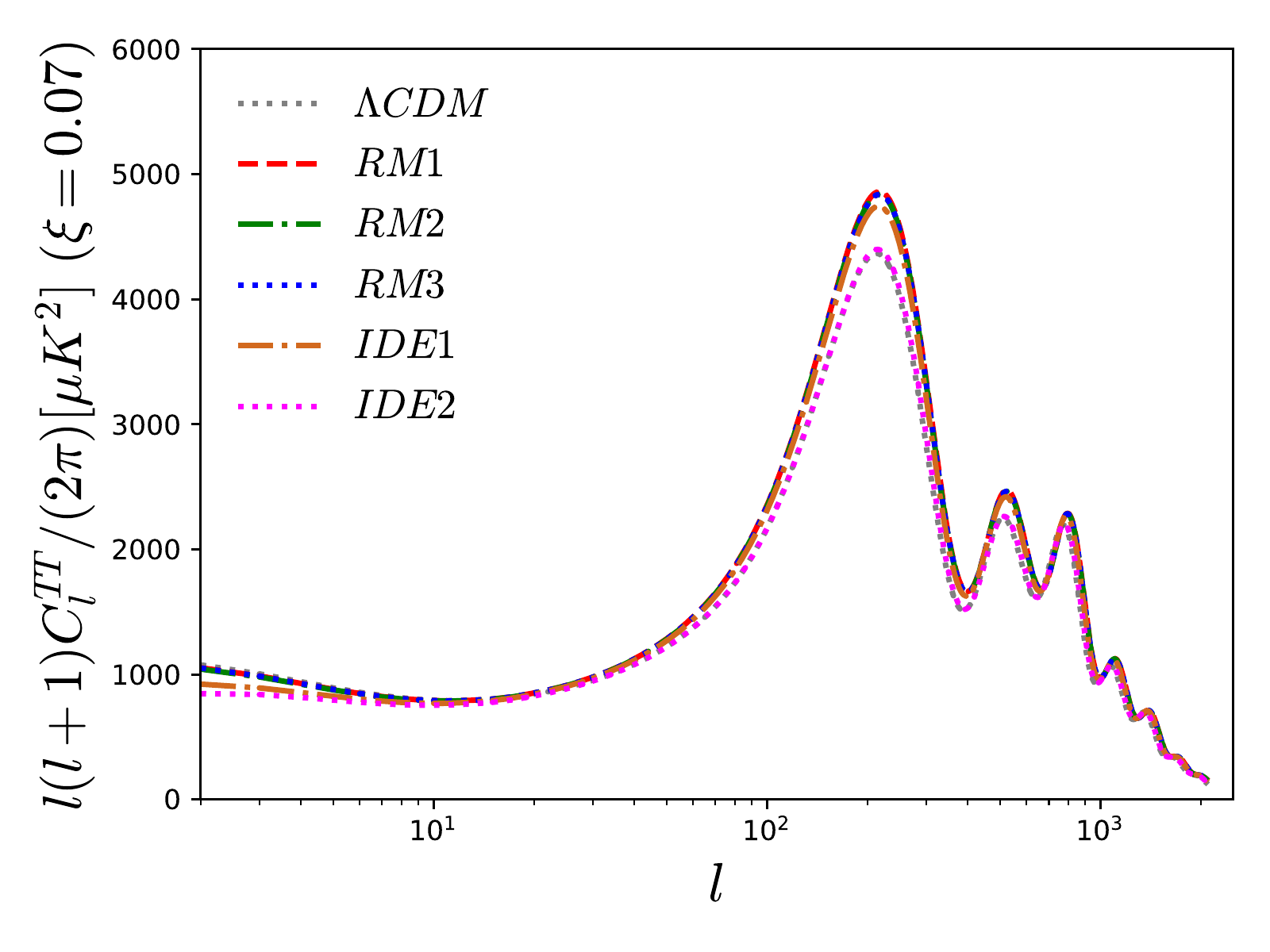}
\caption{Temperature anisotropy in the CMB spectra for different kind of interaction
functions using three different values of the coupling parameter, $\protect%
\xi = 0.01, 0.04, 0.07$ with reference to the base non-interacting 
$\Lambda$CDM model ($\xi = 0$, $w_x = -1$). In agreement with the earlier plots, we have fixed the following parameters, $\Omega_{c0} = 0.28$, $\Omega_{x0} = 0.68$, $\Omega_{r0} = 0.0001$, and $\Omega_{b0} = 1- \Omega_{r0}-\Omega_{c0}-\Omega_{x0} =  0.0399$. }
\label{fig-cmb-compare}
\end{figure}
\begin{figure}
\includegraphics[width=0.4\textwidth]{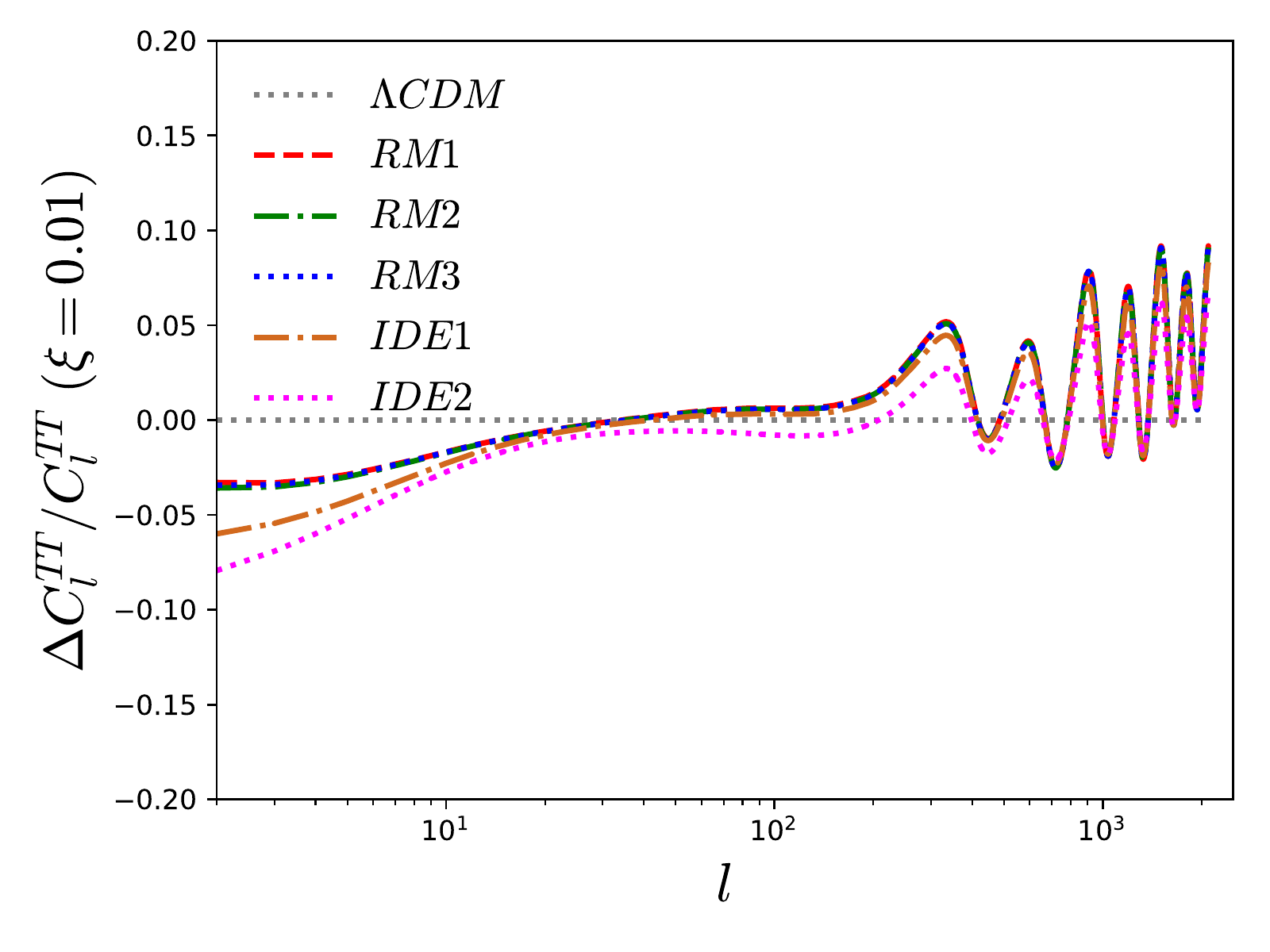} %
\includegraphics[width=0.4\textwidth]{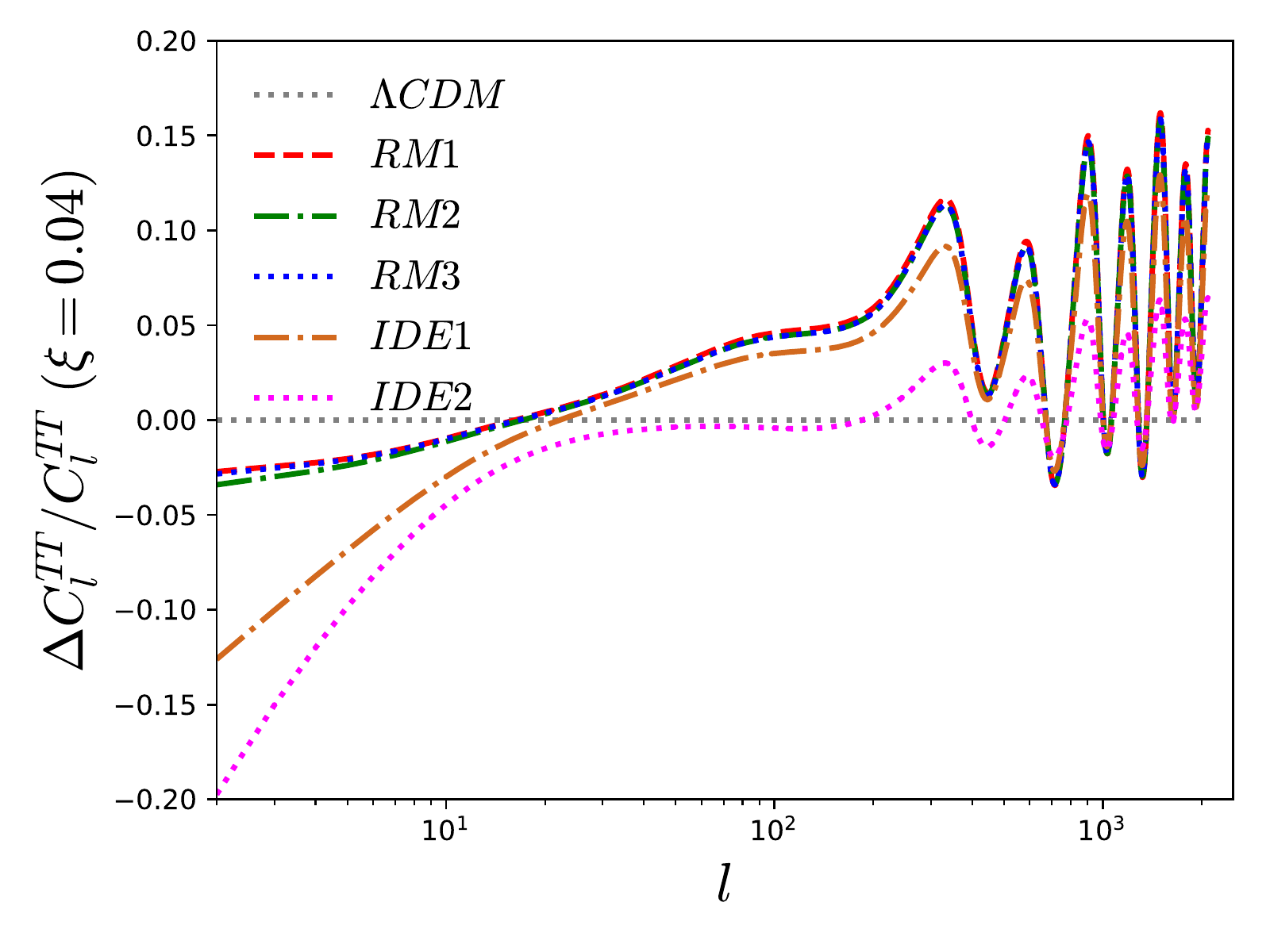} %
\includegraphics[width=0.4\textwidth]{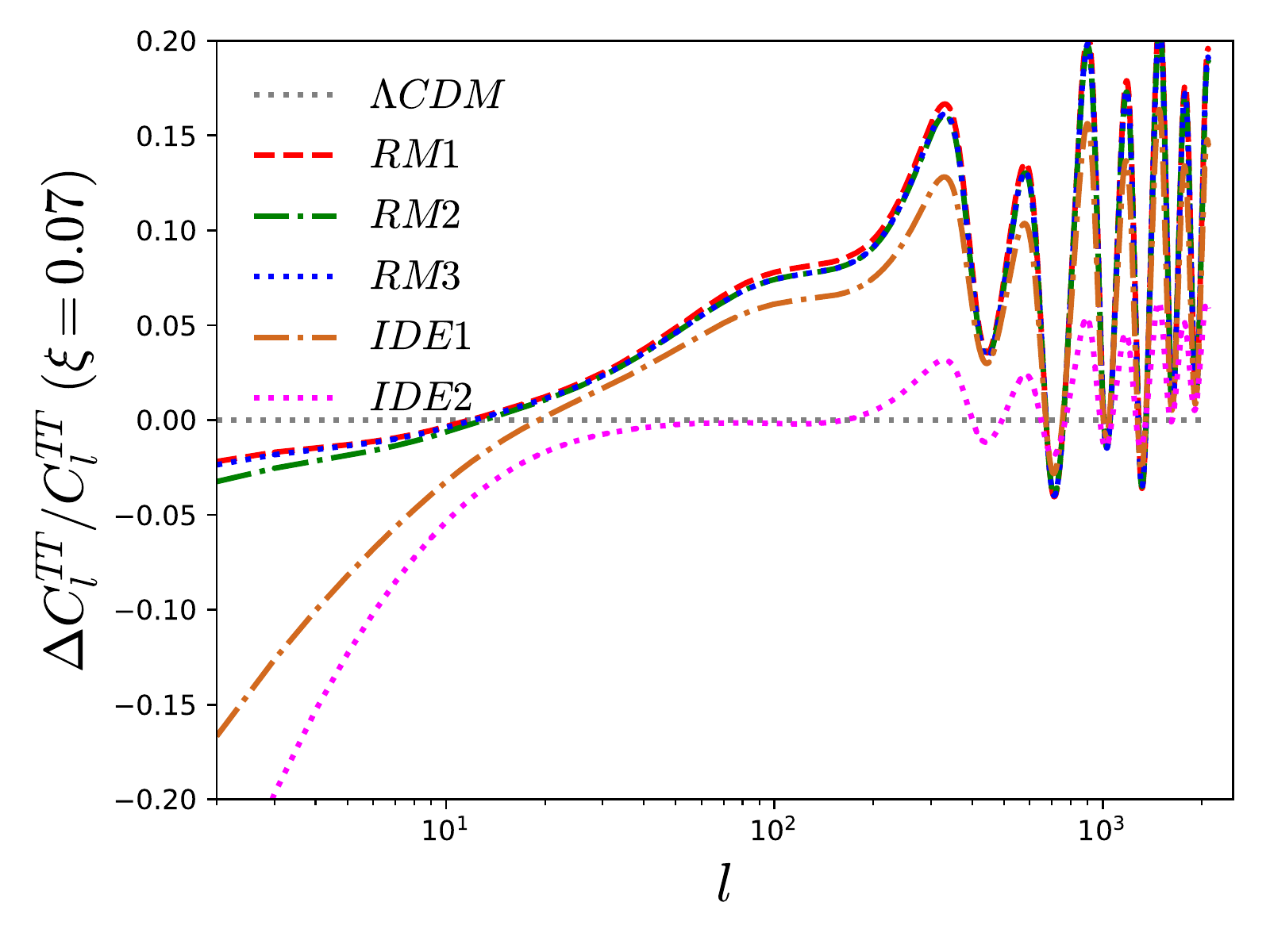}
\caption{Relative deviations of the  IDE models from the base
non-interacting $\Lambda$CDM scenario ($\xi = 0$, $w_x = -1$) via CMB spectra have been shown for three different values of the coupling parameter, as follows: In the left panel we set $\xi =0.01$ for the interacting models; in the middle panel we set $\xi =0.04$ for all the interacting models while in the right panel we set $\xi =0.07$ for all the interacting models. As usual the fixed parameters for all the plots are,  $\Omega_{c0} = 0.28$, $\Omega_{x0} = 0.68$, $\Omega_{r0} = 0.0001$, and $\Omega_{b0} = 1- \Omega_{r0}-\Omega_{c0}-\Omega_{x0} =  0.0399$. }
\label{fig-cmb-compare-ratio}
\end{figure}
\begin{figure}
\includegraphics[width=0.4\textwidth]{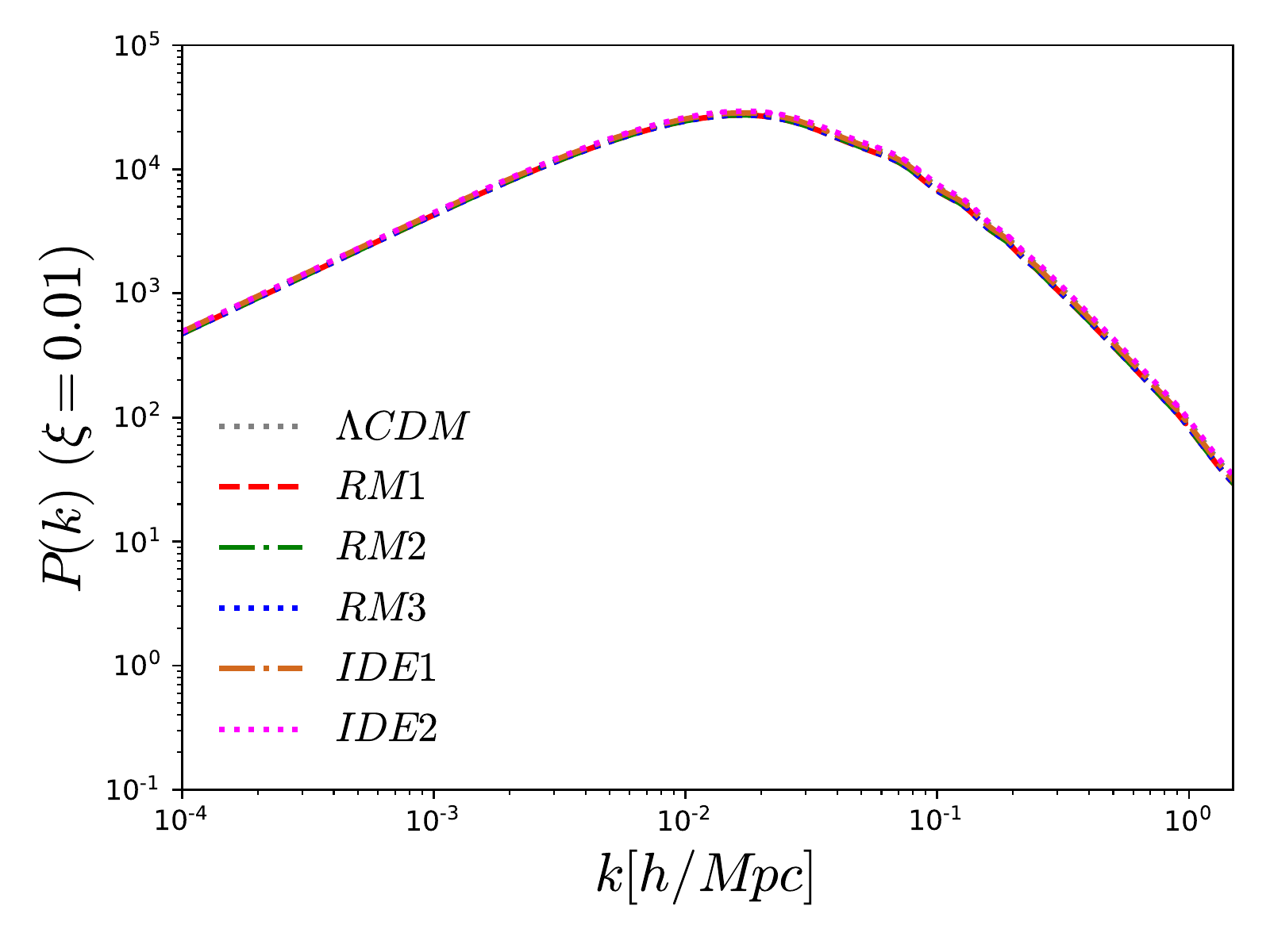} %
\includegraphics[width=0.4\textwidth]{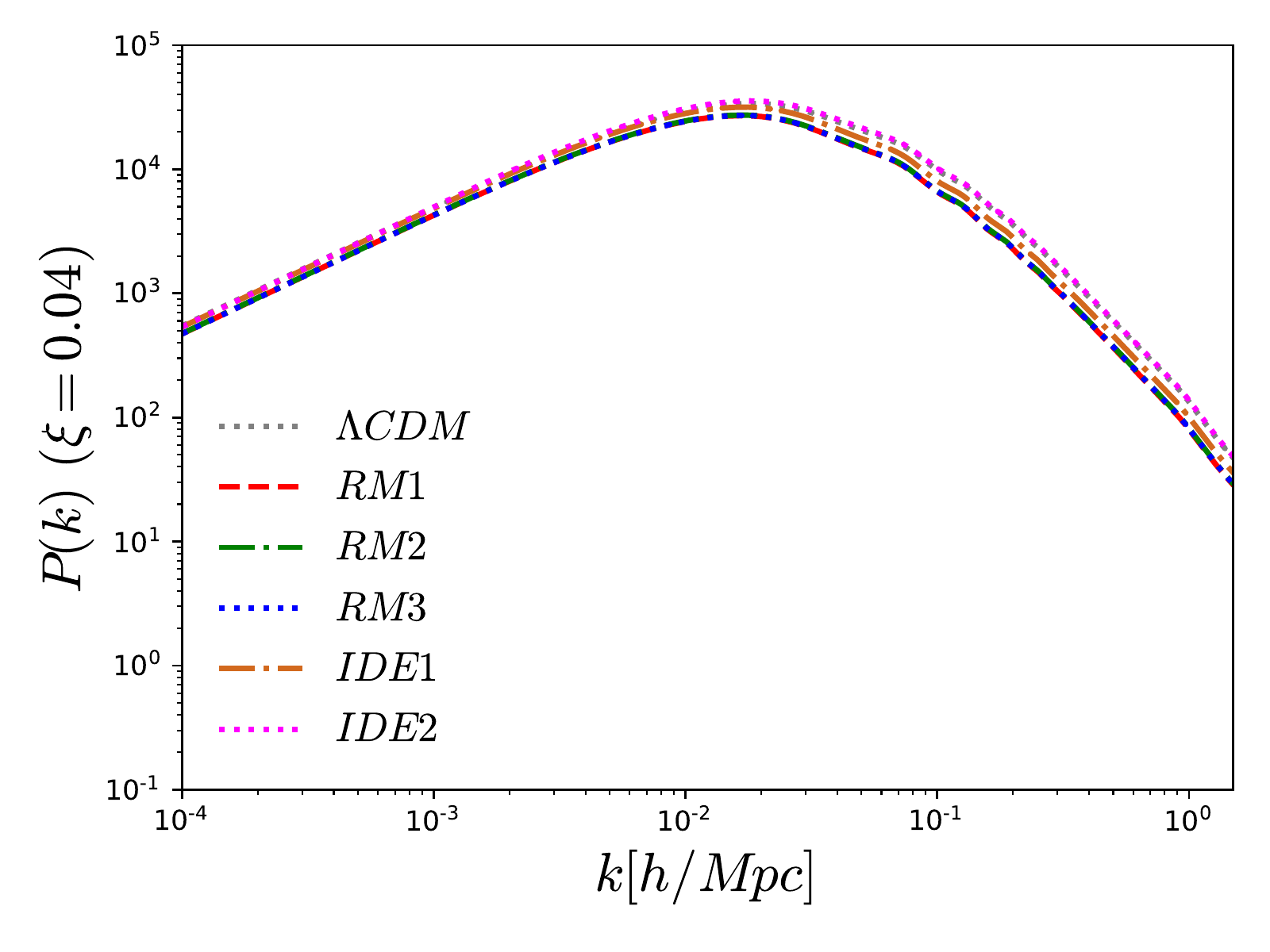} %
\includegraphics[width=0.4\textwidth]{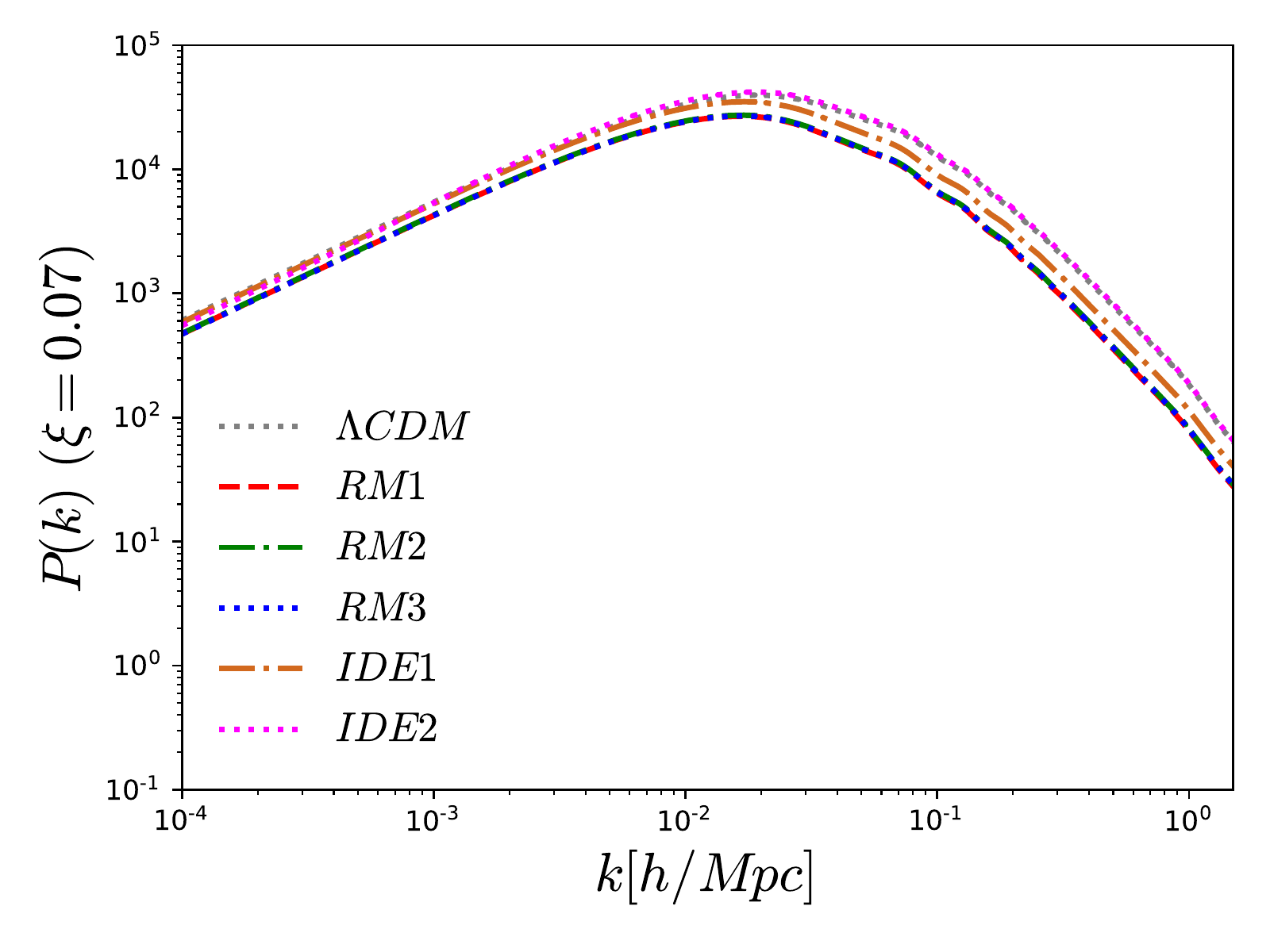}
\caption{Matter power spectra for the interaction functions using three
different values of the coupling parameter, $\protect\xi$, with reference to the base non-interacting  $\Lambda$CDM model ($\xi = 0$, $w_x = -1$). In the left panel we set $\xi =0.01$ for all the interacting models; in the middle panel we set $\xi =0.04$ for all the interacting models while in the right panel we set $\xi =0.07$ for all the interacting models. For all the plots the parameters which have been kept fixed are,  $\Omega_{c0} = 0.28$, $\Omega_{x0} = 0.68$, $\Omega_{r0} = 0.0001$, and $\Omega_{b0} = 1- \Omega_{r0}-\Omega_{c0}-\Omega_{x0} =  0.0399$. }
\label{fig-mpower-compare}
\end{figure}
\begin{figure}
\includegraphics[width=0.4\textwidth]{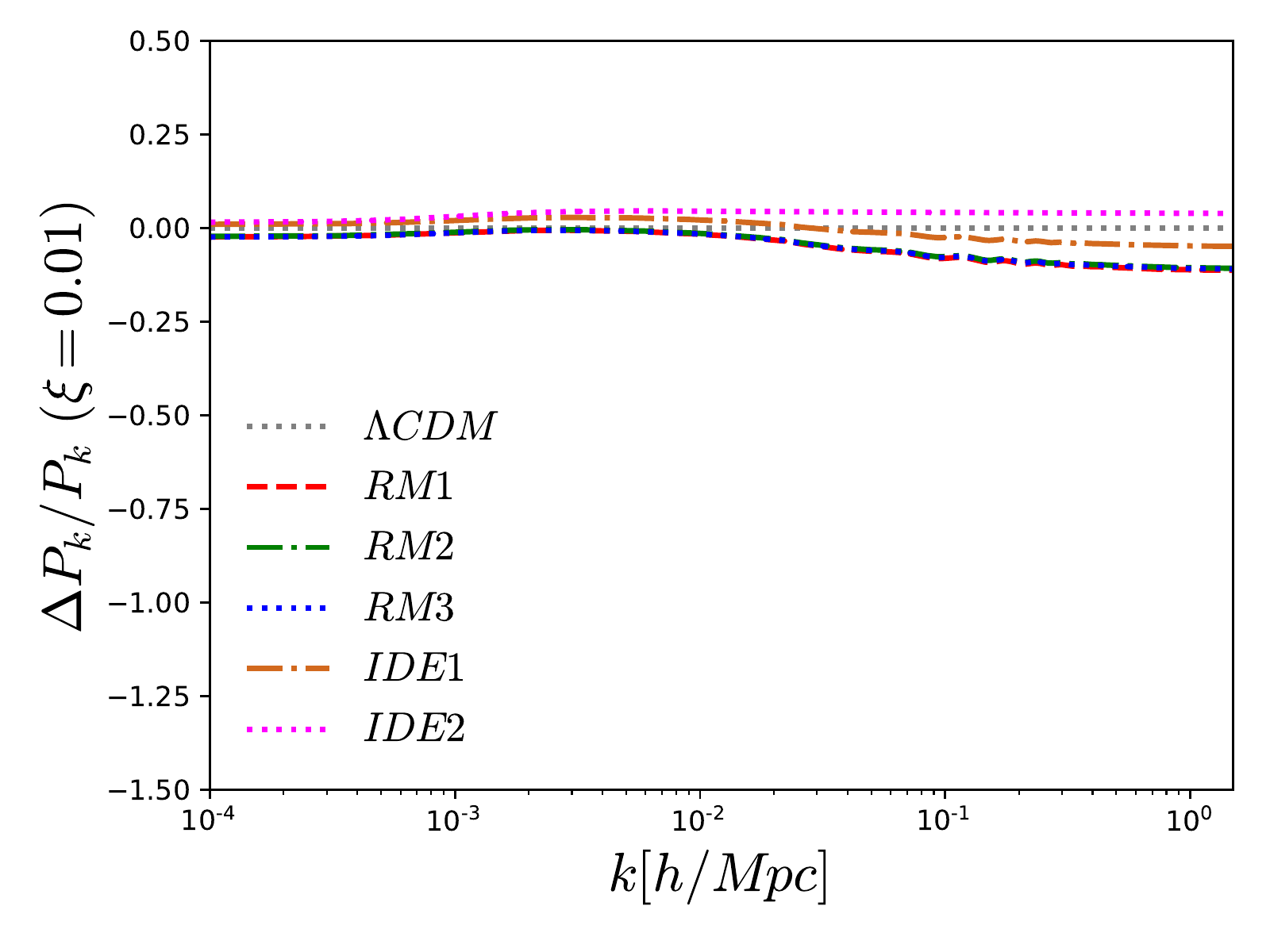} %
\includegraphics[width=0.4\textwidth]{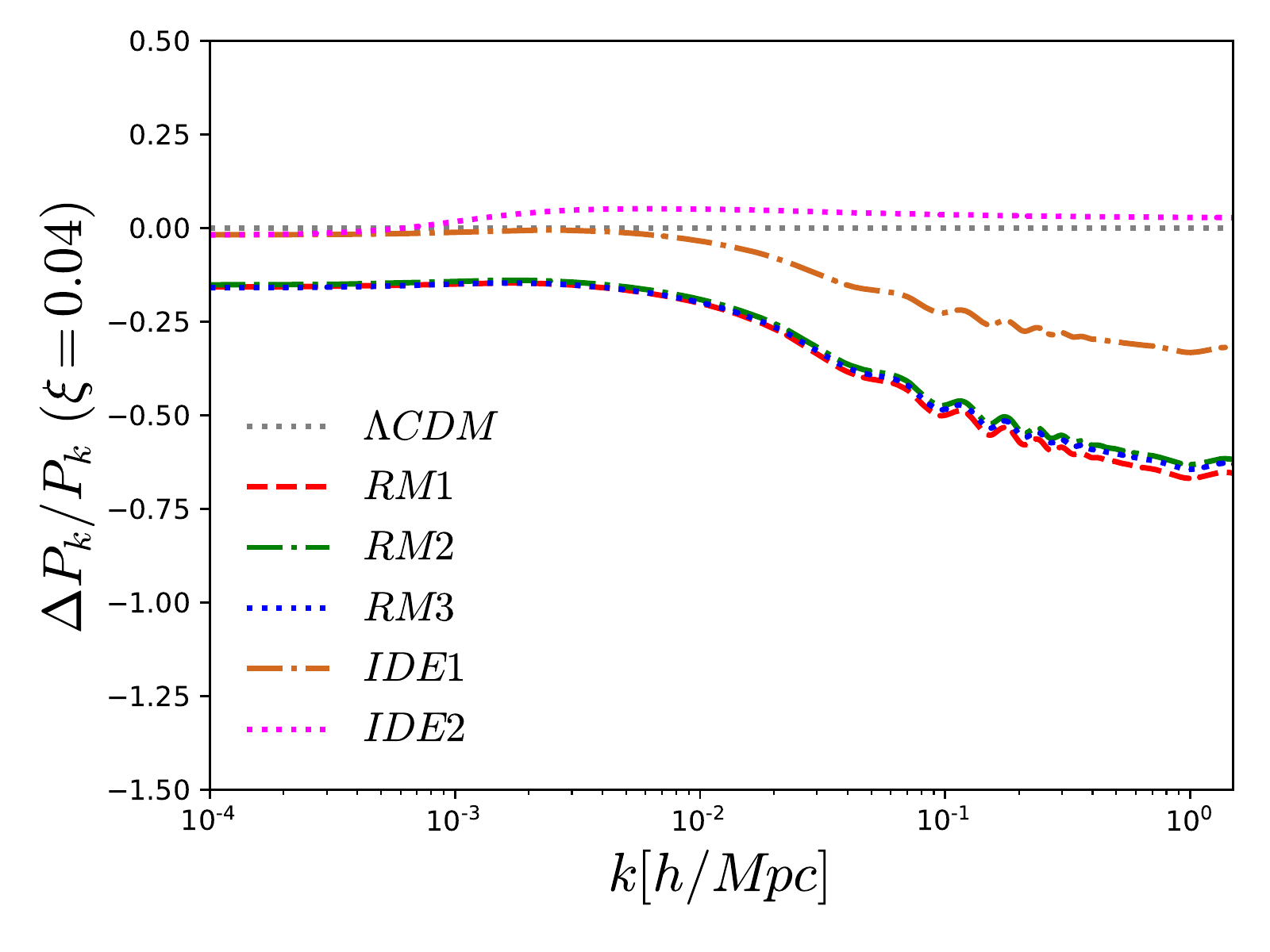} %
\includegraphics[width=0.4\textwidth]{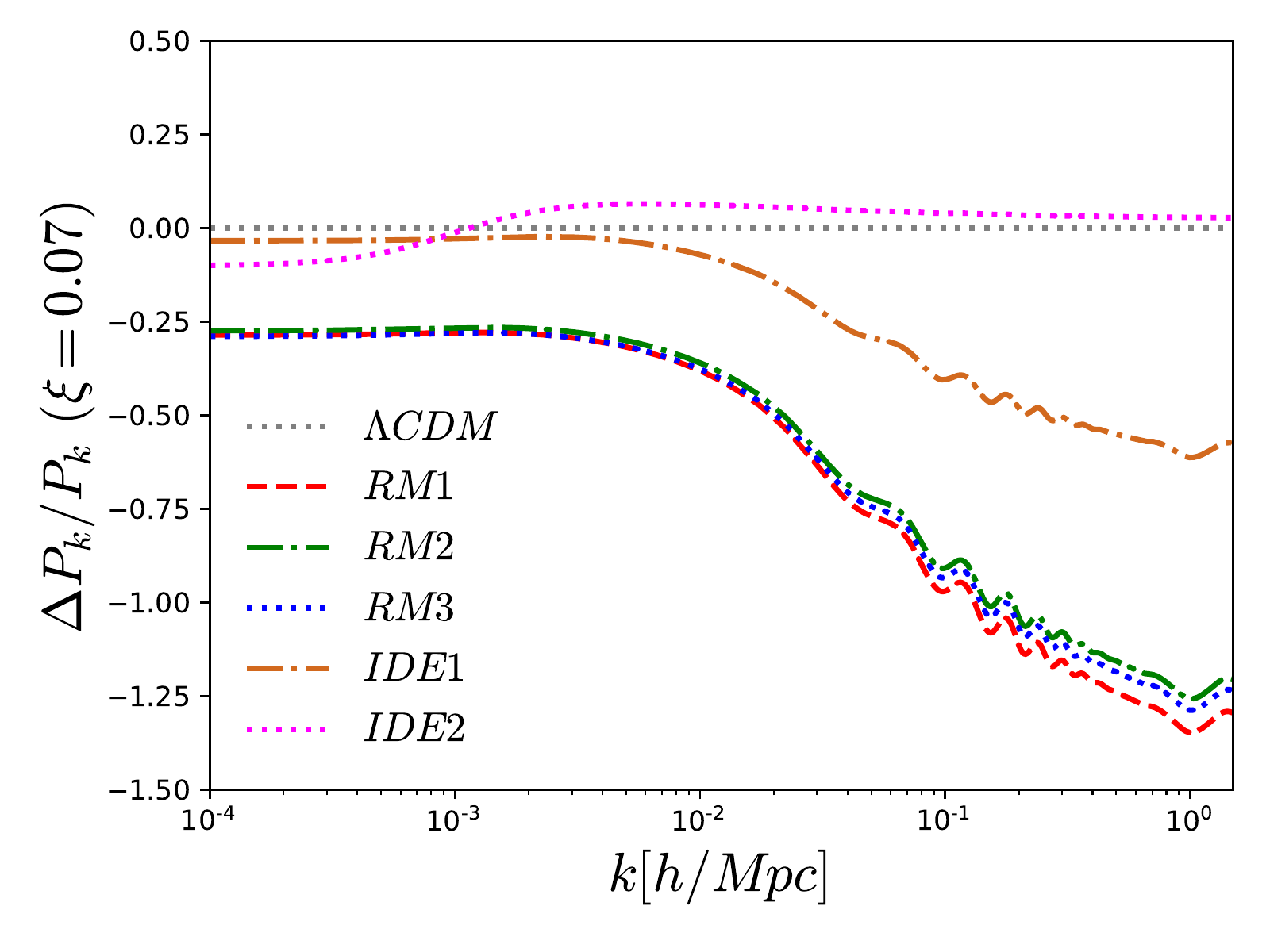}
\caption{Relative deviations of the interacting DE models from the
base non-interacting $\Lambda$CDM scenario ($\xi = 0$, $w_x = -1$) 
via matter power spectra have been shown for three
different values of the coupling parameter, as follows: from left to right
the plots correspond to $\protect\xi =0.01$, $0.04$ and $0.07$. The fixed parameters for all the plots are,  $\Omega_{c0} = 0.28$, $\Omega_{x0} = 0.68$, $\Omega_{r0} = 0.0001$, and $\Omega_{b0} = 1- \Omega_{r0}-\Omega_{c0}-\Omega_{x0} =  0.0399$.  }
\label{fig-mpower-compare-ratio}
\end{figure}

\section{Can we distinguish the present models with some known models?}

\label{sec-distinguish}

This section is devoted to understand the behaviour of the present
interaction models with some already known and well used interacting dark
energy models, when the large scale structure is taken into account. Because
from the observational constraints, the interaction models are not sharply
distinguished from one another. Thus, in order to investigate this issue, we
consider three known interaction models, namely $Q = 3 H \xi (1+w_x) \rho_x$
(Reference Model1), $Q = 3 H \xi (1+w_x) \frac{\rho_c \rho_x}{\rho_c +\rho_x}
$ (Reference Model2) and $Q = 3 H \xi (1+w_x) \frac{\rho_x^2}{(\rho_c
+\rho_x)^2}$ (Reference Model3) where the first model is linear whilst the
last two models are nonlinear. And moreover, all three interaction models
are free from early time instabilities. For convenience, we abbreviate the
reference model as \textit{RM}, thus, the three references models can be
identified as RM1, RM2 and RM3 in the figures that we shall describe now.

We first begin our analysis with Fig. \ref{fig-cmb-compare} where we present
the temperature anisotropy in the CMB spectra for all five (three reference
models plus two proposed models in this work) interaction functions using
three different values of the coupling parameter, namely, $\xi = 0.01$, $0.04
$ and $0.07$ in order to see how the changes in the CMB spectra occur with
the increase of the coupling parameter. The left, middle and right panels of
Fig. \ref{fig-cmb-compare} respectively indicates the scenarios for $\xi
=0.01$, $0.04$ and $0.07$. Looking at the left plot of Fig. \ref{fig-cmb-compare},
one can realize that for $\xi  =0.01$, the IDE models cannot be
distinguished from one another. The difference between the models starts when 
one increases the coupling parameter.  
From three middle and right plots of Fig. \ref{fig-cmb-compare}, 
one can notice that the increase of the coupling
parameter changes in the height of the first acoustic peak in the CMB TT spectra changes and this essentially clarifies the effects of the interaction in the cosmic dynamics.  In fact, as long as the coupling parameter increases, the changes in the lower multipoles (for $l < 10$) are becoming prominent which was absent in the left plot of Fig. \ref{fig-cmb-compare}. The reason of the changes in the CMB spectra are discussed in section \ref{sec-results}. In
Fig. \ref{fig-cmb-compare-ratio}  we show the corresponding residuals
with respect to the non-interacting $\Lambda$CDM scenario for a better 
viewing of the the changes in the CMB TT spectra. From Fig. %
\ref{fig-cmb-compare-ratio}, we find that the present two IDE models are
slightly different from the three reference models and this difference is much pronounced when the coupling parameter increases.  

In a similar fashion, in Fig. \ref{fig-mpower-compare} we show the matter
power spectra for all the interaction models (three reference models
and two models of the present work) as well as for the non-interacting $\Lambda$CDM model using exactly  the same values of the coupling
parameter, namely, $\xi =0.01$, $0.04$ and $0.07$, as used to draw the plots in Fig. \ref{fig-cmb-compare} and Fig. \ref{fig-cmb-compare-ratio}. The left, middle and
right plots of Fig. \ref{fig-mpower-compare} respectively stands for $\xi
=0.01$, $0.04$ and $0.07$. From Fig. \ref{fig-mpower-compare}, one can see that for small values of the coupling
parameter (for example $\xi  =0.01$), 
the five models cannot be differentiated from one another, whilst
for $\xi =0.04$ and $\xi  =0.07$ (see the middle plot and right plot of Fig. \ref{fig-mpower-compare}), the models can be distinguished from one another, and moreover, we find that for IDE2, is slightly different from others but this does not deviate much from $\Lambda$CDM.  From Fig. \ref{fig-mpower-compare-ratio}, this actually becomes clear. From all the plots in Fig. \ref{fig-mpower-compare-ratio}, one can clearly notice that, the present two
IDE models can indeed be distinguished from one another as well as from the reference models. But, an interesting feature that anyone may observe from all three  plots of  Fig. \ref{fig-mpower-compare-ratio} is that even if we allow $\xi$ up to $\xi  =0.07$, IDE2 does not deviate much from $\Lambda$CDM, whilst other interaction models (IDE1 and RM1, RM2, RM3) are clearly distinguished.  

Thus, from all the plots considered for this section, it is pretty clear that both IDE1 and IDE2 can really be considered to be the viable 	contestants  compared to the chosen reference models in this work as well as with other existing  models in the literature.

\section{Bayesian Evidence using Savage-Dickey Density Ratio}

\label{sec-sddr}

We now come to almost end  of this work where we wish to perform the
Bayesian analysis in order to quantify the observational viabilities of the
models relative to some reference model which in this work we assume to be
the non-interacting cosmological model. Usually, there are different ways to
perform the Bayesian analysis of the underlying cosmological models,
however, here we use the Savage-Dickey Density Ratio (SDDR) which is useful
to reduce the computational efforts compared to other known approaches to
calculate the Bayesian evidences. Thus, assuming SDDR, the Bayes factor $%
B_{0i}$ ($i =1$ for IDE1 and $i =2$ for IDE2) can be written as \cite{sddr1,
sddr2, sddr3}

\begin{eqnarray}
B_{0i} = \frac{p(\xi|d, M_i)}{\pi (\xi|M_i)}\Bigg|_{\xi = 0} (\mathrm{SDDR})
\end{eqnarray}
where $M_i$ ($i = 1, 2$) is the concerned interacting DE model; $%
p(\xi|d, M_i)$ is the posterior for $\xi $ for the fixed dataset $d$ and $%
\pi (\xi|M_i)$ is the flat prior on $\xi$ that we have assumed $\xi \in [0,
1]$ (see Table \ref{tab:priors}). In order to quantify the strength of
evidence of the IDE models, we use the values of $\ln B_{0i}$ for which one
can use the Jeffreys' scale as follows: $|\ln B_{0,i}| < 1.0$ is
inconclusive; $|\ln B_{0,i}| = 1.0$ means a positive evidence; $|\ln
B_{0,i}| = 2.5$ gives a moderate evidence and finally $|\ln B_{0,i}| = 5.0$
gives the strong evidence. The values of $\ln B_{0i}$ for the present IDE
models using the SDDR have been shown in Table \ref{tab:SDDR}.  
From the values of $|\ln B_{0,i}|$ (SDDR) obtained for all the
datasets, one can clearly find that $\Lambda$CDM is preferred over the IDE 
models and the result remains true irrespective of the observational datasets we use here. 

\begin{table*}
\begin{tabular}{ccccccccccccc}
\hline 

Data & $|\ln B_{0,1}|$ (IDE1) & $|\ln B_{0,1}|$ (IDE2) \\
\hline 
Planck 2018 & 1.90 & 1.66\\
Planck 2018+BAO & 1.45 & 2.58 \\
Planck 2018+DES & 0.75 & 1.27 \\
Planck 2018+R19 & 1.78 & 2.67\\
Planck 2018+BAO+R19 & 1.34 & 1.26\\

\hline 
\end{tabular}
\caption{We show the values of $|\ln B_{0,1}|$ for IDE1 and IDE2 obtained for all the observational datasets with respect to the reference model $\Lambda$CDM. }
\label{tab:SDDR}
\end{table*}

\section{Summary and conclusions}

\label{sec-conclu}

The cosmology with interaction between DM and DE has
gained a massive attention in the scientific community because the allowance
of an interaction in the dark sector could explain several cosmic
puzzles such as the cosmological constant problem, coincidence problem as
well as some recent observational issues, such as the $H_0$ tension, $\sigma_8$ tension,
and the crossing of the phantom barrier without any scalar field theory. 
However, a drawback of
the interaction models which we should permit is that the interaction rate is not
known, and we do not have any mechanism that could derive the exact rate of
interaction between the dark fluids. That means there is a pure liberty to
choose any kind of interaction model and at the same time there is no
bindings to favor any specific interaction model that have already been well
known for years. This has been the main motivation of this work where we
have shown that there is definitely no reason to go in favor of some
particular interaction model.

Thus, in this work, assuming a spatially flat FLRW universe, we allow an
interaction between DM and DE where the DM fluid
is pressureless and the DE component has a constant barotropic
equation of state. We then propose two new interaction rates in equations (%
\ref{Q1}) and (\ref{Q2}) that around $r = 1 $ if expanded, assume nonlinear
structure. 

We then constrained the interacting scenarios (i.e. IDE1 and IDE2) corresponding to the interaction functions \ref{Q1} and (\ref{Q2} respectively using CMB from Planck 2018, BAO, DES, and a local measurement of $H_0$ from HST (denoted by R19). In  Table \ref{tab:IDE1} (for IDE1) and Table \ref{tab:IDE2} (IDE2) we have summarized the results on the free and derived parameters of both the interacting scenarios. Our analyses clearly show that both IDE1 and IDE2 allow a nonzero interaction in the dark sector. Interestingly, for IDE1 both Planck 2018+BAO and Planck 2018+BAO+R19 show an evidence of $\xi \neq 0$ at more than $2\sigma$ while for IDE2, although $\xi  =0$ is consistent for Planck 2018+BAO but an evidence of $\xi \neq 0$ is exhibited for Planck 2018+BAO+R19. Additionally, we find that both the IDE models can alleviate the $H_0$ tension. In particular, IDE1  perfectly solves the $H_0$ tension within 68\% CL. Within IDE1 scenario, Planck 2018 data alone constrain  $H_0 = 72.67_{-    8.26}^{+    5.43}$ km/s/Mpc (at 68\% CL) which is very close to its local measurement $H_0 = 74.03 \pm 1.42$ km/s/Mpc (at $68\%$ CL)~\cite{Riess:2019cxk}.  This is an interesting property of the interacting dynamics which has
been also explored in some recent works  \cite{Kumar:2017dnp,DiValentino:2017iww, Yang:2018euj,Yang:2018uae, Kumar:2019wfs, Pan:2019gop}. In particular, in  \cite{Kumar:2019wfs, DiValentino:2019ffd}, the authors showed that an interaction in the dark sector is can simultaneously solve both $H_0$ and $\sigma_8$ tensions.    
Moreover, our analyses also report that the dark energy equation of state has a tendency towards the phantom regime. This tendency is strongly supported by Planck 2018+R19 and Planck 2018+BAO+R19 where $w_x< -1$ at more than $2\sigma$. 

We discussed the direct effects of the coupling parameter on the large scale structure of our Universe through the changes in the CMB TT and matter power spectra (see Figs. \ref{fig-ide1-TT-mower} and \ref{fig-ide2-TT-mower}). From the analyses we find that with the increase of the coupling parameter, IDE1 presents mild deviation from the non-interacting $\Lambda$CDM model compared to IDE2.   

As the models are completely new, we wanted to  compare them with some
known interaction models having both linear and nonlinear structures. Such investigations enable us to understand the qualitative 
differences between the IDE models already existing in the literature. 
Our analyses report that for small coupling parameter, the present models are extremely hard to distinguish from the known reference models (see Fig. \ref{fig-cmb-compare} and Fig. \ref{fig-mpower-compare}) although a mild difference always exists between them, see Fig. \ref{fig-cmb-compare-ratio} and Fig. \ref{fig-mpower-compare-ratio}. But the differences between the models are actually prominent for higher values of the coupling parameter.  But, overall, except IDE2 which has some violent nature in the large scales (only pronounced clearly from Fig. \ref{fig-mpower-compare-ratio}), IDE1 is very
close to the reference models. Therefore, based on the present analyses, IDE1 might be considered to be a competitor of the existing IDE models in the literature.

Finally, we performed the Bayesian evidence analysis through the
Savage-Dickey Density Ratio which is easy and useful to reduce the
computational efforts. In Table \ref{tab:SDDR} we have summarized the results 
of Bayesian evidence analysis considering $\Lambda$CDM as the reference model. Our analyses predict that still $\Lambda$CDM is preferred over the IDE models.

However, as a closing remark, we would like to comment that the present two interacting functions are really interesting. In particular, IDE1 driven by the sign changeable nonlinear interaction function should further be investigated in light of future cosmological observations.      
The most simplest and elegant work could  be its extension by considering the dynamical DE equation of state which extends the parameters space compared to the parameters space of the present interacting scenarios. Apart from that there are many other directions within this context that can be considered. We believe that the current scientific community might be interested to survey some of them along with us.

\section*{ACKNOWLEDGMENTS}
The authors thank the referee for some very important comments and suggestions that certainly helped to improve the quality of the manuscript.  SP was supported by the Mathematical Research Impact-Centric Support Scheme (MATRICS), File No. MTR/2018/000940, given by the Science and Engineering Research Board (SERB), Govt. of India.  WY acknowledges the financial support from the National Natural Science Foundation of China under Grants No. 11705079 and No. 11647153. We thank R. C. Nunes for some nice discussions and E. Di Valentino
for helping in the Bayesian evidence part.

\bibliographystyle{unsrt}
\bibliography{references.bib}

\begin{thebibliography}{10}

\bibitem{Wetterich-ide1}
Christof Wetterich.
\newblock {The Cosmon model for an asymptotically vanishing time dependent
  cosmological 'constant'}.
\newblock {\em Astron. Astrophys.}, 301:321--328, 1995.

\bibitem{Amendola-ide1}
Luca Amendola.
\newblock {Coupled quintessence}.
\newblock {\em Phys. Rev.}, D62:043511, 2000.

\bibitem{Billyard:2000bh}
Andrew~P. Billyard and Alan~A. Coley.
\newblock {Interactions in scalar field cosmology}.
\newblock {\em Phys. Rev.}, D61:083503, 2000.

\bibitem{Zimdahl:2001ar}
Winfried Zimdahl and Diego Pav\'{o}n.
\newblock {Interacting quintessence}.
\newblock {\em Phys. Lett.}, B521:133--138, 2001.

\bibitem{Amendola-ide2}
Luca Amendola and Claudia Quercellini.
\newblock {Tracking and coupled dark energy as seen by WMAP}.
\newblock {\em Phys. Rev.}, D68:023514, 2003.

\bibitem{Pavon:2005yx}
Diego Pav\'{o}n and Winfried Zimdahl.
\newblock {Holographic dark energy and cosmic coincidence}.
\newblock {\em Phys. Lett.}, B628:206--210, 2005.

\bibitem{Barrow:2006hia}
John~D. Barrow and T.~Clifton.
\newblock {Cosmologies with energy exchange}.
\newblock {\em Phys. Rev.}, D73:103520, 2006.

\bibitem{He:2008tn}
Jian-Hua He and Bin Wang.
\newblock {Effects of the interaction between dark energy and dark matter on
  cosmological parameters}.
\newblock {\em JCAP}, 0806:010, 2008.

\bibitem{Valiviita:2008iv}
Jussi Valiviita, Elisabetta Majerotto, and Roy Maartens.
\newblock {Instability in interacting dark energy and dark matter fluids}.
\newblock {\em JCAP}, 0807:020, 2008.

\bibitem{delCampo:2008sr}
Sergio del Campo, Ramon Herrera, and Diego Pav\'{o}n.
\newblock {Toward a solution of the coincidence problem}.
\newblock {\em Phys. Rev.}, D78:021302, 2008.

\bibitem{delCampo:2008jx}
Sergio del Campo, Ramon Herrera, and Diego Pav\'{o}n.
\newblock {Interacting models may be key to solve the cosmic coincidence
  problem}.
\newblock {\em JCAP}, 0901:020, 2009.

\bibitem{Gavela:2009cy}
M.~B. Gavela, D.~Hernandez, L.~Lopez~Honorez, O.~Mena, and S.~Rigolin.
\newblock {Dark coupling}.
\newblock {\em JCAP}, 0907:034, 2009.

\bibitem{Majerotto:2009np}
Elisabetta Majerotto, Jussi Valiviita, and Roy Maartens.
\newblock {Adiabatic initial conditions for perturbations in interacting dark
  energy models}.
\newblock {\em Mon. Not. Roy. Astron. Soc.}, 402:2344--2354, 2010.

\bibitem{Clemson:2011an}
Timothy Clemson, Kazuya Koyama, Gong-Bo Zhao, Roy Maartens, and Jussi
  Valiviita.
\newblock {Interacting Dark Energy -- constraints and degeneracies}.
\newblock {\em Phys. Rev.}, D85:043007, 2012.

\bibitem{Avelino:2012tc}
P.~P. Avelino and H.~M.~R. da~Silva.
\newblock {Effective dark energy equation of state in interacting dark energy
  models}.
\newblock {\em Phys. Lett.}, B714:6--10, 2012.

\bibitem{Harko:2012za}
Tiberiu Harko and Francisco S.~N. Lobo.
\newblock {Irreversible thermodynamic description of interacting dark
  energy-dark matter cosmological models}.
\newblock {\em Phys. Rev.}, D87(4):044018, 2013.

\bibitem{Sun:2013pda}
Cheng-Yi Sun and Rui-Hong Yue.
\newblock {Stable large-scale perturbations in interacting dark-energy model}.
\newblock {\em JCAP}, 1308:018, 2013.

\bibitem{Pan:2013rha}
Supriya Pan and Subenoy Chakraborty.
\newblock {Will there be again a transition from acceleration to deceleration
  in course of the dark energy evolution of the universe?}
\newblock {\em Eur. Phys. J.}, C73:2575, 2013.

\bibitem{Yang:2014gza}
Weiqiang Yang and Lixin Xu.
\newblock {Cosmological constraints on interacting dark energy with
  redshift-space distortion after Planck data}.
\newblock {\em Phys. Rev.}, D89(8):083517, 2014.

\bibitem{yang:2014vza}
Weiqiang Yang and Lixin Xu.
\newblock {Testing coupled dark energy with large scale structure observation}.
\newblock {\em JCAP}, 1408:034, 2014.

\bibitem{Nunes:2014qoa}
Rafael~C. Nunes and Edesio~M. Barboza.
\newblock {Dark matter-dark energy interaction for a time-dependent EoS
  parameter}.
\newblock {\em Gen. Rel. Grav.}, 46:1820, 2014.

\bibitem{Pan:2014afa}
Supriya Pan and Subenoy Chakraborty.
\newblock {A cosmographic analysis of holographic dark energy models}.
\newblock {\em Int. J. Mod. Phys.}, D23(11):1450092, 2014.

\bibitem{Yang:2014hea}
Weiqiang Yang and Lixin Xu.
\newblock {Coupled dark energy with perturbed Hubble expansion rate}.
\newblock {\em Phys. Rev.}, D90(8):083532, 2014.

\bibitem{Pan:2012ki}
Supriya Pan, Subhra Bhattacharya, and Subenoy Chakraborty.
\newblock {An analytic model for interacting dark energy and its observational
  constraints}.
\newblock {\em Mon. Not. Roy. Astron. Soc.}, 452(3):3038--3046, 2015.

\bibitem{Nunes:2016dlj}
Rafael~C. Nunes, Supriya Pan, and Emmanuel~N. Saridakis.
\newblock {New constraints on interacting dark energy from cosmic
  chronometers}.
\newblock {\em Phys. Rev.}, D94(2):023508, 2016.

\bibitem{Kumar:2016zpg}
Suresh Kumar and Rafael~C. Nunes.
\newblock {Probing the interaction between dark matter and dark energy in the
  presence of massive neutrinos}.
\newblock {\em Phys. Rev.}, D94(12):123511, 2016.

\bibitem{Yang:2016evp}
Weiqiang Yang, Hang Li, Yabo Wu, and Jianbo Lu.
\newblock {Cosmological constraints on coupled dark energy}.
\newblock {\em JCAP}, 1610(10):007, 2016.

\bibitem{Caprini:2016qxs}
Chiara Caprini and Nicola Tamanini.
\newblock {Constraining early and interacting dark energy with gravitational
  wave standard sirens: the potential of the eLISA mission}.
\newblock {\em JCAP}, 1610(10):006, 2016.

\bibitem{Pan:2016ngu}
S.~Pan and G.~S. Sharov.
\newblock {A model with interaction of dark components and recent observational
  data}.
\newblock {\em Mon. Not. Roy. Astron. Soc.}, 472(4):4736--4749, 2017.

\bibitem{Sharov:2017iue}
German~S. Sharov, Subhra Bhattacharya, Supriya Pan, Rafael~C. Nunes, and
  Subenoy Chakraborty.
\newblock {A new interacting two fluid model and its consequences}.
\newblock {\em Mon. Not. Roy. Astron. Soc.}, 466(3):3497--3506, 2017.

\bibitem{Kumar:2017dnp}
Suresh Kumar and Rafael~C. Nunes.
\newblock {Echo of interactions in the dark sector}.
\newblock {\em Phys. Rev. D}, 96(10):103511, 2017.

\bibitem{Yang:2017yme}
Weiqiang Yang, Narayan Banerjee, and Supriya Pan.
\newblock {Constraining a dark matter and dark energy interaction scenario with
  a dynamical equation of state}.
\newblock {\em Phys. Rev. D}, 95(12):123527, 2017.
\newblock [Addendum: Phys. Rev.D96,no.8,089903(2017)].

\bibitem{Yang:2017ccc}
Weiqiang Yang, Supriya Pan, and David~F. Mota.
\newblock {Novel approach toward the large-scale stable interacting dark-energy
  models and their astronomical bounds}.
\newblock {\em Phys. Rev. D}, 96(12):123508, 2017.

\bibitem{Shahalam:2017fqt}
M.~Shahalam, S.~D. Pathak, Shiyuan Li, R.~Myrzakulov, and Anzhong Wang.
\newblock {Dynamics of coupled phantom and tachyon fields}.
\newblock {\em Eur. Phys. J.}, C77(10):686, 2017.

\bibitem{Kumar:2017bpv}
Suresh Kumar and Rafael~C. Nunes.
\newblock {Observational constraints on dark matter–dark energy scattering
  cross section}.
\newblock {\em Eur. Phys. J.}, C77(11):734, 2017.

\bibitem{DiValentino:2017iww}
Eleonora Di~Valentino, Alessandro Melchiorri, and Olga Mena.
\newblock {Can interacting dark energy solve the $H_0$ tension?}
\newblock {\em Phys. Rev.}, D96(4):043503, 2017.

\bibitem{Yang:2017zjs}
Weiqiang Yang, Supriya Pan, and John~D. Barrow.
\newblock {Large-scale Stability and Astronomical Constraints for Coupled
  Dark-Energy Models}.
\newblock {\em Phys. Rev. D}, 97(4):043529, 2018.

\bibitem{Pan:2017ent}
Supriya Pan, Ankan Mukherjee, and Narayan Banerjee.
\newblock {Astronomical bounds on a cosmological model allowing a general
  interaction in the dark sector}.
\newblock {\em Mon. Not. Roy. Astron. Soc.}, 477(1):1189--1205, 2018.

\bibitem{Yang:2018euj}
Weiqiang Yang, Supriya Pan, Eleonora Di~Valentino, Rafael~C. Nunes, Sunny
  Vagnozzi, and David~F. Mota.
\newblock {Tale of stable interacting dark energy, observational signatures,
  and the $H_0$ tension}.
\newblock {\em JCAP}, 1809(09):019, 2018.

\bibitem{Yang:2018xlt}
Weiqiang Yang, Supriya Pan, Ramón Herrera, and Subenoy Chakraborty.
\newblock {Large-scale (in) stability analysis of an exactly solved coupled
  dark-energy model}.
\newblock {\em Phys. Rev.}, D98(4):043517, 2018.

\bibitem{Yang:2018uae}
Weiqiang Yang, Ankan Mukherjee, Eleonora Di~Valentino, and Supriya Pan.
\newblock {Interacting dark energy with time varying equation of state and the
  $H_0$ tension}.
\newblock {\em Phys. Rev.}, D98(12):123527, 2018.

\bibitem{Yang:2018ubt}
Weiqiang Yang, Supriya Pan, Lixin Xu, and David~F. Mota.
\newblock {Effects of anisotropic stress in interacting dark matter-dark energy
  scenarios}.
\newblock {\em Mon. Not. Roy. Astron. Soc.}, 482(2):1858--1871, 2019.

\bibitem{Yang:2018pej}
Weiqiang Yang, Supriya Pan, and Andronikos Paliathanasis.
\newblock {Cosmological constraints on an exponential interaction in the dark
  sector}.
\newblock {\em Mon. Not. Roy. Astron. Soc.}, 482(1):1007--1016, 2019.

\bibitem{vonMarttens:2018iav}
R.~von Marttens, L.~Casarini, D.~F. Mota, and W.~Zimdahl.
\newblock {Cosmological constraints on parametrized interacting dark energy}.
\newblock {\em Phys. Dark Univ.}, 23:100248, 2019.

\bibitem{vonMarttens:2018bvz}
Rodrigo von Marttens, Valerio Marra, Luciano Casarini, J.~E. Gonzalez, and
  Jailson Alcaniz.
\newblock {Null test for interactions in the dark sector}.
\newblock {\em Phys. Rev.}, D99(4):043521, 2019.

\bibitem{Paliathanasis:2019hbi}
Andronikos Paliathanasis, Supriya Pan, and Weiqiang Yang.
\newblock {Dynamics of nonlinear interacting dark energy models}.
\newblock {\em Int. J. Mod. Phys.}, D28(12):1950161, 2019.

\bibitem{Yang:2019bpr}
Weiqiang Yang, Supriya Pan, Eleonora Di~Valentino, Bin Wang, and Anzhong Wang.
\newblock {Forecasting Interacting Vacuum-Energy Models using Gravitational
  Waves}.
\newblock {\em 1904.11980}, 2019.

\bibitem{Yang:2019vni}
Weiqiang Yang, Sunny Vagnozzi, Eleonora Di~Valentino, Rafael~C. Nunes, Supriya
  Pan, and David~F. Mota.
\newblock {Listening to the sound of dark sector interactions with
  gravitational wave standard sirens}.
\newblock {\em JCAP}, 1907(07):037, 2019.

\bibitem{Yang:2019uzo}
Weiqiang Yang, Olga Mena, Supriya Pan, and Eleonora Di~Valentino.
\newblock {Dark sectors with dynamical coupling}.
\newblock {\em Phys. Rev.}, D100(8):083509, 2019.

\bibitem{Pan:2019gop}
Supriya Pan, Weiqiang Yang, Eleonora Di~Valentino, Emmanuel~N. Saridakis, and
  Subenoy Chakraborty.
\newblock {Interacting scenarios with dynamical dark energy: Observational
  constraints and alleviation of the $H_0$ tension}.
\newblock {\em Phys. Rev.}, D100(10):103520, 2019.

\bibitem{Yang:2019uog}
Weiqiang Yang, Supriya Pan, Rafael~C. Nunes, and David~F. Mota.
\newblock {Dark calling Dark: Interaction in the dark sector in presence of
  neutrino properties after Planck CMB final release}.
\newblock {\em 1910.08821}, 2019.

\bibitem{DiValentino:2019ffd}
Eleonora Di~Valentino, Alessandro Melchiorri, Olga Mena, and Sunny Vagnozzi.
\newblock {Interacting dark energy after the latest Planck, DES, and $H_0$
  measurements: an excellent solution to the $H_0$ and cosmic shear tensions}.
\newblock {\em 1908.04281}, 2019.

\bibitem{DiValentino:2019jae}
Eleonora Di~Valentino, Alessandro Melchiorri, Olga Mena, and Sunny Vagnozzi.
\newblock {Non-minimal dark sector physics and cosmological tensions}.
\newblock {\em 1910.09853}, 2019.

\bibitem{vonMarttens:2019ixw}
Rodrigo von Marttens, Lucas Lombriser, Martin Kunz, Valerio Marra, Luciano
  Casarini, and Jailson Alcaniz.
\newblock {Dark degeneracy I: Dynamical or interacting dark energy?}
\newblock {\em 1911.02618}, 2019.

\bibitem{Papagiannopoulos:2019kar}
G.~Papagiannopoulos, Pavlina Tsiapi, Spyros Basilakos, and Andronikos
  Paliathanasis.
\newblock {Dynamics and cosmological evolution in {$\Lambda$}-varying
  cosmology}.
\newblock {\em 1911.12431}, 2019.

\bibitem{Bolotin:2013jpa}
Yu.~L. Bolotin, A.~Kostenko, O.~A. Lemets, and D.~A. Yerokhin.
\newblock {Cosmological Evolution With Interaction Between Dark Energy And Dark
  Matter}.
\newblock {\em Int. J. Mod. Phys.}, D24(03):1530007, 2015.

\bibitem{Wang:2016lxa}
B.~Wang, E.~Abdalla, F.~Atrio-Barandela, and D.~Pav\'{o}n.
\newblock {Dark Matter and Dark Energy Interactions: Theoretical Challenges,
  Cosmological Implications and Observational Signatures}.
\newblock {\em Rept. Prog. Phys.}, 79(9):096901, 2016.

\bibitem{Pan:2020zza}
Supriya Pan, German~S. Sharov, and Weiqiang Yang.
\newblock {Field theoretic interpretations of interacting dark energy scenarios
  and recent observations}.
\newblock {\em 2001.03120}, 2020.

\bibitem{Lewis:1999bs}
Antony Lewis, Anthony Challinor, and Anthony Lasenby.
\newblock {Efficient computation of CMB anisotropies in closed FRW models}.
\newblock {\em Astrophys. J.}, 538:473--476, 2000.

\bibitem{Lewis:2002ah}
Antony Lewis and Sarah Bridle.
\newblock {Cosmological parameters from CMB and other data: A Monte Carlo
  approach}.
\newblock {\em Phys. Rev.}, D66:103511, 2002.

\bibitem{Wei:2010cs}
Hao Wei.
\newblock {Cosmological Constraints on the Sign-Changeable Interactions}.
\newblock {\em Commun. Theor. Phys.}, 56:972--980, 2011.

\bibitem{Sun:2010vz}
Cheng-Yi Sun and Rui-Hong Yue.
\newblock {New Interaction between Dark Energy and Dark Matter Changes Sign
  during Cosmological Evolution}.
\newblock {\em Phys. Rev.}, D85:043010, 2012.

\bibitem{Guo:2017deu}
Juan-Juan Guo, Jing-Fei Zhang, Yun-He Li, Dong-Ze He, and Xin Zhang.
\newblock {Probing the sign-changeable interaction between dark energy and dark
  matter with current observations}.
\newblock {\em Sci. China Phys. Mech. Astron.}, 61(3):030011, 2018.

\bibitem{Arevalo:2019axj}
Fabiola Arevalo, Antonella Cid, Luis~P. Chimento, and Patricio Mella.
\newblock {On sign-changeable interaction in FLRW cosmology}.
\newblock {\em Eur. Phys. J.}, C79(4):355, 2019.

\bibitem{Pan:2019jqh}
Supriya Pan, Weiqiang Yang, Chiranjeeb Singha, and Emmanuel~N. Saridakis.
\newblock {Observational constraints on sign-changeable interaction models and
  alleviation of the $H_0$ tension}.
\newblock {\em Phys. Rev.}, D100(8):083539, 2019.

\bibitem{Mukhanov}
V.~F. Mukhanov, H.~A. Feldman, and R.~H. Brandenberger.
\newblock {\em Phys. Rept.}, 215:203, 1992.

\bibitem{Ma:1995ey}
Chung-Pei Ma and Edmund Bertschinger.
\newblock {Cosmological perturbation theory in the synchronous and conformal
  Newtonian gauges}.
\newblock {\em Astrophys. J.}, 455:7--25, 1995.

\bibitem{Malik:2008im}
Karim~A. Malik and David Wands.
\newblock {Cosmological perturbations}.
\newblock {\em Phys. Rept.}, 475:1--51, 2009.

\bibitem{Aghanim:2018oex}
N.~Aghanim et~al.
\newblock {Planck 2018 results. VIII. Gravitational lensing}.
\newblock {\em 1807.06210}, 2018.

\bibitem{Aghanim:2019ame}
N.~Aghanim et~al.
\newblock {Planck 2018 results. V. CMB power spectra and likelihoods}.
\newblock {\em 1907.12875}, 2019.

\bibitem{Beutler:2011hx}
Florian Beutler, Chris Blake, Matthew Colless, D.~Heath Jones, Lister
  Staveley-Smith, Lachlan Campbell, Quentin Parker, Will Saunders, and Fred
  Watson.
\newblock {The 6dF Galaxy Survey: Baryon Acoustic Oscillations and the Local
  Hubble Constant}.
\newblock {\em Mon. Not. Roy. Astron. Soc.}, 416:3017--3032, 2011.

\bibitem{Ross:2014qpa}
Ashley~J. Ross, Lado Samushia, Cullan Howlett, Will~J. Percival, Angela Burden,
  and Marc Manera.
\newblock {The clustering of the SDSS DR7 main Galaxy sample - I. A 4 per cent
  distance measure at $z = 0.15$}.
\newblock {\em Mon. Not. Roy. Astron. Soc.}, 449(1):835--847, 2015.

\bibitem{Alam:2016hwk}
Shadab Alam et~al.
\newblock {The clustering of galaxies in the completed SDSS-III Baryon
  Oscillation Spectroscopic Survey: cosmological analysis of the DR12 galaxy
  sample}.
\newblock {\em Mon. Not. Roy. Astron. Soc.}, 470(3):2617--2652, 2017.

\bibitem{Cheng:2019bkh}
Gong Cheng, Yinzhe Ma, Fengquan Wu, Jiajun Zhang, and Xuelei Chen.
\newblock {Testing interacting dark matter and dark energy model with
  cosmological data}.
\newblock {\em 1911.04520}, 2019.

\bibitem{Krause:2017ekm}
E.~Krause et~al.
\newblock {Dark Energy Survey Year 1 Results: Multi-Probe Methodology and
  Simulated Likelihood Analyses}.
\newblock {\em 1706.09359}, 2017.

\bibitem{Troxel:2017xyo}
M.~A. Troxel et~al.
\newblock {Dark Energy Survey Year 1 results: Cosmological constraints from
  cosmic shear}.
\newblock {\em Phys. Rev.}, D98(4):043528, 2018.

\bibitem{Abbott:2017wau}
T.~M.~C. Abbott et~al.
\newblock {Dark Energy Survey year 1 results: Cosmological constraints from
  galaxy clustering and weak lensing}.
\newblock {\em Phys. Rev.}, D98(4):043526, 2018.

\bibitem{Riess:2019cxk}
Adam~G. Riess, Stefano Casertano, Wenlong Yuan, Lucas~M. Macri, and Dan
  Scolnic.
\newblock {Large Magellanic Cloud Cepheid Standards Provide a 1\% Foundation
  for the Determination of the Hubble Constant and Stronger Evidence for
  Physics beyond $\Lambda$CDM}.
\newblock {\em Astrophys. J.}, 876(1):85, 2019.

\bibitem{Aghanim:2018eyx}
N.~Aghanim et~al.
\newblock {Planck 2018 results. VI. Cosmological parameters}.
\newblock {\em 1807.06209}, 2018.

\bibitem{sddr1}
R.~Trotta.
\newblock Bayes in the sky: Bayesian inference and model selection in
  cosmology.
\newblock {\em Contemp. Phys.}, 49:71, 2008.

\bibitem{sddr2}
R.~Trotta.
\newblock Applications of bayesian model selection to cosmological parameters.
\newblock {\em Mon. Not. R. Astron. Soc.}, 378:72, 2007.

\bibitem{sddr3}
I.~Verdinelli and L.~Wasserman.
\newblock Computing bayes factors using a generalization of the savage-dickey
  density ratio.
\newblock {\em J. Am. Stat. Assoc.}, 90:614, 1995.

\bibitem{Kumar:2019wfs}
Suresh Kumar, Rafael~C. Nunes, and Santosh~Kumar Yadav.
\newblock {Dark sector interaction: a remedy of the tensions between CMB and
  LSS data}.
\newblock {\em Eur. Phys. J.}, C79(7):576, 2019.

\end{thebibliography}

\end{document}